\documentclass[a4paper]{article}

\usepackage{graphicx}% Include figure files

\usepackage{soul}
\usepackage{graphicx}
\usepackage{epstopdf, epsfig}
\usepackage{amsmath}
\usepackage[table]{xcolor}
\usepackage{subcaption}
\usepackage{tabularx}
\usepackage{soul}
\usepackage{makecell}
\usepackage{multirow}

\newcommand{\soptitle}{On the Association of Kinematics, Spanwise Instability and Growth of Secondary Vortex Structures in the Wake of Oscillating Foils}

\colorlet{Purple}{blue!40!red}
\begin{document}

%\maketitle %\maketitle must follow title, authors, abstract and \pacs

%\preprint{}
%\renewcommand{\thefootnote}{\roman{footnote}}
%\title{On the association of kinematics, spanwise instability and growth of secondary vortex structures in the wake of oscillating foils} %Title of paper

%\title{On the association of kinematics, spanwise instability and growth of secondary vortex structures in the wake of oscillating foils\footnote{This work is under review in \textit{Proceedings of Royal Society A}}}

%\thispagestyle{empty}

\begin{center}
\Large \bf{\soptitle}
\vspace{0.1in}
%\normalsize{\textcolor{darkblue}{Muhammad Khalid, PhD}}
\end{center}

\begin{center}
{Suyash Verma$^1$, Muhammad Saif Ullah Khalid$^{1,2}$, and Arman Hemmati$^{1\star}$}\\
\vspace{0.1in}
\end{center}
\begin{center}
$^1$Department of Mechanical Engineering, University of Alberta, Edmonton, Alberta T6G 2R3, Canada\\
$^2$Department of Mechanical and Mechatronics Engineering, Lakehead University, Thunder Bay, ON P7B 5E1, Canada\\
\vspace{0.05in}
$^\star$\small{Corresponding Author, Email: arman.hemmati@ualberta.ca}
\end{center}

%Prepared by Mehran Masoumifar @ UAlberta 2021


%\email[]{mmasoumi@ualberta.ca}
%\thanks{}
%\affiliation{Department of Mechanical Engineering, University of Alberta, Edmonton, Alberta T6G 2R3, Canada %\\This line break forced with \textbackslash\textbackslash


\date{\today}

\begin{abstract}

Three-dimensional wake of an oscillating foil with combined heaving and pitching motion is numerically evaluated at a range of chord-based Strouhal number (0.32 $\le St_c \le$ 0.56) and phase offset (90$^\circ$ $\le \phi \le$ 270$^\circ$) at $Re=$ 8000. The changes in $\phi$ and $St_c$ reflect a unique route of transition in  mechanisms that govern the origin of spanwise instabilities and growth of secondary wake structures. At lower $St_c$, heave dominated kinematics demonstrates a strong secondary leading edge vortex ($LEV$) as the source of growing spanwise instability on the primary $LEV$, followed by an outflux of streamwise vorticity filaments from the secondary $LEV$. With increasing heave domination, the origin of stronger spanwise instability is governed by a counter-rotating trailing edge vortex ($TEV$) and $LEV$ that leads to growth of streamwise secondary structures. A decreasing heave domination ultimately coincides with an absence of strong $LEV$ undulations and secondary structures. The consistent transition routes are represented on a phase-space map, where a progression of spanwise instability and growth of secondary structures becomes evident within regimes of decreased heave domination. The increasing strength of circulation for the primary $LEV$, with increasing $St_c$, provides a crucial reasoning for this newly identified progression. 

\end{abstract}

%\pacs{}% insert suggested PACS numbers in braces on next line

\section{INTRODUCTION}
\label{Section:1}

Achieving fast decay of vortical structures is crucial in modern high-speed aircraft to reduce the risk of incidents during take-off or landing, while also improving stealth for tactical submarines through reduced noise propagation \cite{1,2,6}. Such aspects are also useful in technologies that improve the propulsive lift and thrust generation of bio-inspired robotic swimmers during marine surveillance \cite{37}, operations of conventional helicopter rotors  \cite{49}, and flight performance of micro aerial vehicles (MAV$s$) \cite{50}. In the past, vortex decay processes required fundamental understanding of their formation and advection through experiments. These studies conducted assessments on equal or unequal strength counter-rotating vortical pairs \cite{1,2,3,4}. The observations suggested novel interaction mechanisms of paired structures that eventually developed wavy undulations in spanwise direction, which were then characterized by a particular wavelength and periodicity \cite{2,4}. Detailed theoretical and evolution aspects of vortex instabilities had been considered, which helped out understanding disintegration of vortical structures in wakes \cite{5}. On a similar note, origins of wake instabilities garnered sufficient attention besides characterizing their spatio-temporal development stages during the evolution process \cite{8,7}. This further revealed intricate physical connection to the formation of secondary vortical structures as well \cite{8}, whose role in promoting wake three-dimensionality  was immense. Primarily, such secondary vortical structures had been largely characterized in wakes of stationary and oscillating bluff bodies including cylinders (with varying cross-sectional shapes) and hydrofoils or airfoils \cite{8,10,11,12,13,15}. Currently, association of secondary wake structures with prevailing spanwise instability on the primary vortex still require more attention, specifically for foils oscillating in complex multi-degree of freedom motion \cite{16}. This study extends our knowledge by characterizing the origins of secondary wake structures, including their inherent association with the governing spanwise instability and prescribed coupled kinematics of an oscillating foil.

Elliptic instability had been primarily associated with counter-rotating pairs of spanwise coherent structures (or rollers), either with equal or unequal strength \cite{1,2,4}. Experiments and numerical studies that modeled such instabilities focused on isolated pair of vortices, wherein the temporal evolution of spatial dislocations on rollers was quite vivid \cite{1,2,3,4}. Crow's linear stability model effectively described the nature of equal strength counter-rotating vortex pair \cite{22}, where the early growth phase of instability coincided with vortex pair merger at specific spanwise intervals. This subsequently led to vortex ring-like arrangement \cite{22}. This model was further extended to a system of multiple rollers, including pairs that constituted unequal strength counter-rotating spanwise coherent structures \cite{2}. Klien et al. \cite{23} assessed the wrapping mechanism for a weak vortex filament around the stronger paired roller, which eventually yielded a vortex loop. Stability analysis for paired vortices was conducted by Ortega and Savas \cite{24}, wherein, the qualitative observations supported the previous analysis of Klien et al. \cite{23}. A follow-up study also provided a quantitative assessment in terms of circulation strength ($\Gamma$), internal strain-field distribution, and evolution mechanisms of sinusoidal instability. This revealed spatial wavelength of lower magnitude \cite{2}, compared to Crow's estimation for equal strength vortex pair \cite{22}. Bristol et al. \cite{3} extended the analysis to a co-rotating vortex pair that depicted formation of vorticity bridges on account of the elliptic instability that subsequently contributed to vortex merger in the wake. 

Besides the mentioned fundamental studies on  spanwise instabilities of isolated vortex pairs,   their dominant contribution to wake three-dimensionality had been well observed for stationary bluff bodies, such as a circular cylinder \cite{7,9,10} or the blunt trailing edge airfoil \cite{12,25}. Experimental visualizations suggested that the growth of streamwise coherent structures, called ribs, were also associated with spanwise instability modes that possessed different spatio-temporal characteristics \cite{7}. Mode A, in case of stationary cylinder, was characterized by core vorticity outflux from the rollers that consequently led to tongue like dislocations in the wake \cite{7,8}. These later elongated into pairs of counter-rotating streamwise filaments with alternating periodicity and a spanwise wavelength ($\lambda_{z}$) equal to four times the cylinder diameter \cite{7}. Mode B, in contrast, featured streamwise filaments  characterized by $\lambda_{z}$ on the order of cylinder diameter \cite{7}. These thin strands were associated with the straining of  streamwise vorticity that existed in the braid region between two consecutive rollers \cite{7,9,10}. Mittal and Balachander \cite{8} highlighted the formation of secondary hairpin-like structures that also appeared as spatial dislocation on spanwise rollers, shed in the wake of stationary circular cylinder. Upon evolution, these dislocated formations elongated and formed rib pairs \cite{8}. These studies \cite{7,8} therefore confirmed that the intermediary structures played a pivotal role in governing three-dimensionality of the wake. Extension of this work on cylinders with varying cross-sections (i.e.square) also exhibited other instability modes \cite{11}, while characterizing their association with secondary vortex pairs. Even for the case of a blunt trailing edge ($BTE$) airfoil, experiments \cite{25} and computational \cite{12} studies utilizing Floquet stability analysis, highlighted the existence of instability modes (Mode S), whose spatio-temporal features of secondary wake structures were different compared to previously identified modes \cite{7}. Extensive knowledge and understanding obtained in cases of stationary bluff bodies motivated fluid dynamicists to expand the instability assessments onto rigid bodies that perform prescribed oscillations \cite{13,15,16,26,27,51,54}. These outlined different aspects of secondary wake structures and how they influenced the wake three-dimensional features.

Kinematics prescribed on foils generally included single degree of freedom oscillations in the form of either pitching or heaving \cite{13,15,26,28,29,30,32}. The aspects of three-dimensional wakes for such oscillating bodies were initially evaluated for a pitching foil with low aspect ratios \cite{28,29,30}. Visbal \cite{26} initially expanded on the secondary vorticity outflux observed near the leading edge of a heaving foil at high Reynolds numbers ($Re \approx$ 10$^4$). Existence of long and short wavelength instability modes and the associated secondary vortex arrangement were also examined \cite{13}, where the spatio-temporal features of modes differed in terms of spanwise wavelength and streamwise periodicity. Deng et al. \cite{13} performed Floquet analysis for the case of pitching foil that highlighted a direct correspondence of secondary vortex pair formation, in conjunction with asymmetric arrangement of primary rollers shed behind the foil. Sun et al. \cite{15} discussed features of Mode A, Mode B, and Mode S for a heaving foil that were characterized by $\lambda_{z}\approx$ 1.05, 0.2, and 0.39, respectively. Some studies also presented limited assessments on foils with combined pitching and heaving oscillations \cite{16,31}. Floquet stability analysis of Moriche et al. \cite{31} on a foil with combined heaving and pitching motion provided details on the three-dimensional wake transition, along with associated impacts on aerodynamic force generation. Although the observations showed minimal effects on forces, the simulations revealed that the onset of three-dimensionality on the infinite span oscillating wing was linked to the bending of trailing edge vortex ($TEV$) \cite{31}. Cheireghin et al. \cite{32} observed the existence of sinusoidal undulation on the shed $LEV$ filament in the wake of a high aspect ratio heaving swept wing. The origins, however, remained unclear and they were speculated to be either an instability of oscillating shear flow, mixing layer, or the vortex filament itself \cite{32}. It was also noted by Cheireghin et al. \cite{32} that increasing circulation of $LEV$ at high reduced frequencies ($k$) led to stronger deformations, which also coincided with significant effects on lift and bending moments. The presence of secondary structures and their association with the $LEV$ dynamics was, however, not evident in the study of Cheireghin et al.\cite{32}. Verma and Hemmati \cite{16} investigated the three-dimensional wake of an infinite span foil with combined heaving and pitching oscillations, at $Re_c =$ 8000, which confirmed that despite changes in spatial arrangement of rollers in the wake, spanwise instability on the leading and trailing edge vortical structures closely resembled cooperative instability \cite{1,2,3,4} of the counter-rotating or co-rotating vortex pairs \cite{16}. The wavelength ($\lambda_{z}$), estimated for $LEV$ filaments, were in strong agreement with previous experimental studies \cite{1,2,4}. Verma $\&$ Hemmati \cite{16} further reported that tongue-like vortex dislocations and conjoint horseshoe-hairpin secondary structures had an apparent association with the cooperative instability \cite{1,2,3,4}, in terms of similar magnitude for the estimated $\lambda_{z}$. While the presence of secondary wake structures provided few unique insights into wakes of an oscillating foil, discussion with respect to onset mechanisms of instability and their association to prescribed kinematics was fairly limited \cite{16}. Son et al. \cite{51} recently provided comparative assessments of $LEV$ dynamics on an oscillating finite aspect ratio wing, and an airfoil of infinite span, with prescribed heaving motion at $Re =$ 10$^4$. Their experimental and computational observations were in agreement with the hypothesis of Verma $\&$ Hemmati \cite{16}, in terms of the counter-rotating rollers being responsible for the resulting spanwise undulations of $LEV$ filament. Their study \cite{51} also found a dependence of $LEV$ spanwise undulations on the reduced frequency ($k$), which contributed to differences in circulation ratio of $LEV$ and $TEV$ within one oscillation cycle. This coincided with the occurrence of Crow's instability mechanism \cite{22} triggered by either a $LEV-TEV$ pair (at $k=$ 2 and 3), or an $LEV$ paired with its image $LEV$ structure that generated on account of vortex-foil interaction \cite{26} (at $k=$ 1) \cite{51}. A similar qualitative assessment of changes in $LEV$ instability features that were associated with strong vortex-foil interaction, with an increase in the aspect ratio ($A.R$) of the heaving foil, was recently completed by Hammer et al. \cite{54} at $Re =$ $2$x$10^5$. Despite this higher $Re$, the qualitative and quantitative observations were in agreement with instability aspects reported at lower $Re$ \cite{16,26,32,51}. The effects were dominant as $A.R$ increased from 4 to 8, but above 8, the changes were not noticeable. Although an association of $LEV$ instability with $k$ and $A.R$ of oscillating foil were apparent in the study of Son et al.\cite{51} and Hammer et al.\cite{54}, respectively, they were limited to heaving and a single motion or geometrical parameter (i.e. $k$ or $A.R$). This hints at the need for an adequate extension of this particular association to more complex prescribed kinematics, and its relationship to eventual growth of secondary wake structures. 

Here, we provide detailed insights on specific mechanisms that govern the onset of a spanwise instability on primary rollers, and its association with the growth of secondary wake structures. The work particularly expands on the results of Verma $\&$ Hemmati \cite{16,21}, by looking at a wider kinematic parameter space (i.e chord based Strouhal number, $St_c$, and phase offset between heave and pitch, $\phi$) for an oscillating foil in combined heaving and pitching motion. As identified in our previous study \cite{16}, spanwise undulations observed on rollers near the leading edge held key details for the induced spanwise instability, that consequently promoted the growth of secondary wake structures. Thus, to ensure that a strong $LEV$ generation is captured in our numerical study, we specifically focus on the oscillations that govern heave dominated coupled kinematics \cite{33}. A brief preliminary discussion was recently presented by Verma $\&$ Hemmati \cite{57}, which outlined the association of spanwise wake instability and the growth of secondary wake structures at $St_c=$ 0.32 and $\phi=$ 90$^\circ$. This paper elaborates their findings \cite{57}, while also revealing the effects on spanwise instability and secondary wake structures, as kinematics slowly undergoes transition and enters a pitch-dominated regime of foil motion. The methodology and problem description in  Section \ref{Section:2} provide further details on the prescribed kinematics, along with the numerical procedure employed in the current study. This is followed by a comprehensive discussion of novel findings in the Section \ref{Section:3}, and a summary of the current ongoing work in Section \ref{Section:4}. 

\section{Problem Description}
\label{Section:2}

{The flow around an infinitely span foil with a maximum thickness ($D$) to chord length ($c$) ratio of $D/c=0.1$ is examined numerically for a range of chord based ($St_{c} = fc/U_{\infty} = 0.32 - 0.56$) and amplitude based Strouhal numbers (0.05 $\le St_{A} \le$ 0.4). The foil cross-section shown in zoomed snippet of Figure \ref{fig:foil_geom} resembles a teardrop hydrofoil shape, which was used in recent experimental investigations \cite{34,35}. These studies focused on propulsive performance and scaling relations of underwater propulsors, and suggested that the teardrop foil geometry represented a simplified model for fish tail fins \cite{34}. The Reynolds number is $Re = U_{\infty} c/\nu =$ 8000, which is consistent with previous studies \cite{16} and agrees closely with the biological characteristics of swimming fish and bio-inspired flights \cite{36,37}. Here, $U_\infty$ and $\nu$ represent the freestream velocity and  fluid's kinematic viscosity, respectively. 
	\begin{figure}
		\centering
		\includegraphics[width=11.95cm,height=5.95cm]{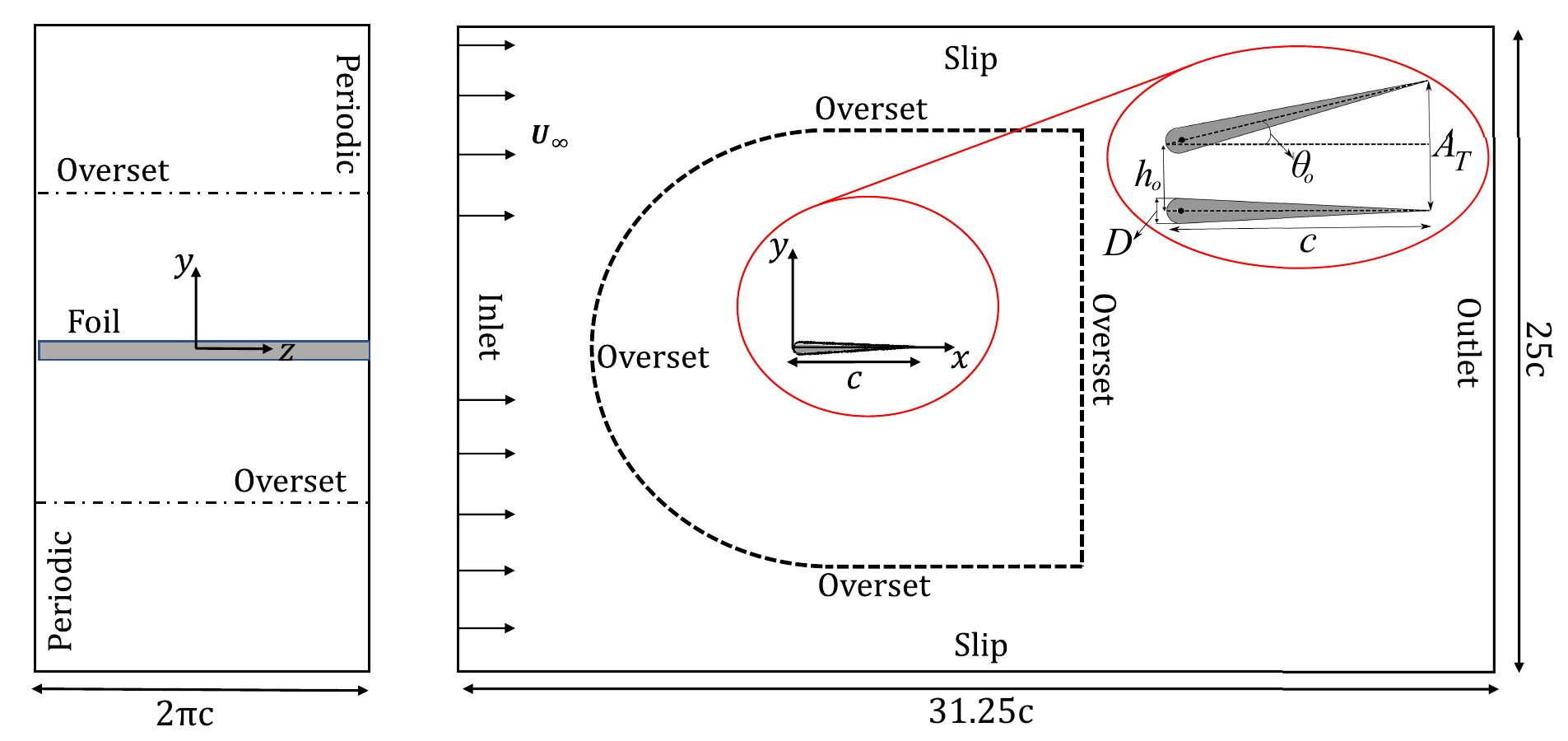}
		\caption{Schematic of the foil geometry and motion.}
		\label{fig:foil_geom}
	\end{figure} 
	The prescribed foil kinematics is characterized by a coupled heaving and pitching motion, where the location of pitch axis is approximately 0.05$c$ from the leading edge. Figure \ref{fig:foil_geom} marks the heave and pitch amplitudes as $h_o$ and $\theta_o$, respectively. The resultant trailing edge amplitude is also shown as $A_T$. The motion profiles of heave ($h$) and pitch ($\theta$), where pitching has a phase advancement (or offset) of $\phi$ relative to heaving, are represented using: 
	
	\begin{equation}\label{eq:ht}
	h(t)=h_{o} \sin (2 \pi f t)
	\end{equation}
	\begin{equation}\label{eq:theta_t}
	\theta(t)=\theta_{o} \sin (2 \pi f t+\phi)
	%\vspace{2mm}
	\end{equation} 
	
	%For this study, we have a fixed $h_o/c$ and $\theta_o$ equal to 0.25 and 10$^\circ$, respectively. 
	In order to present a broader association of spanwise instability, secondary wake structures and kinematics, we vary the phase offset ($\phi$) between heaving and pitching motion in the range of 90$^\circ$ and 270$^\circ$. This leads to changes in $A_T$ relative to a fixed $h_o/c$ ( = 0.25) and $\theta$ = (10$^\circ$) \cite{33}. $\phi$ below 225$^\circ$ are suggestive of a heave-dominated kinematics, based on the relatively higher $h_L$ compared to $A_T$ \cite{18}. At $\phi \ge$ 225$^\circ$, the oscillating foil kinematics starts transitioning towards pitch dominated motion settings \cite{18}. On account of variation in $\phi$ and $A_T$, Strouhal number ($St_A = 2fA_{T}/U_{\infty}$) is varied in the range specified earlier, where the maximum $St_{A}$ corresponds to both $\phi=$ 90$^\circ$ and 270$^\circ$, while the least $St_A$ is observed at $\phi=$ 180$^\circ$ \cite{33}. Anderson et al.\cite{36} indicated that significant transitions in the wake of flapping foils were observable at 0.2 $< St_{A} <$ 0.4. This also coincides with the range corresponding to optimal propulsive efficiency in swimming mammals \cite{37,38}. 
	In addition to above considerations, the most propulsively efficient phase offset is 90$^\circ$ according to Anderson et al.\cite{39}, which was also employed by Zheng et al.\cite{40} and Lagopoulos et al.\cite{41}. This phase offset corresponds to $\phi=$ 270$^\circ$ following the reference coordinate system employed by Van Buren et al.\cite{34}. %This study would follow a similar coordinate system as used in \cite{VanBuren2019}.
	
	\subsection{Numerical details}
	
	The continuity %(Eq. \ref{eq:con}) 
	and Navier-Stokes equations 
	%(Eq. \ref{eq:NS}) 
	were solved directly using OpenFOAM, which is a numerical package based on the finite-volume method. It is extensively used for simulating wake dynamics of oscillating foils and panels \cite{16,17,18,19,20,33,42,43,40,44}.
	%\begin{figure}
	%	\centering
	%	\hspace{-0.02in}	\includegraphics[width=11.95cm,height=5.95cm]{BC_1_New-eps-converted-to.pdf}%
	%	\caption{Schematics of the computational domain (not to scale) with boundary conditions.}
	%	\label{fig:Domain}
	%\end{figure}
	The oscillatory foil dynamics was modeled using Overset Grid Assembly (OGA) method, based on a stationary background grid and a moving overset grid that were merged for the simulation \cite{45}. Extensive details of the method can be found in Verma $\&$ Hemmati \cite{16,20}. 
	
	The computational domain is also presented in Figure \ref{fig:foil_geom}, which highlights the C-type overset boundary containing the foil. The boundary conditions at the inlet are prescribed a uniform fixed velocity (Dirichlet) and a zero normal gradient (Neumann) for pressure. At the outlet, a zero-gradient outflow boundary condition is implied \cite{13}. The top and bottom walls are further prescribed a slip boundary condition that effectively model open-channel or free-surface flows, and closely resemble the experimental and computational conditions of Van Buren et al.\cite{34} and Hemmati et al.\cite{43}, respectively. At the foil boundary, a no-slip condition for velocity and a zero-gradient condition for pressure is ensured. The periodic boundary condition is further implemented on the side boundaries, coinciding with the spanwise extent of the foil. This provides an effective way to model an infinitely span case without the end or tip effects. {The spanwise length of the foil ($\approx$ $\pi c$) follows the assessments conducted in recent studies \cite{55,57}, which showed imminent presence of three-dimensionality with a similar range of kinematics assumed here. A spanwise domain length analysis was further conducted here to establish the insensitivity of the wavelength for spanwise instability on the spanwise length of the computational domain.} Figure \ref{fig:Mesh} further exhibits non-homogeneous hexahedral grid, comprising of a C-type grid to model the foil geometry and motion. This region overlaps a rectangular background grid. Higher grid refinements in critical overlapping regions ensured accurate simulation of the near body vortex shedding and wake interactions. The grid further expands towards the domain boundaries with an expansion factor of less than $1.02$. The flow is solved using \textit{overPimpleDyMFoam} solver, which integrates the functionality of OGA and PIMPLE algorithms. Second order accurate backward and central difference schemes are employed for temporal and spatial discretizations, respectively \cite{42}. The momentum equations are solved using Preconditioned Biconjugate Gradient (\textit{PbiCGSTAB}) method, whereas Preconditioned Conjugate Gradient (\textit{PCG}) method is employed for the Poisson equation \cite{13,40}. The simulations are completed using the Cedar and Narval high performance clusters, operated by Digital Research Alliance of Canada. The parallel decomposition and assignment of computational domain utilizes 96 CPUs with a total of 190 GB memory and 1440 simulation hours per case.
	
	\begin{figure}
		\centering
		\hspace{-0.3in}	\includegraphics[width=9.95cm,height=5.05cm]{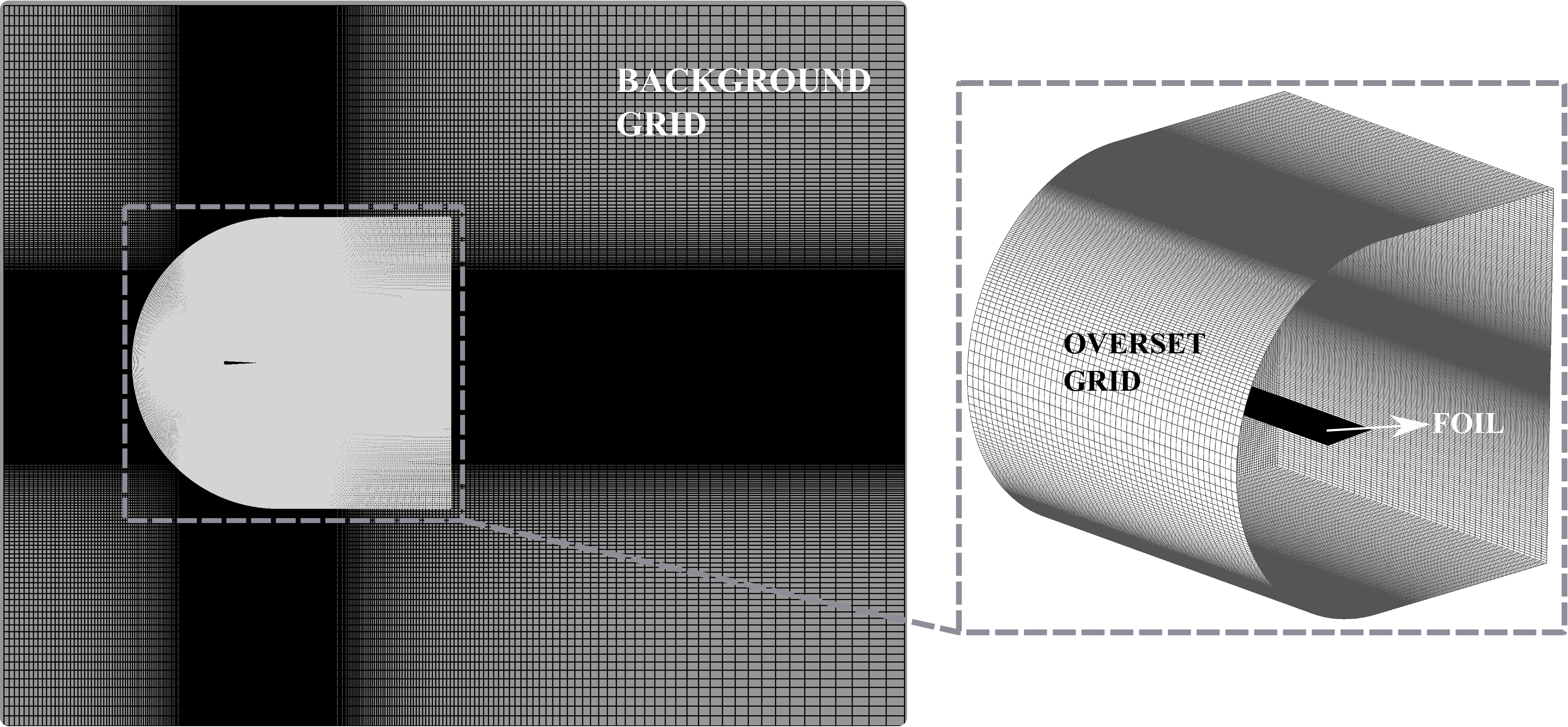}%
		\caption{Schematics of the grid and Overset Grid Assembly method.}
		\label{fig:Mesh}
	\end{figure}
	
	\begin{table}
		\caption{Grid refinement details for the current study. $N_{total}$ represents the sum of hexahedral elements in background grid and overset grid.}
		\begin{center}
			\def~{\hphantom{0}}
			\begin{tabular}{lccccc||cccc}
				Study  & $N_{total}$ & $\overline{C_{T}}$  & $C_{L}^{rms}$ & $\epsilon_T$ & $\epsilon_{L}^{rms}$ & $x=$ & $1c$ &  $2.5c$  &  $5c$\\[3pt]
				\hline
				Grid1 & $8.4\times10^6$ &  0.64 & 2.86 & 0.084 & 0.010 & $\delta^*=$ & 7.3 &  16.9 & 55.2 \\
				Grid2  & $1.7\times10^7$ & 0.60 & 2.84 & 0.017 & 0.003 & $\delta^*=$ & 3.7  & 8.5 & 15.8  \\
				Grid3  & $3.1\times10^7$ & 0.59 & 2.83 & 0.001 & - & $\delta^*=$ & 1.8 & 4.2 & 7.9  \\
				Exp.  & - & 0.59 & - & - & - & $\delta^*=$ & - & - & -  \\
			\end{tabular}
			\label{tab:grid_verification}
		\end{center}
	\end{table}
	
	A spatial convergence analysis is completed at $Re=$8000, $h_{o}/c=$0.25, $\theta_{o}=$15$^\circ$, $\phi=$270$^\circ$ and $St_c=$0.67. This enables comparative evaluation of the numerical results with respect to experiments of Van Buren et al.\cite{34}. 
	Table \ref{tab:grid_verification} and Figure \ref{fig:Grid_Convergence} summarize the grid convergence results involving three grids, Grid1, Grid2 and Grid3. The ratio ($\delta^*$) of minimum grid size element ($\Delta x$) to Kolmogorov scale ($\eta$) is kept approximately below 10, within the critical region near the foil ($x <$ 2.5$c$), specifically for Grid2 and Grid3 (see Table \ref{tab:grid_verification}), where the origin of spanwise instability and secondary structures are speculated to emerge and grow as the wake evolves \cite{16,51}. Here, Kolmogorov scale is calculated based on kinematic viscosity ($\nu$) and the dissipation rate of turbulence kinetic energy ($\epsilon$), computed by $\eta\approx (\nu^3/\epsilon)^{0.25}$ \cite{46}. The coefficient of thrust ($\overline{C_T}=\bar{T} /0.5 \rho s c U_{\infty}^{2}$) and root-mean-square ($rms$) of lift coefficient ($C_L= L /0.5 \rho s c U_{\infty}^{2}$), which is averaged over the final 10 oscillating cycles following the statistical convergence, are used as a quantitative estimate for spatial grid convergence. The relative error in prediction of $\overline{C_{T}}$ ($\epsilon_T=|\overline{C_{T}}_{,exp}-\overline{C_{T}}| / \overline{C_{T}}_{,exp}$), calculated with respect to the experimental results \cite{34}, was below 5$\%$ for Grid2. Similarly, $\epsilon_{L}^{rms} (=|C_{L,Grid3}^{rms}-C_{L}^{rms}| / C_{L,Grid3}^{rms})$, calculated with respect to the finest grid (Grid3) was below 0.1$\%$. The corresponding experimental results for $C_L$ are not yet available. 
	\begin{figure}
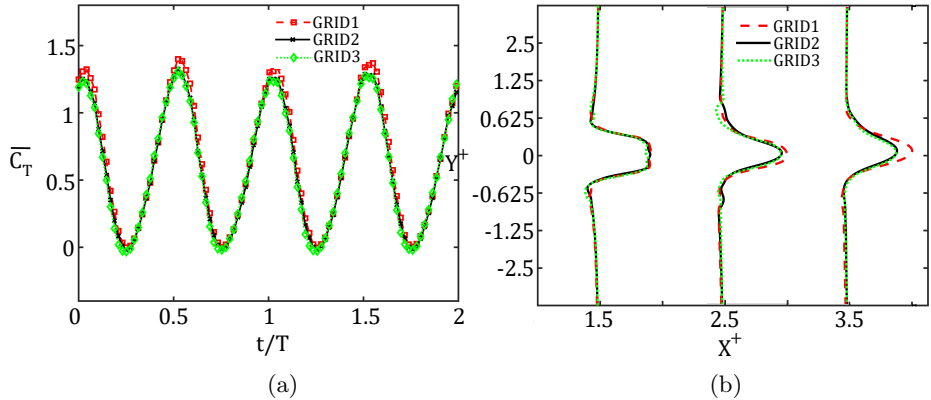

		\centering
		\begin{minipage}{0.47\textwidth}
			\centering
			\subcaptionbox{\hspace*{-2.75em}}{%
				\hspace{0.0in}	\includegraphics[width=6.5cm,height=5.0cm]{Grid_Con1_New-eps-converted-to.pdf}%
			}\qquad
		\end{minipage}\hfill
		\begin{minipage}{0.47\textwidth}
			\centering
			\subcaptionbox{\hspace*{-0.3em}}{%
				\hspace{-0.30in}	\includegraphics[width=6.75cm,height=5.0cm]{Grid_Con2_New-eps-converted-to.pdf}
			}\qquad
		\end{minipage}\\
		\caption{Comparison of (a) unsteady variation of $C_T$, and (b) cross-stream velocity profiles at increasing streamwise distance ($x^+$) for three grids.}
		\label{fig:Grid_Convergence}
	\end{figure}
	%\begin{figure}
	%	\centering
	%	\begin{minipage}{0.47\textwidth}
	%		\centering
	%		\subcaptionbox{\hspace*{-2.75em}}{%
	%			\hspace{0.0in}	\includegraphics[width=6.5cm,height=5.25cm]{Validate_1_New-eps-converted-to.pdf}%
	%		}\qquad
	%	\end{minipage}\hfill
	%	\begin{minipage}{0.47\textwidth}
	%		\centering
	%		\subcaptionbox{\hspace*{-0.3em}}{%
	%			\hspace{-0.25in}	\includegraphics[width=6.75cm,height=5.25cm]{Validate_2_New-eps-converted-to.pdf}
	%		}\qquad
	%	\end{minipage}\\
	%	\caption{Comparison of (a) numerically obtained variation of $\overline{C_{T}}$ and (b) unsteady contributions of transverse force and pitch moment to the total input power of the oscillating foil with experiments of \cite{34}.}
	%	\label{fig:Validate_Exp}
	%\end{figure}
	Figure \ref{fig:Grid_Convergence}(a) shows the instantaneous $C_T$ for the three grids, which demonstrates the insensitivity of propulsive performance to the spatial grid resolution. Similar agreement also holds in terms of wake characteristics, where the cross-stream distribution of mean streamwise velocity ($\overline{U_{x}^+}$) at increasing streamwise distance $(X^+)$ (Figure \ref{fig:Grid_Convergence}b) are relatively similar for successively refined grids. This provides sufficient confidence to use Grid2 for our analysis. The time-step size ($\delta t$) is also set in agreement with Verma $\&$ Hemmati\cite{16}. The ratio of eddy turnover time ($\tau_{\eta}$) to the set $\delta t$ for the smallest dissipative eddy \cite{46} yield at least 100 time-steps, which thus ensures that our simulations have adequate resolution to capture essential turbulent characteristics \cite{47}. Extensive details for verification and validation of the numerical solver with respect to the domain size, spatial and temporal grid, OGA solver and boundary conditions can be found in Hemmati et al.\cite{43} and Verma $\&$ Hemmati\cite{16,20}. %We find close agreement of computationally predicted $\overline{C_{T}}$ with experiments \cite{34} for 0$<h_{o}/c<$0.25, $\theta_{o}=$15$^\circ$, $\phi=$270$^\circ$ and $St_c=$0.67 in Figure \ref{fig:Validate_Exp}(a). Figure \ref{fig:Validate_Exp}(b) shows the comparison of instantaneous contribution of the transverse force ($P_{F_Y}$) and pitch moment ($P_{M_Z}$) to the total power ($P_o$) in one oscillation cycle in our numerical simulations and experiments \cite{34}, respectively. 
	%Sample equations.
	%
	%%%% Numbered equation
	%\begin{align}\label{1.1}
	%\begin{split}
	%\frac{\partial u(t,x)}{\partial t} &= Au(t,x) \left(1-\frac{u(t,x)}{K}\right)-B\frac{u(t-\tau,x) w(t,x)}{1+Eu(t-\tau,x)},\\
	%\frac{\partial w(t,x)}{\partial t} &=\delta \frac{\partial^2w(t,x)}{\partial x^2}-Cw(t,x)+D\frac{u(t-\tau,x)w(t,x)}{1+Eu(t-\tau,x)},
	%\end{split}
	%\end{align}
	%
	%\begin{align}\label{1.2}
	%\begin{split}
	%\frac{dU}{dt} &=\alpha U(t)(\gamma -U(t))-\frac{U(t-\tau)W(t)}{1+U(t-\tau)},\\
	%\frac{dW}{dt} &=-W(t)+\beta\frac{U(t-\tau)W(t)}{1+U(t-\tau)}.
	%\end{split}
	%\end{align}
	%
	%%%%% Unnumbered equation
	%\begin{eqnarray}
	%\frac{\partial(F_1,F_2)}{\partial(c,\omega)}_{(c_0,\omega_0)} = \left|
	%\begin{array}{ll}
	%\frac{\partial F_1}{\partial c} &\frac{\partial F_1}{\partial \omega} \\\noalign{\vskip3pt}
	%\frac{\partial F_2}{\partial c}&\frac{\partial F_2}{\partial \omega}
	%\end{array}\right|_{(c_0,\omega_0)}\notag\\
	%=-4c_0q\omega_0 -4c_0\omega_0p^2 =-4c_0\omega_0(q+p^2)>0.
	%\end{eqnarray}
	
	%%%%%%%%%%%%%%% End of first page %%%%%%%%%%%%%%%%%%%%%

\section{RESULTS AND DISCUSSION}
\label{Section:3}

We begin with elucidating the onset of spanwise instability on the $LEV$s and associated growth of secondary wake structures at a low chord-based Strouhal number ($St_c =$ 0.32), while $\phi$ varies from 90$^\circ$ to 270$^\circ$. The transition in growth mechanisms that leads to the onset of secondary wake structures is described, which coincide with the changes of kinematics from heave to pitch domination. Moreover, the progression of secondary structure growth at kinematics governed by onset of pitch domination ($\approx$270$^\circ$) is further evaluated at an increasing $St_c$. Quantitative assessments of circulation strength of primary $LEV$s reveals physical reasoning behind the observed progression. 

\subsection{Transition in growth mechanisms of secondary wake structures}

\begin{figure}
	\centering
	\begin{minipage}{0.5\textwidth}
		\centering
		\hspace{-0.15in}\includegraphics[width=5.0cm,height=1.0cm]{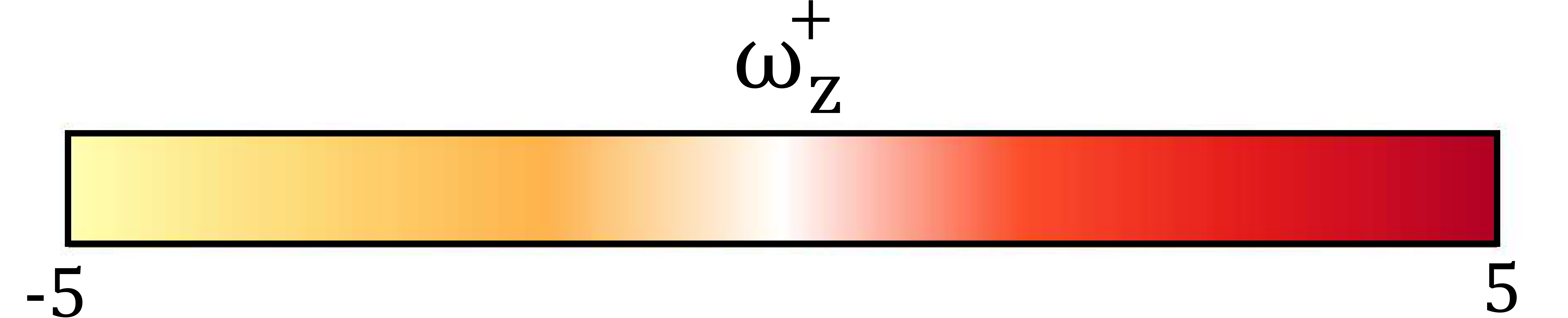}%
	\end{minipage}\\
	\vspace{0.1in}
	\begin{minipage}{0.25\textwidth}
		\centering
		\subcaptionbox{\hspace*{-2.75em}}{%
			\hspace{-0.05in}\includegraphics[width=3.5cm,height=3.4cm]{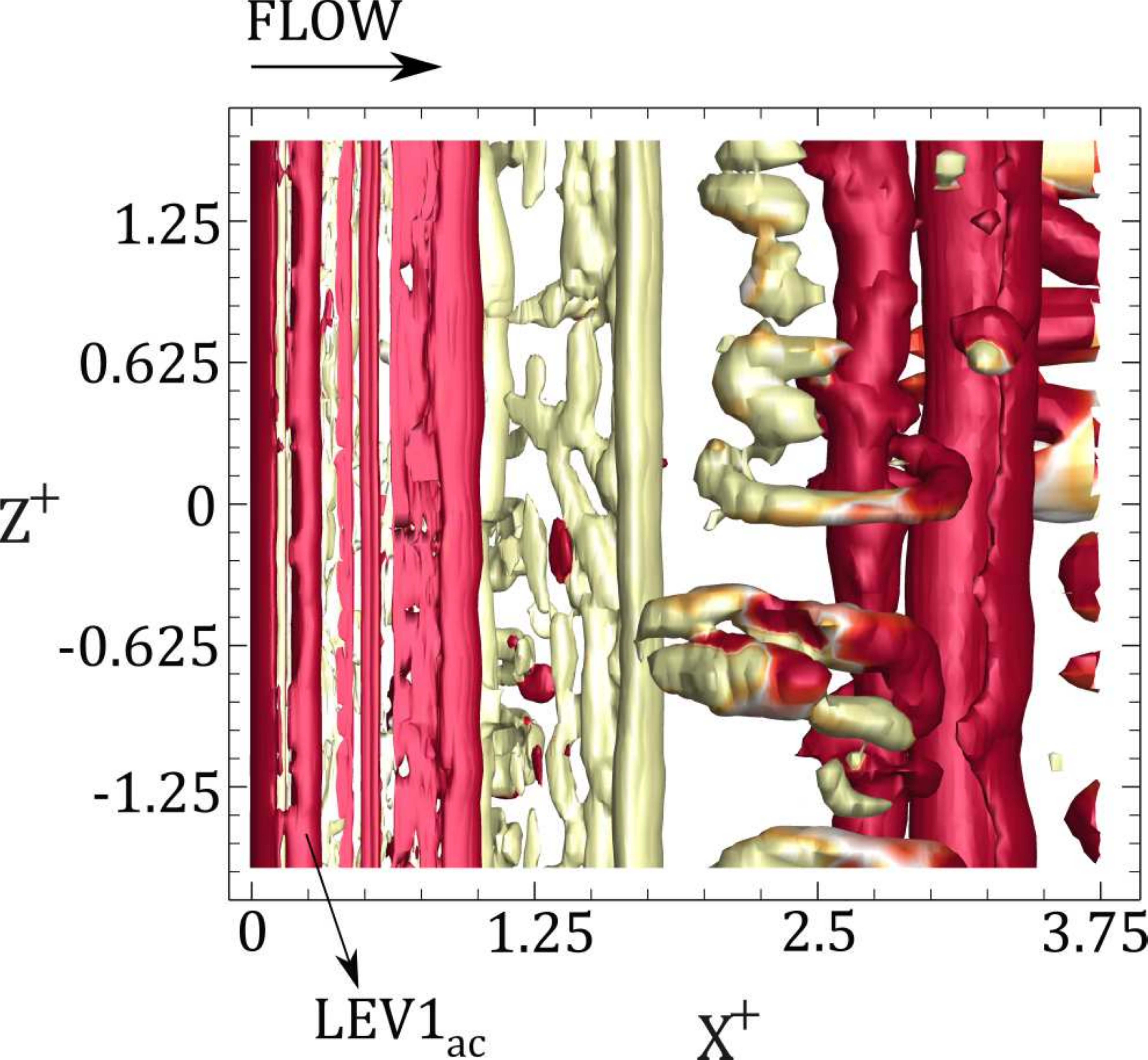}%
		}
	\end{minipage}\hfill
	\begin{minipage}{0.25\textwidth}
		\centering
		\subcaptionbox{\hspace*{-2.75em}}{%
			\hspace{-0.00in}\includegraphics[width=3.2cm,height=3.4cm]{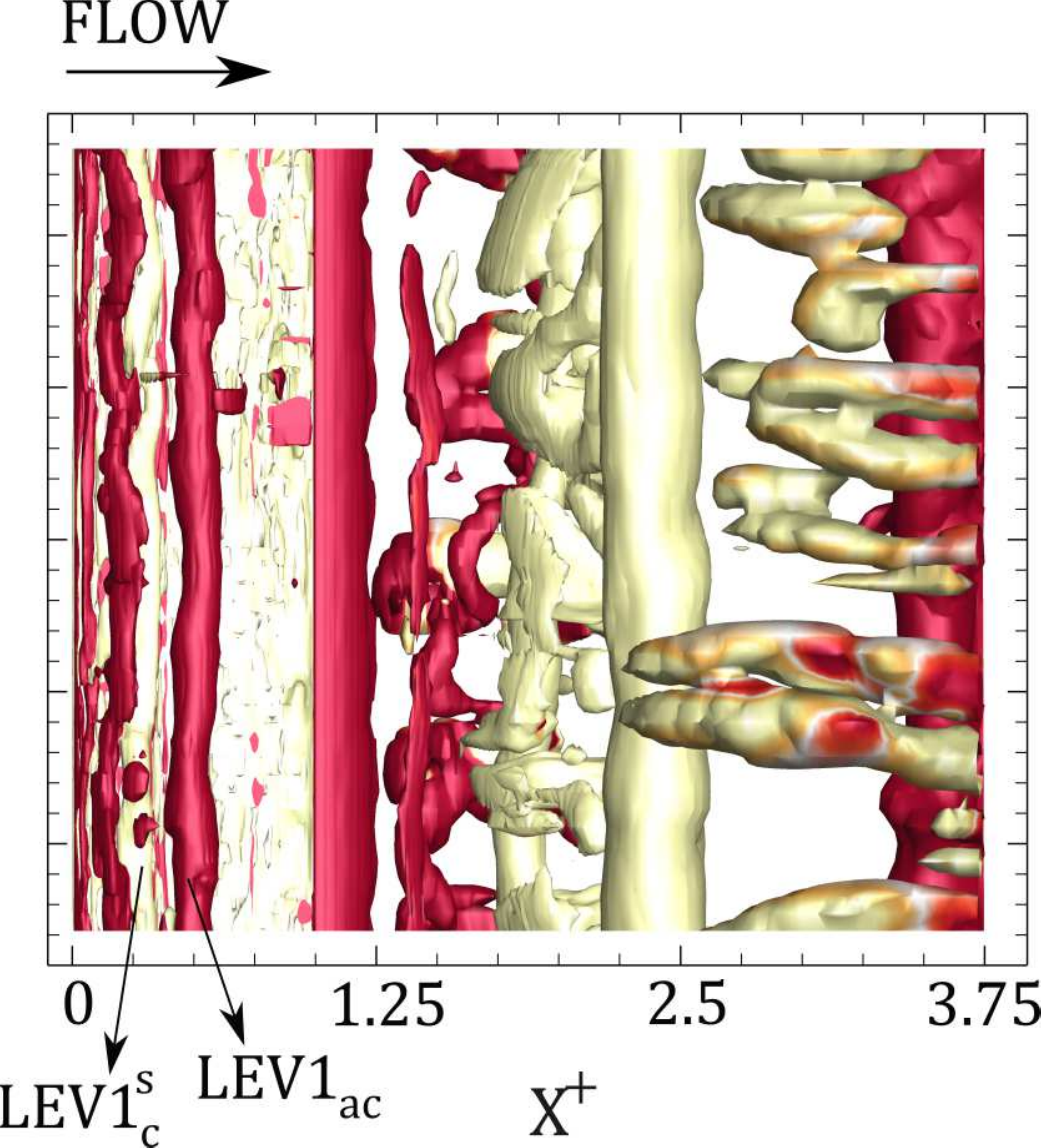}%
		}
	\end{minipage}\hfill
	\begin{minipage}{0.25\textwidth}
		\centering
		\subcaptionbox{\hspace*{-2.75em}}{%
			\hspace{-0.05in}\includegraphics[width=3.2cm,height=3.4cm]{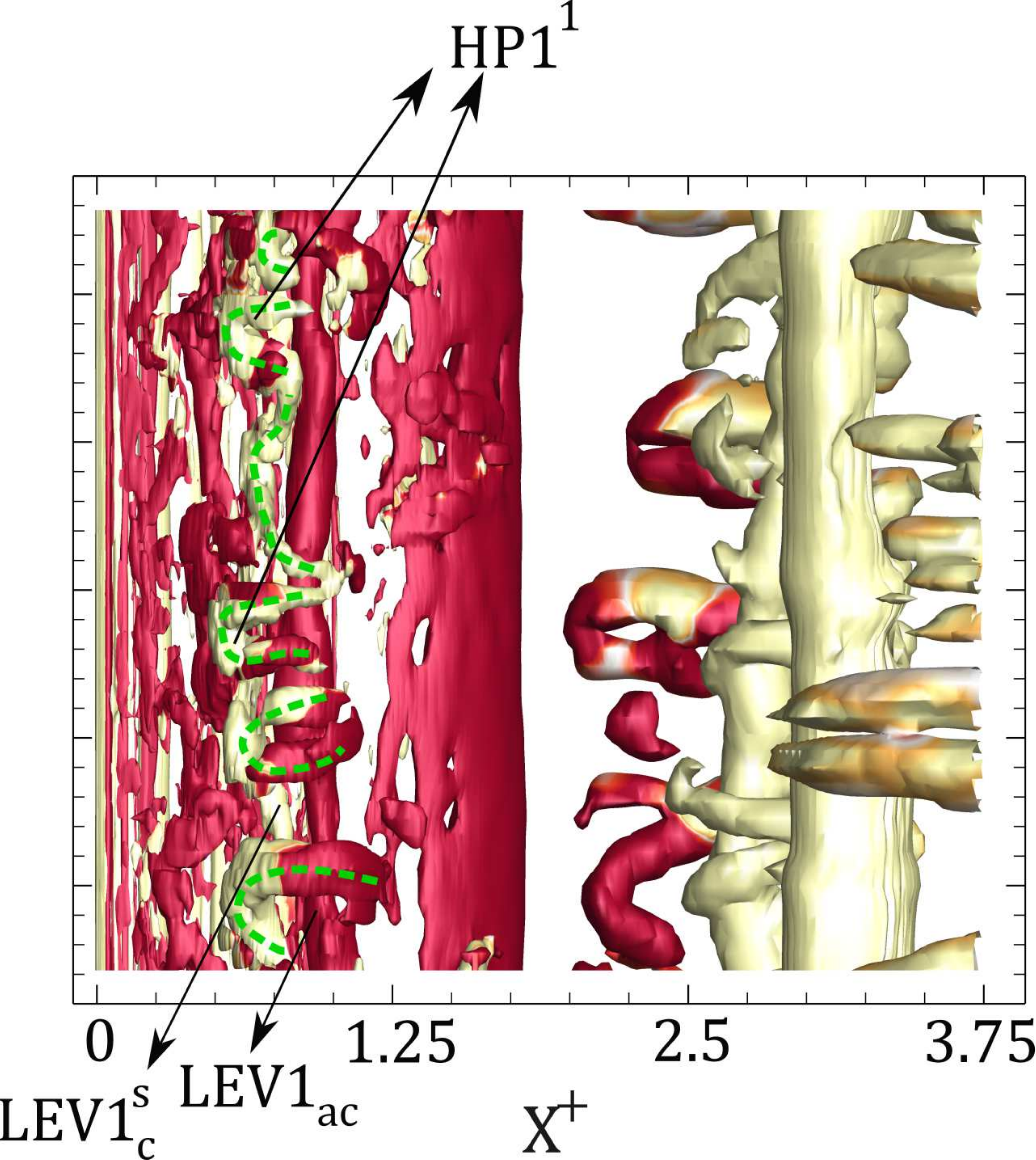}%
		}
	\end{minipage}\hfill
	\begin{minipage}{0.25\textwidth}
		\centering
		\subcaptionbox{\hspace*{-2.75em}}{%
			\hspace{-0.05in}\includegraphics[width=3.2cm,height=3.4cm]{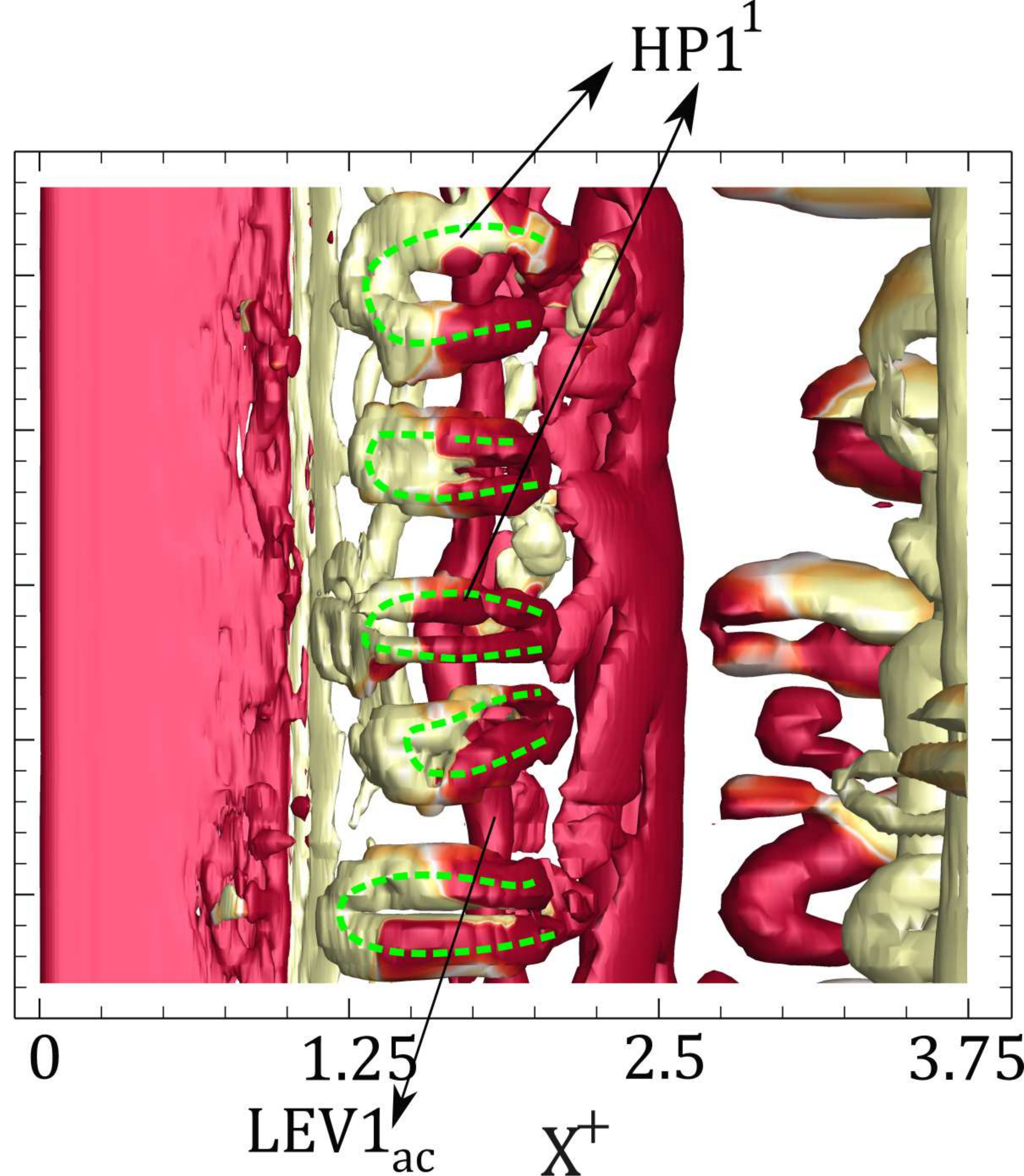}%
		}
	\end{minipage}\\
	\begin{minipage}{0.25\textwidth}
		\centering
		\subcaptionbox{\hspace*{-2.75em}}{%
			\hspace{-0.05in}\includegraphics[width=3.5cm,height=3.4cm]{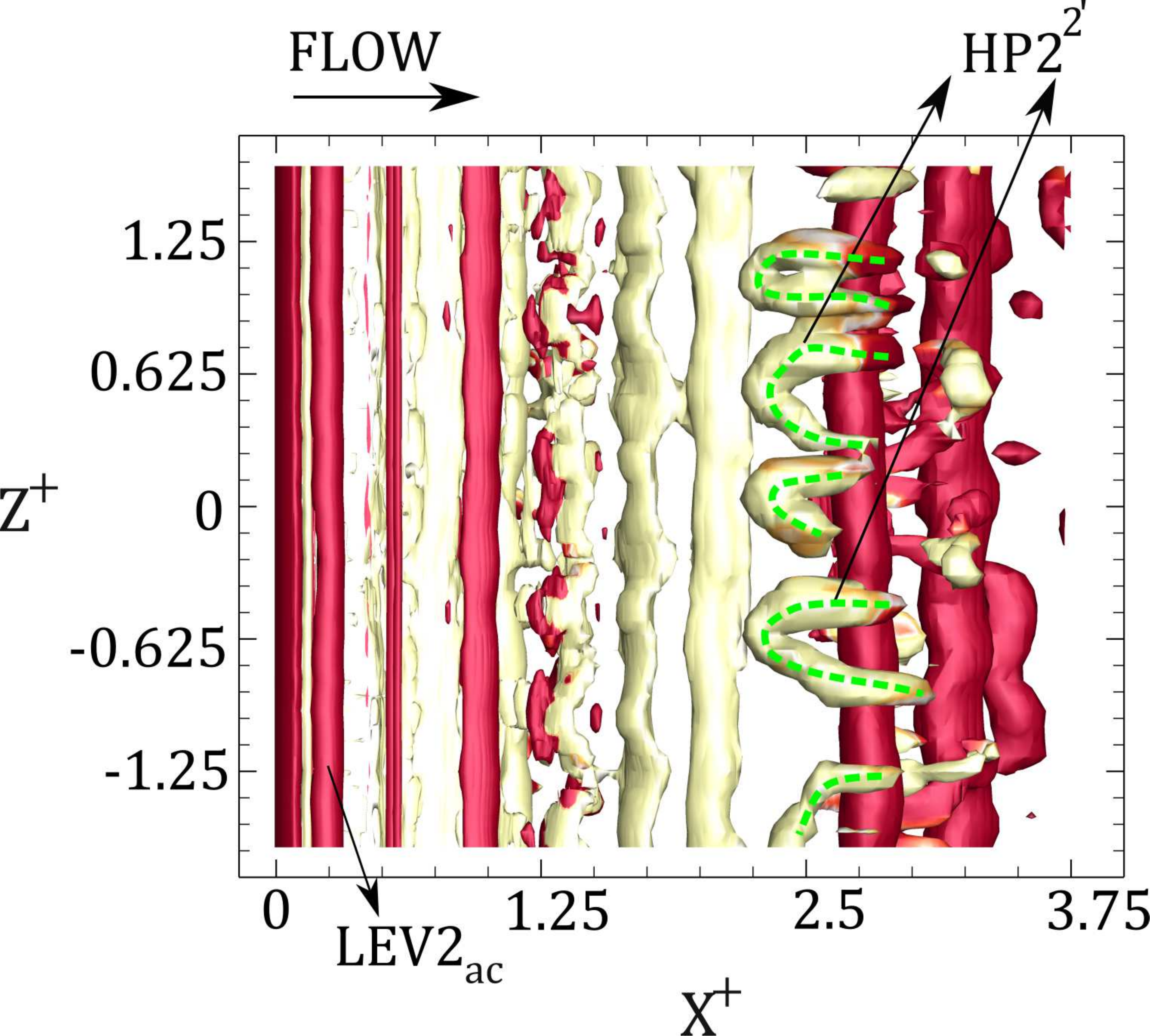}%
		}
	\end{minipage}\hfill
	\begin{minipage}{0.25\textwidth}
		\centering
		\subcaptionbox{\hspace*{-2.75em}}{%
			\hspace{-0.00in}\includegraphics[width=3.2cm,height=3.4cm]{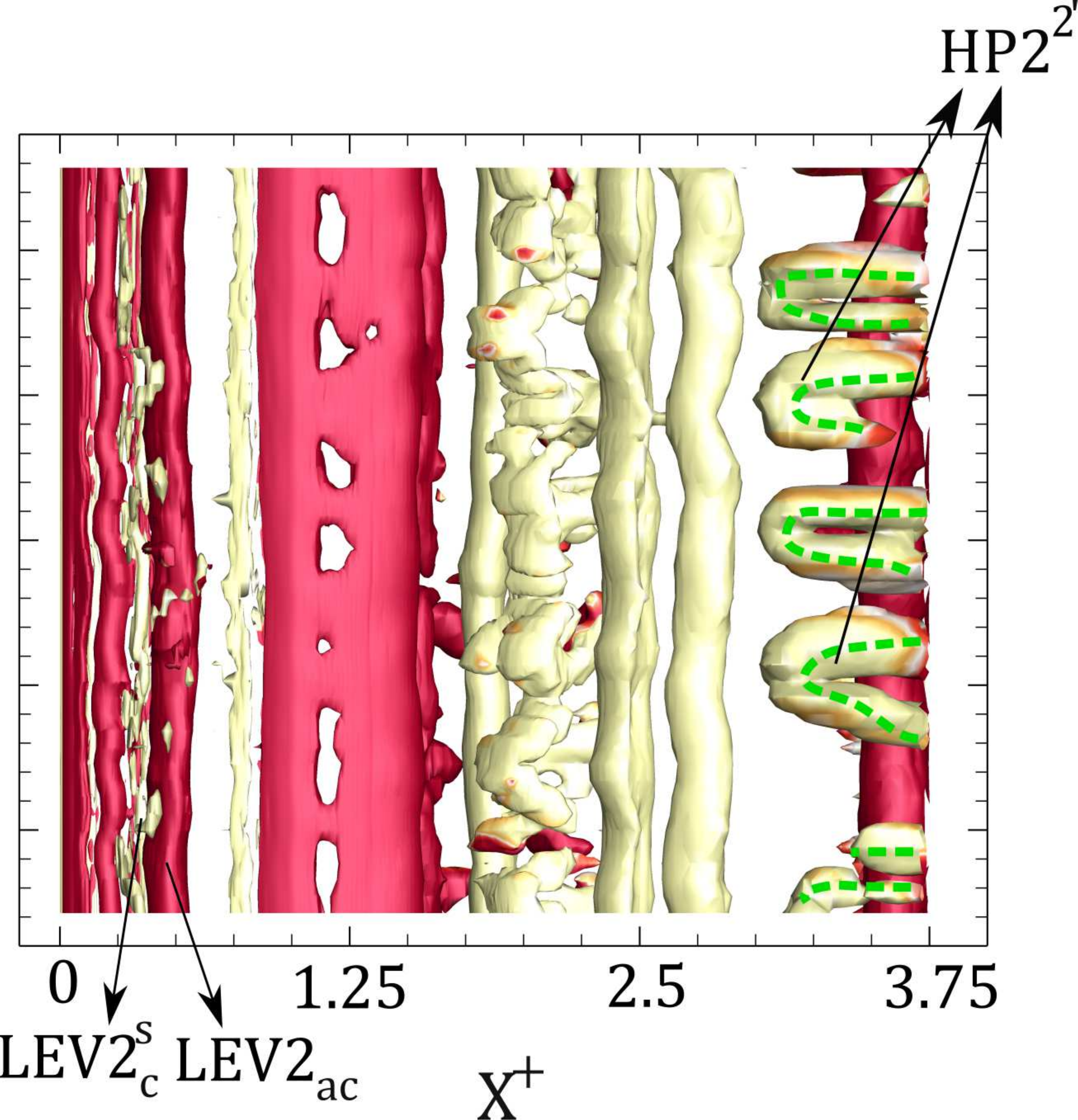}%
		}
	\end{minipage}\hfill
	\begin{minipage}{0.25\textwidth}
		\centering
		\subcaptionbox{\hspace*{-2.75em}}{%
			\hspace{-0.05in}\includegraphics[width=3.2cm,height=3.4cm]{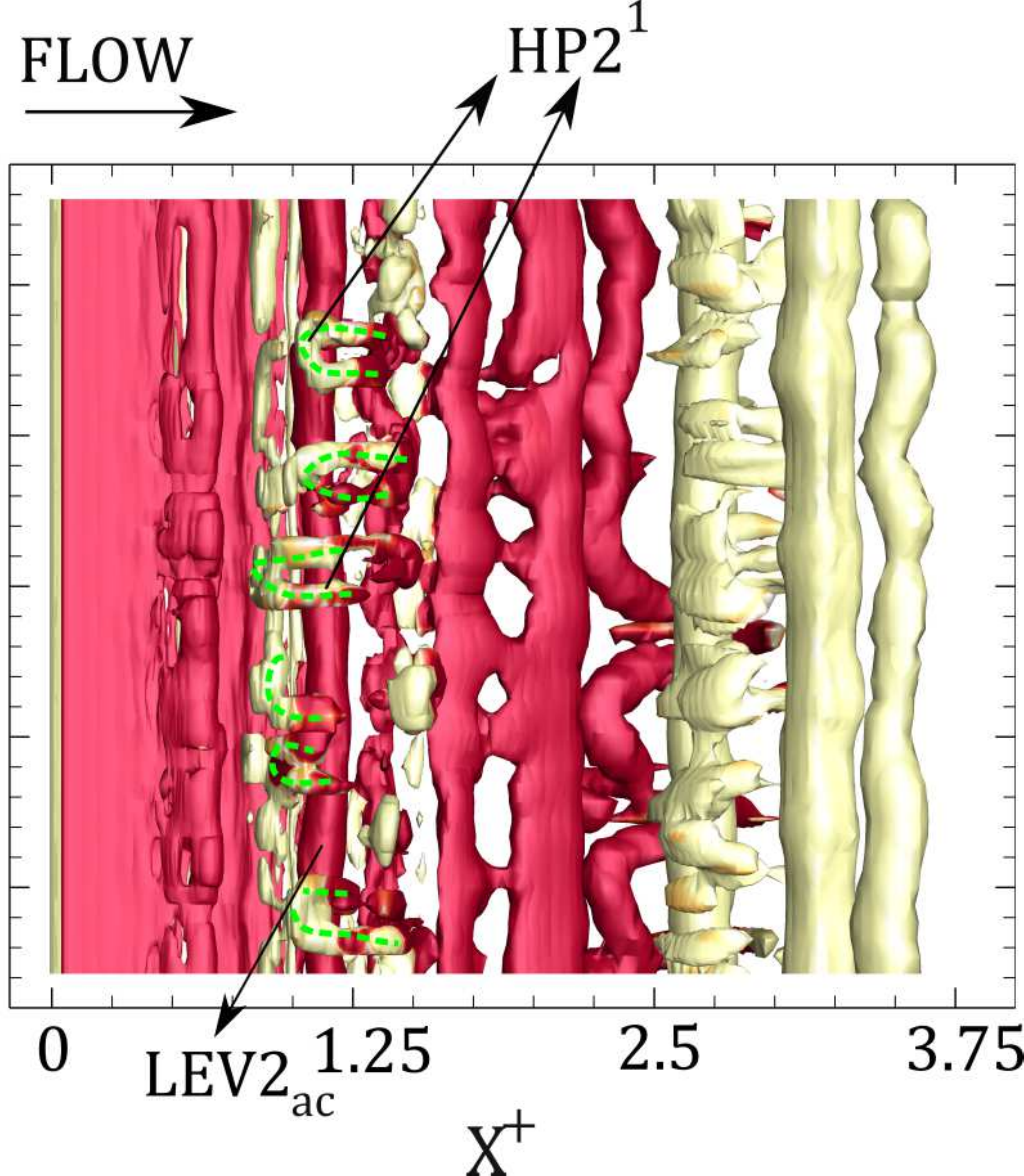}%
		}
	\end{minipage}\hfill
	\begin{minipage}{0.25\textwidth}
		\centering
		\subcaptionbox{\hspace*{-2.75em}}{%
			\hspace{-0.05in}\includegraphics[width=3.2cm,height=3.4cm]{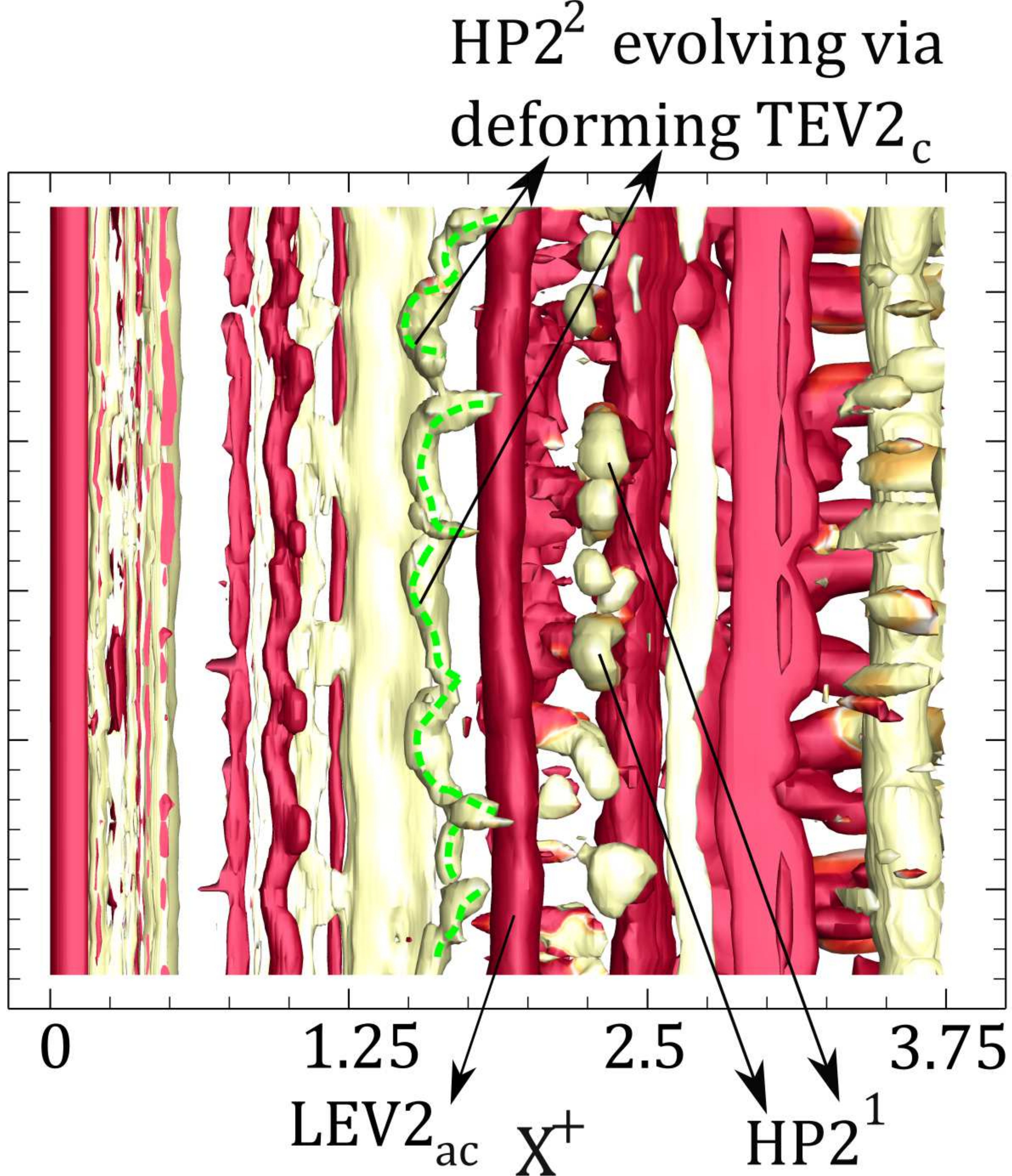}%
		}
	\end{minipage}\\
	\begin{minipage}{0.5\textwidth}
		\centering
		\subcaptionbox{\hspace*{-8.75em}}{%
			\hspace{0.75in}\includegraphics[width=3.6cm,height=3.4cm]{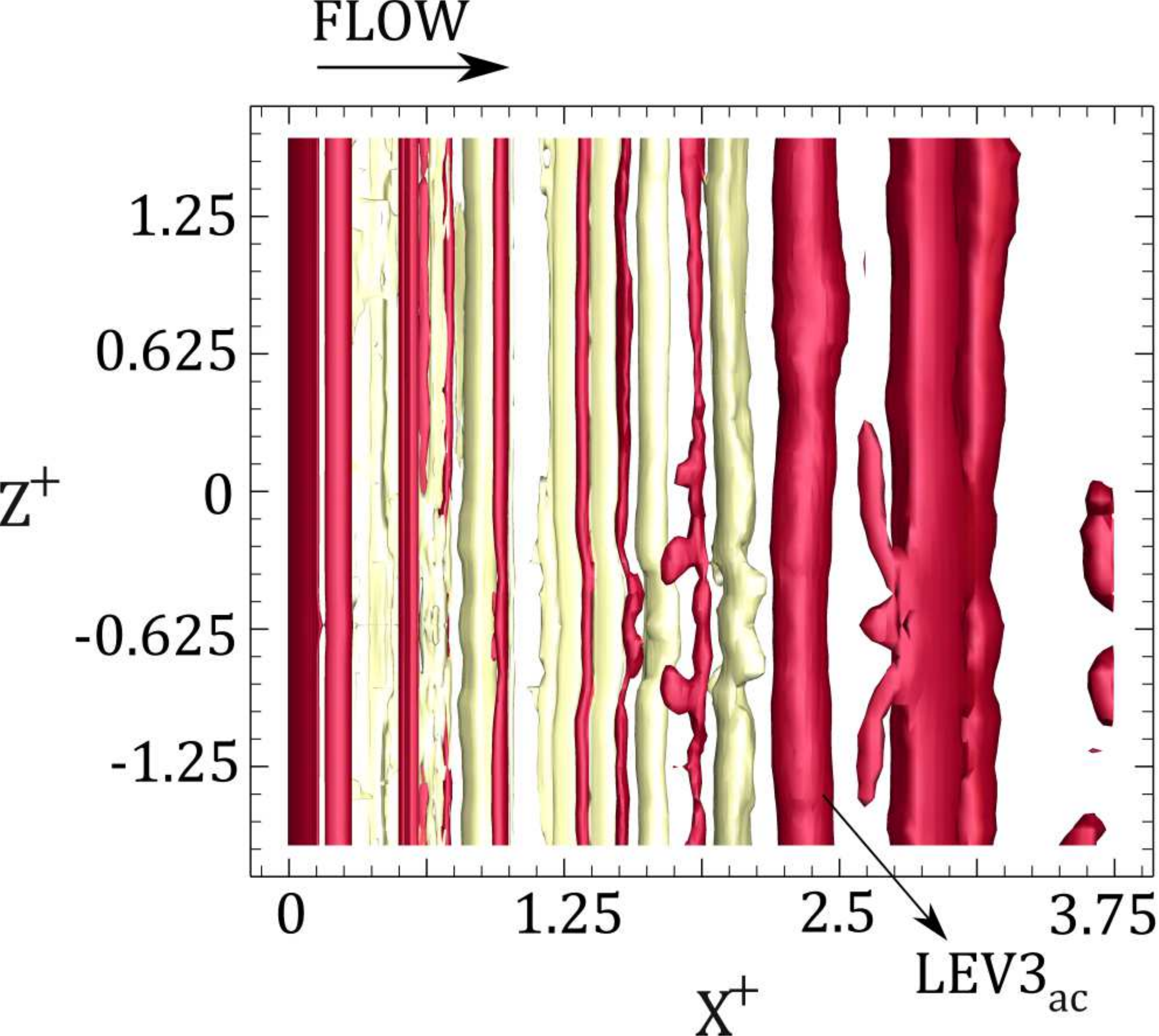}%
		}
	\end{minipage}\hfill
	\begin{minipage}{0.5\textwidth}
		\centering
		\subcaptionbox{\hspace*{2.75em}}{%
			\hspace{-0.75in}\includegraphics[width=3.6cm,height=3.4cm]{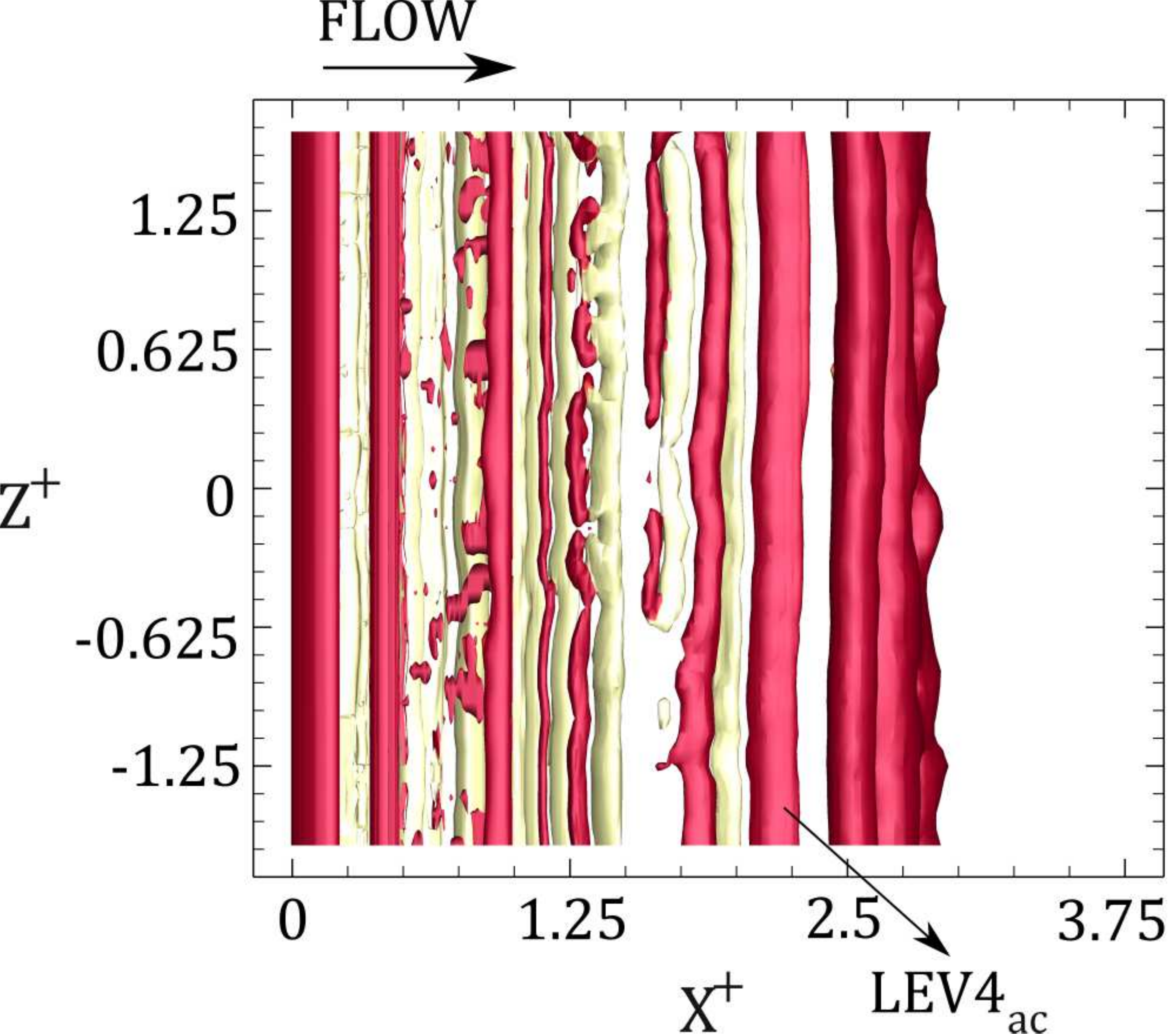}%
		}
	\end{minipage}\\
	\caption{Wakes corresponding to (a-d) $\phi=$ 90$^\circ$, (e-h) $\phi=$ 180$^\circ$, (i) $\phi=$ 225$^\circ$ and (j) $\phi=$ 270$^\circ$, at $St_{c}=$ 0.32. The oscillation time varies from $t^+=$ 0 (a,e) to 0.75 (d,h) in increments of 0.25 for the cases of $\phi=$ 90$^\circ$ and 180$^\circ$. $t^+$ corresponds to 0.75 for $\phi=$ 225$^\circ$ and 270$^\circ$. Each stage is represented using iso-surfaces of $\lambda_2(= -0.05)$, which are colored based on $|\omega_{z}^+|=$ 5.}
	\label{fig:Lambda2_Present_Phi}
\end{figure}

We first describe the growth mechanisms that primarily govern the onset of secondary wake structures at $St_c =$ 0.32. Figures \ref{fig:Lambda2_Present_Phi}(a-d) and \ref{fig:Lambda2_Present_Phi}(e-h) present the three-dimensional instantaneous wake structures for the cases corresponding to $\phi=$ 90$^\circ$ and 180$^\circ$, respectively. The normalized time instants ($t^{+} $= $t/T$) in Figure \ref{fig:Lambda2_Present_Phi} vary from $t^+ =$ 0 to 0.75, in increments of 0.25. The primary vortex (or rollers marked as $LEV1_{ac}$ and $LEV2_{ac}$) and coherent streamwise secondary structures in the form of elongated hairpin-like formations (marked by $HP$) are clearly identifiable for both cases. These wake structures are represented using iso-surfaces of $\lambda_2$ criterion \cite{48}, which is normalized with respect to $U_{\infty}$ and $c$, i.e. $\lambda_2^+$ =$ \lambda c^{2}/U_{\infty}^2$. This method had been proven effective and accurate to extract vortex cores \cite{48}.

At $\phi=$ 90$^\circ$ (see Figure \ref{fig:Lambda2_Present_Phi}(a)), as the primary $LEV_1^{ac}$ gains strength and grows in size, a secondary roller ($LEV1_s^{c}$) begins to form in close vicinity of $LEV_1^{ac}$. The outflux of counter-rotating vorticity via vortex-foil interaction is likely the cause of this secondary $LEV$ formation, as was also suggested by Visbal \cite{26} and Son et al.\cite{51}. However, as $LEV1_s^{c}$ further evolve downstream, the cooperative elliptic instability \cite{1,2} starts developing on these counter-rotating vortex structures of unequal strength, which then leads to undulations of both $LEV$ filaments (see Figure \ref{fig:Lambda2_Present_Phi}(b)). This is followed by a secondary streamwise vorticity outflux via $LEV1_s^{c}$, owing to its relatively weak strength compared to $LEV_1^{ac}$, which subsequently reveals weak hairpin-like structures (marked as $HP_1$ in Figure \ref{fig:Lambda2_Present_Phi}(c)). 
The legs of these hairpin-like structures elongate to form ribs, downstream of the foil trailing edge due to the braid vorticity straining, similar to the mechanism outlined by Mittal $\&$ Balachander \cite{8} in the wake of a stationary circular cylinder. An interaction of primary and secondary $LEV$ rollers is the governing mechanism for the growth of secondary structures at $\phi=$ 90$^\circ$, which coincides with heave dominated kinematics. We identify this mechanism by label $``1"$ in this study. The spanwise hairpin-like arrangement that originate via mechanism $``1"$ will therefore be referred as $HPn^{1}$, where $n$ is used for vortex numbering in the wake, corresponding to different settings of kinematics.  %The trailing edge structure $TEV_1^{c}$ is also seen in its early stages of growth. %(see Figure \ref{fig:Phi_90_Evolution}(e)). 
%As it grows in strength and size, the strong primary $LEV$ and secondary hairpin-like structures have already advected sufficiently downstream in the wake. Therefore, there are no mutual interactions with the developing $TEV$ near the foil trailing edge. 

The stages of the wake evolution at $\phi=$ 180$^\circ$ are now discussed to understand the distinct changes in characteristics of spanwise instability and secondary structures, as kinematics change compared to $\phi=$ 90$^\circ$. Verma $\&$ Hemmati \cite{33} reported that there is a clear decrease in trailing edge amplitude relative to the leading edge, which coincided with a decrease in peak effective angle of attack ($\alpha_{o}$) within an oscillation cycle. However, both cases still represent a heave-dominated kinematic regime for oscillating foils in the coupled motion \cite{18,33}. 

Figures \ref{fig:Lambda2_Present_Phi}(e-h) depicts three-dimensional iso-surface ($\lambda_2 =$ -0.05) plots at every quarter of the oscillation cycle. At $t^+=$ 0 (Figure \ref{fig:Lambda2_Present_Phi}(e)), development of a primary roller ($LEV_{2}^{ac}$) is observed near the leading edge while hairpin-like structures (marked $HP2^\prime$) are evident downstream of the trailing edge. These secondary structures are formed through a similar process of core vorticity outflux and subsequent shear straining of streamwise filaments that emanate from rollers during the oscillation cycle \cite{7,8}. However, it is still important to investigate if these streamwise vorticity filaments remain associated with a secondary $LEV$, as observed in the case of $\phi=$ 90$^\circ$. At $t^+=$ 0.25 (Figure \ref{fig:Lambda2_Present_Phi}(f)), a thin neighboring filament ($LEV_{2,s}^c$) forms besides $LEV_{2}^{ac}$,  although in a partially diffused form. Verma $\&$ Hemmati \cite{57} recently attributed this effect to the reduction in normalized circulation, $\Gamma^+ = \Gamma/(U_{\infty} c)$, of the primary $LEV$ ($LEV_{2}^{ac}$) at $\phi=$ 180$^\circ$, compared to $LEV_{1}^{ac}$ observed at $\phi=$ 90$^\circ$. This contributes to a reduction in intensity of vortex-foil interaction that subsequently decrease the outflux of secondary spanwise vorticity. %The quantitative assessment of $\Gamma^+$ is also discussed below. 
Spanwise undulation of primary roller ($LEV_{2}^{ac}$) emerge at $t^+=$ 0.25 (Figure \ref{fig:Lambda2_Present_Phi}(f)), owing to the cooperative instability induced by the neighboring paired vortex $LEV_{2,s}^c$. As the primary and secondary $LEV$ pair approach the trailing edge at approximately $t^+=$ 0.5 (see Figure \ref{fig:Lambda2_Present_Phi}(g)), the outgrowth of thin streamwise hairpin-like filaments ($HP_2^1$) is evident. However at $t^+=$ 0.75 (Figure \ref{fig:Lambda2_Present_Phi}(h)), the coincident growth of $TEV_{2}^c$ from the bottom part of the foil inhibits the continued elongation and straining process of $HP_2^1$ filaments identified in Figure \ref{fig:Lambda2_Present_Phi}(g). Thus, it is important to note here that although the growth of streamwise filaments from a secondary $LEV$ is consistent with the observations at $\phi=$ 90$^\circ$, the suppression mechanism by a growing TEV actually leads to distinct changes in spatio-temporal evolution of secondary hairpin-like structures in the wake. This is evident in Figure \ref{fig:Lambda2_Present_Phi}(g), where only fragmented $HP2^1$ filaments are visible. At the same instant, the cooperative elliptic instability still retains its dominance due to the presence of counter-rotating vortex rollers ($LEV_{2}^{ac}$ and $TEV_2^c$). This subsequently promotes core vorticity outflux from $TEV_{2}^c$, which then demonstrates valley and bulge like features \cite{8} along the spanwise direction. Such dislocations eventually evolve into dominant secondary hairpin-like structures ($HP2^2$), as identified at the beginning of oscillation cycle (see Figure \ref{fig:Lambda2_Present_Phi}(e)), which appears to be stronger in the wake compared to $HP2^1$. 
%The detailed evaluation of the evolution stages certainly provides a fundamental reasoning behind the spatio-temporal delay observed in the formation of secondary wake structures in Figure \ref{fig:Lambda2_Present_Phi}(b), compared to their formation at $\phi=$ 90$^\circ$ in Figure \ref{fig:Lambda2_Present_Phi}(a). 
From this reasoning, we infer that the presence of spanwise cooperative instability is consistent during distinct evolution mechanisms discussed within the heave dominated kinematics at $\phi=$ 90$^\circ$ and 180$^\circ$, respectively. However, the origins of dominant secondary wake structures does not necessarily associate with the presence of secondary $LEV$ roller (mechanism $``1"$) alone, despite the heave dominated kinematics. At $\phi=$ 180$^\circ$, a growing $TEV$ and its paired interaction with primary $LEV$ provides a second mechanism that can govern the growth of secondary wake structures. This will now be referred as mechanism $``2"$.  

Interestingly, increasing $\phi$ further towards 225$^\circ$ and 270$^\circ$ reveals the absence of secondary hairpin-like structures, as shown in Figures \ref{fig:Lambda2_Present_Phi}(i) and \ref{fig:Lambda2_Present_Phi}(j), respectively. This also coincides with the onset of pitch domination for a foil in coupled heaving and pitching oscillation. {However, the presence of spanwise undulations on the primary rollers ($LEV_3^{ac}$ and $LEV_4^{ac}$) in the wake of foils is still identifiable.} These small-amplitude undulations are attributed to the weak foil-vortex interactions that occur during the $LEV$ advection over the foil boundary. The much weaker secondary $LEV$ is observed to either diffuse or merge with the shear layer separating from the bottom part of the foil.% as observed earlier at $\phi=$ 180$^\circ$ (see Figure \ref{fig:Lambda2_180_Mech}(d)). 
The entire temporal evolution of vortex structures within one oscillation cycle are not presented here for brevity {(see supplementary videos for reference)}, since the observations that concerned growth and advection of the primary $LEVs$ and $TEVs$ are mostly similar to $\phi=$ 180$^\circ$. Disappearance of secondary wake structures also coincides with decreased circulation strength of the primary rollers as discussed by Verma $\&$ Hemmati \cite{57}. Cheireghin et al.\cite{32} further hinted in their study that a larger deformation of the $LEV$ filament could be attributed to an increased circulation strength. However, no plausible association with the growth of secondary wake structures was discussed in their study \cite{32}. 

{We now quantitatively assess the variation of pressure gradient over the foil surface as the primary $LEV$ approaches the trailing edge, under kinematics characterized by a transition from heave- to pitch-dominated motion. This provides further explanation in addition to the decreasing circulation strength of primary $LEV$ \cite{57} for the transition from mechanism $``1"$ to $``2"$ detailed above. Particularly, streamwise pressure gradient along the chordwise extent of the foil demonstrate an association between the growth of secondary vortex, and the subsequent core vorticity outflux observed in mechanism $``1"$ (which leads to formation of streamwise hairpin-like filaments). To proceed, we utilize the methodology explained by Obabco and Cassel \cite{56} to identify the mechanism of secondary recirculation zone formation in the vicinity of a larger vortex core. In brief, Obabco and Cassel \cite{56} highlighted that large scale vortical interactions were characterized by a rapid change in the streamwise pressure gradient, which coincided with a growth of secondary re-circulation zone \cite{56}. Particularly, the region characterized by the primary re-circulation zone experiences a sudden streamwise compression due to a rapid change in  pressure gradient, that favors and accelerates the growth of a new secondary re-circulation zone \cite{56}. We assess this correspondence of streamwise pressure gradient and growth of secondary re-circulation region, which in our study will represent the secondary vortex formation observed at $\phi=$ 90$^\circ$ and 180$^\circ$ in Figure \ref{fig:Lambda2_Present_Phi}(b) and \ref{fig:Lambda2_Present_Phi}(f), respectively. The coincident association of surface pressure distribution and secondary hairpin-like growth through mechanism $``1"$ will also establish a  novel finding for the wake evolution of oscillating foils.
	
	\begin{figure}
		\centering
		\begin{minipage}{0.3\textwidth}
			\centering
			\subcaptionbox{\hspace*{-2.75em}}{%
				\hspace{-0.05in}\includegraphics[width=4.65cm,height=3.6cm]{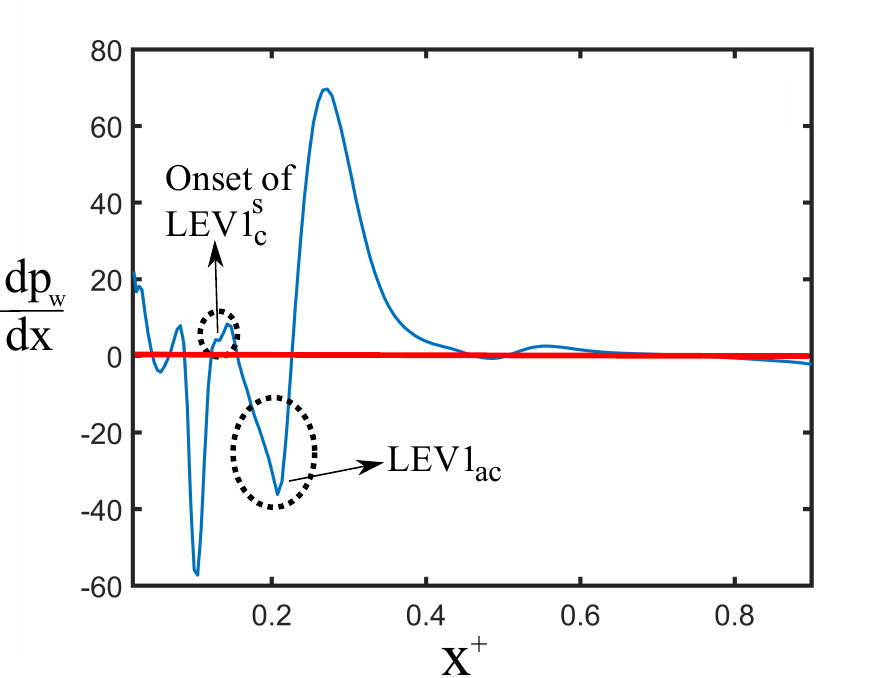}%
			}
		\end{minipage}\hfill
		\begin{minipage}{0.3\textwidth}
			\centering
			\subcaptionbox{\hspace*{-2.75em}}{%
				\hspace{-0.1in}\includegraphics[width=4.65cm,height=3.6cm]{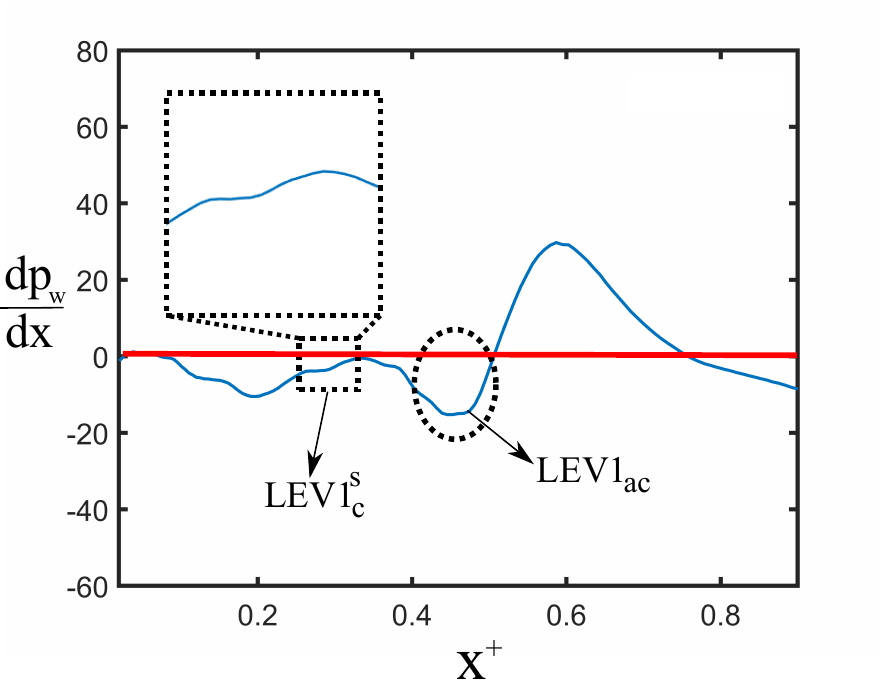}%
			}
		\end{minipage}\hfill
		\begin{minipage}{0.3\textwidth}
			\centering
			\subcaptionbox{\hspace*{-2.75em}}{%
				\hspace{-0.1in}\includegraphics[width=4.65cm,height=3.6cm]{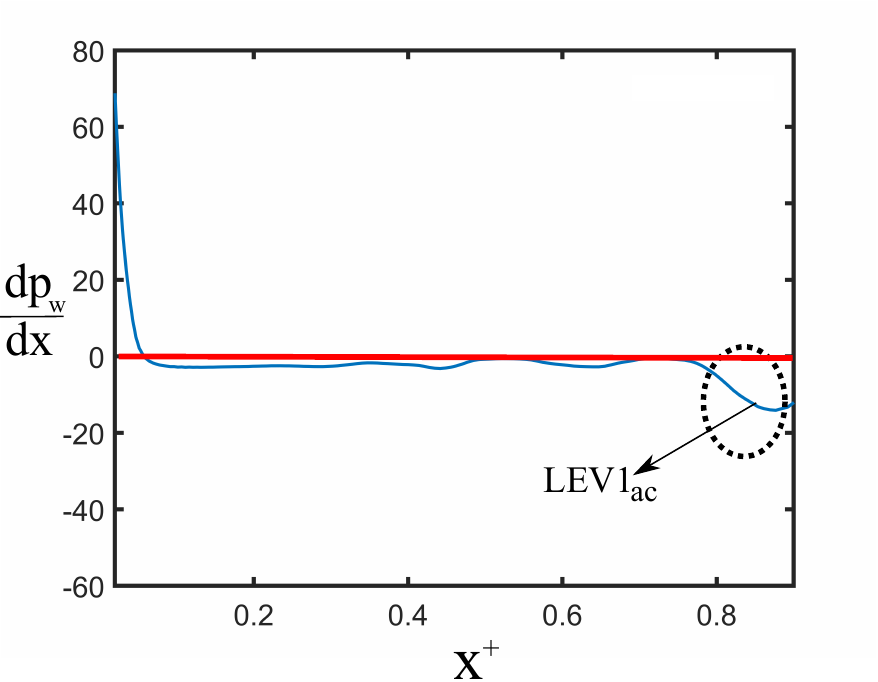}%
			}
		\end{minipage}\\
		\begin{minipage}{0.3\textwidth}
			\centering
			\subcaptionbox{\hspace*{-2.75em}}{%
				\hspace{-0.05in}\includegraphics[width=4.65cm,height=3.6cm]{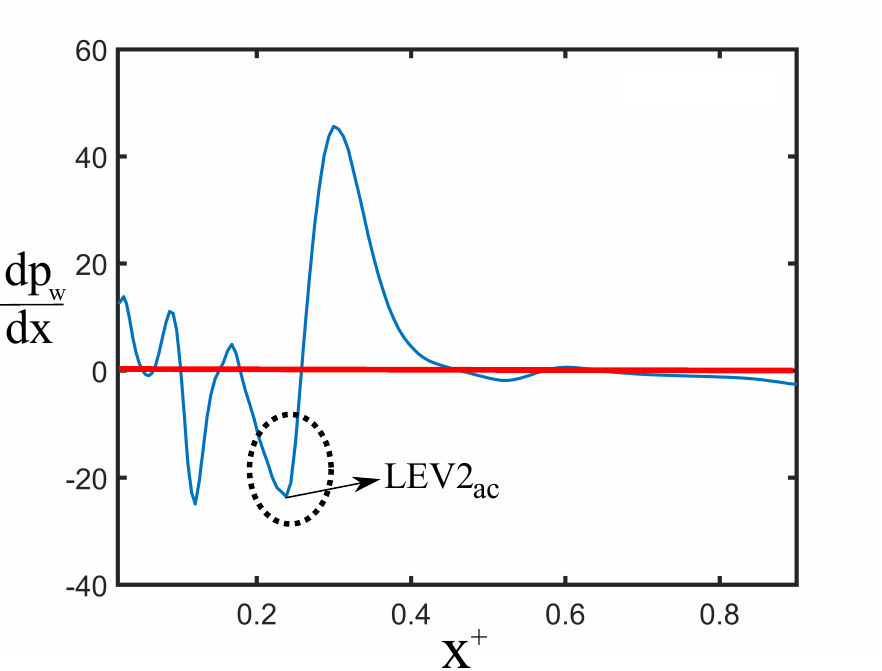}%
			}
		\end{minipage}\hfill
		\begin{minipage}{0.3\textwidth}
			\centering
			\subcaptionbox{\hspace*{-2.75em}}{%
				\hspace{-0.1in}\includegraphics[width=4.65cm,height=3.6cm]{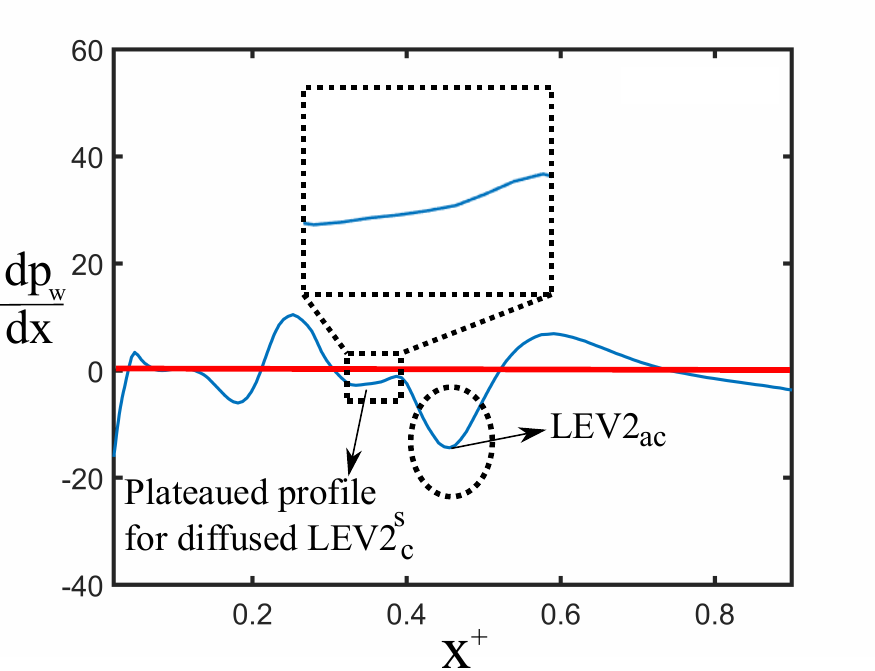}%
			}
		\end{minipage}\hfill
		\begin{minipage}{0.3\textwidth}
			\centering
			\subcaptionbox{\hspace*{-2.75em}}{%
				\hspace{-0.1in}\includegraphics[width=4.65cm,height=3.6cm]{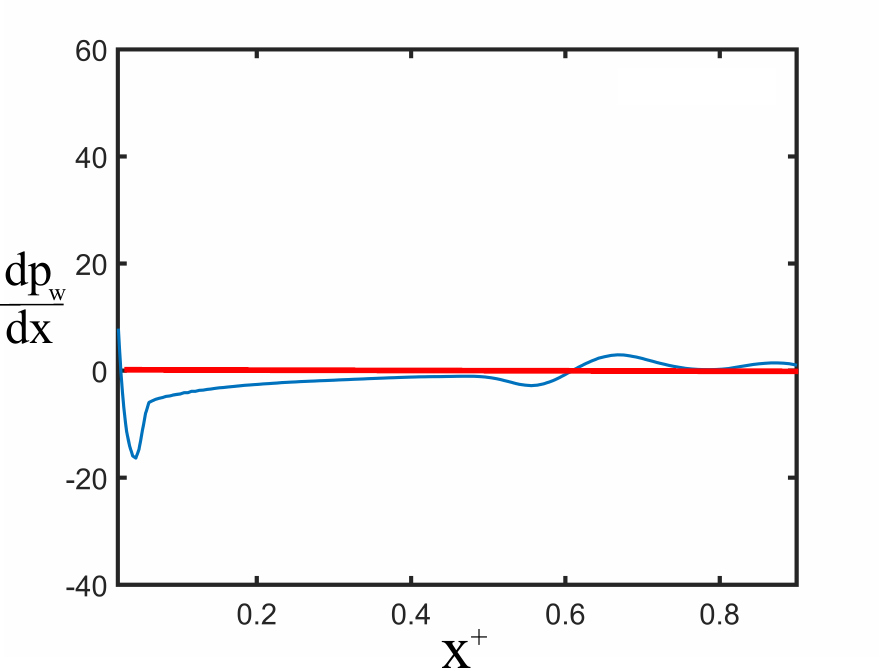}%
			}
		\end{minipage}\\
		\begin{minipage}{0.3\textwidth}
			\centering
			\subcaptionbox{\hspace*{-2.75em}}{%
				\hspace{-0.05in}\includegraphics[width=4.65cm,height=3.6cm]{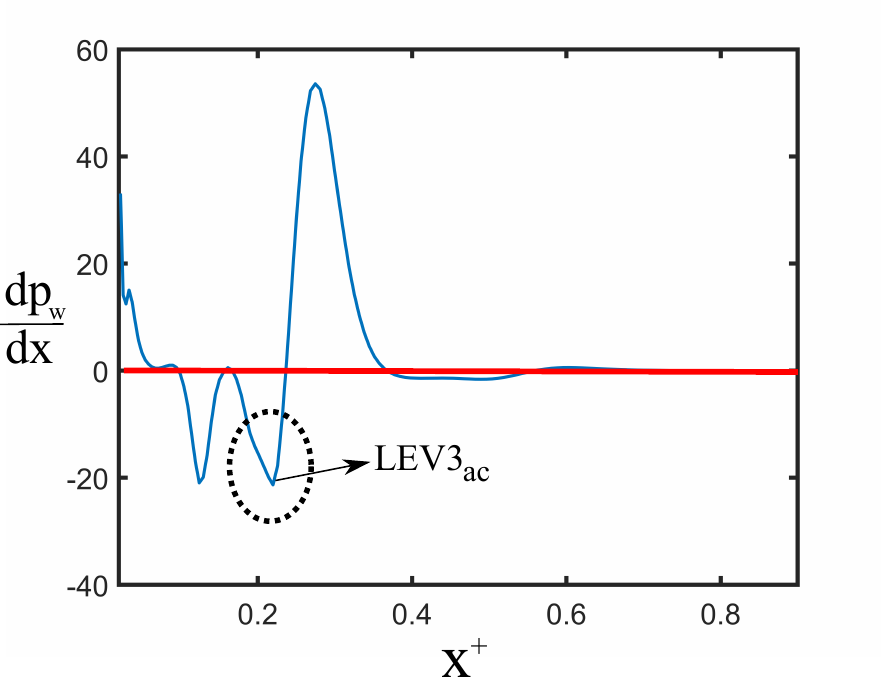}%
			}
		\end{minipage}\hfill
		\begin{minipage}{0.3\textwidth}
			\centering
			\subcaptionbox{\hspace*{-2.75em}}{%
				\hspace{-0.1in}\includegraphics[width=4.65cm,height=3.6cm]{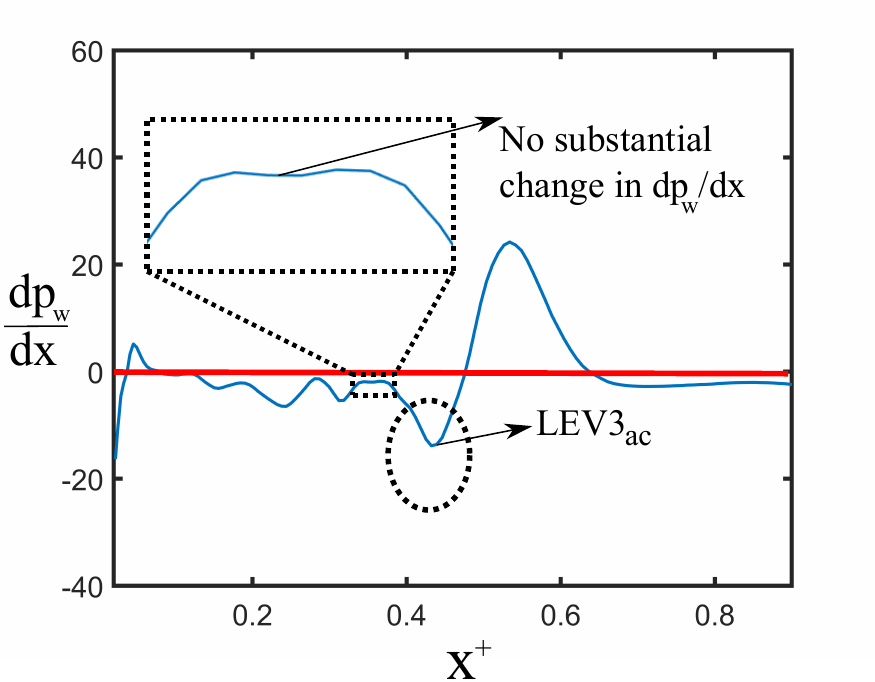}%
			}
		\end{minipage}\hfill
		\begin{minipage}{0.3\textwidth}
			\centering
			\subcaptionbox{\hspace*{-2.75em}}{%
				\hspace{-0.1in}\includegraphics[width=4.65cm,height=3.6cm]{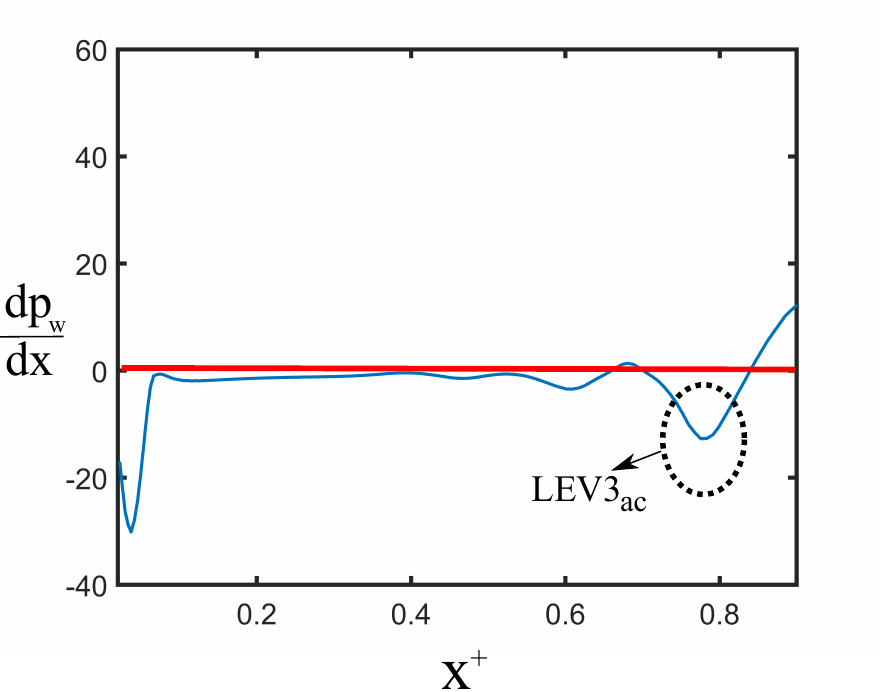}%
			}
		\end{minipage}\\
		\begin{minipage}{0.3\textwidth}
			\centering
			\subcaptionbox{\hspace*{-2.75em}}{%
				\hspace{-0.05in}\includegraphics[width=4.65cm,height=3.6cm]{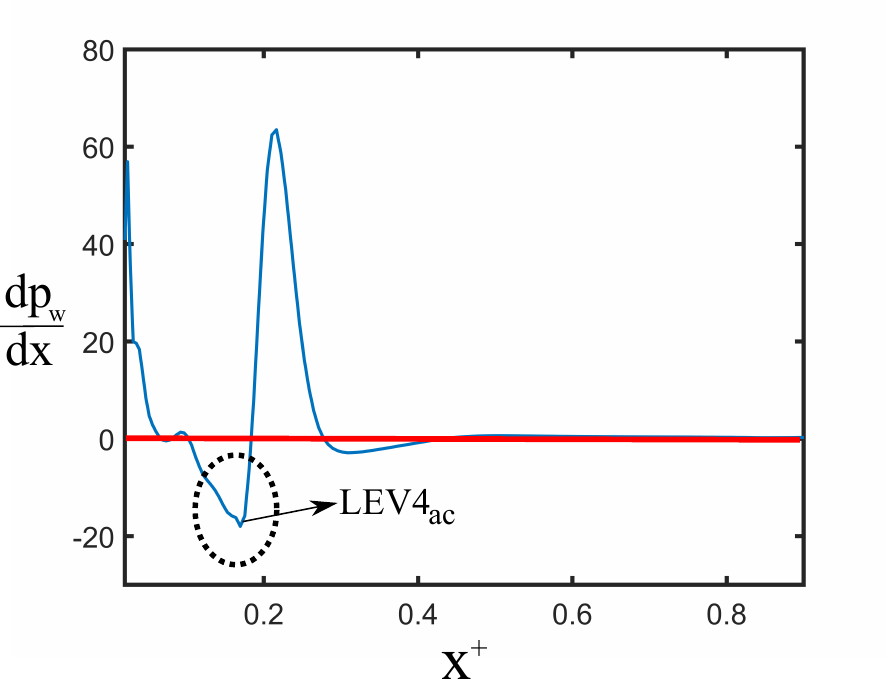}%
			}
		\end{minipage}\hfill
		\begin{minipage}{0.3\textwidth}
			\centering
			\subcaptionbox{\hspace*{-2.75em}}{%
				\hspace{-0.1in}\includegraphics[width=4.65cm,height=3.6cm]{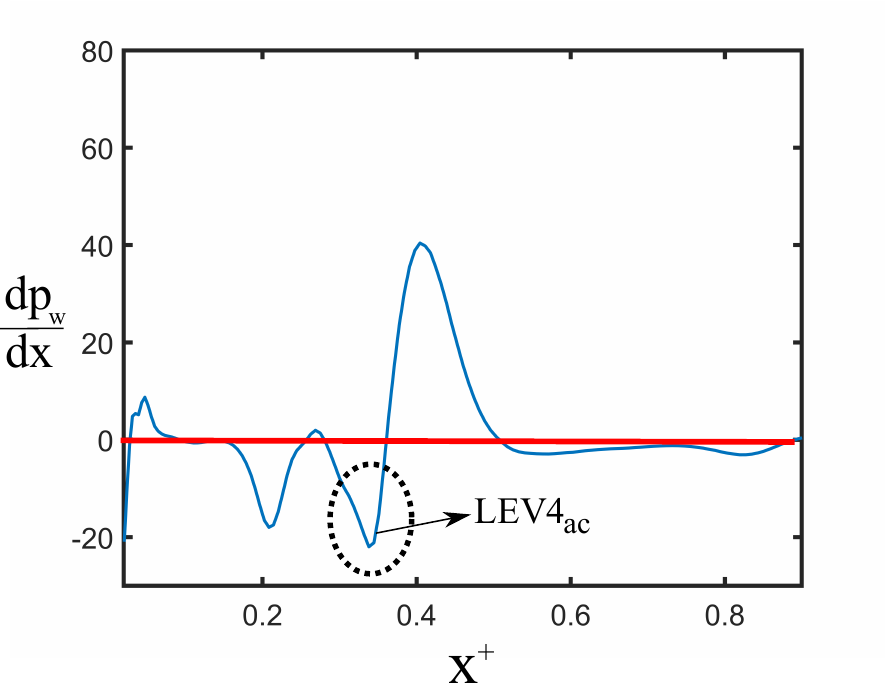}%
			}
		\end{minipage}\hfill
		\begin{minipage}{0.3\textwidth}
			\centering
			\subcaptionbox{\hspace*{-2.75em}}{%
				\hspace{-0.1in}\includegraphics[width=4.65cm,height=3.6cm]{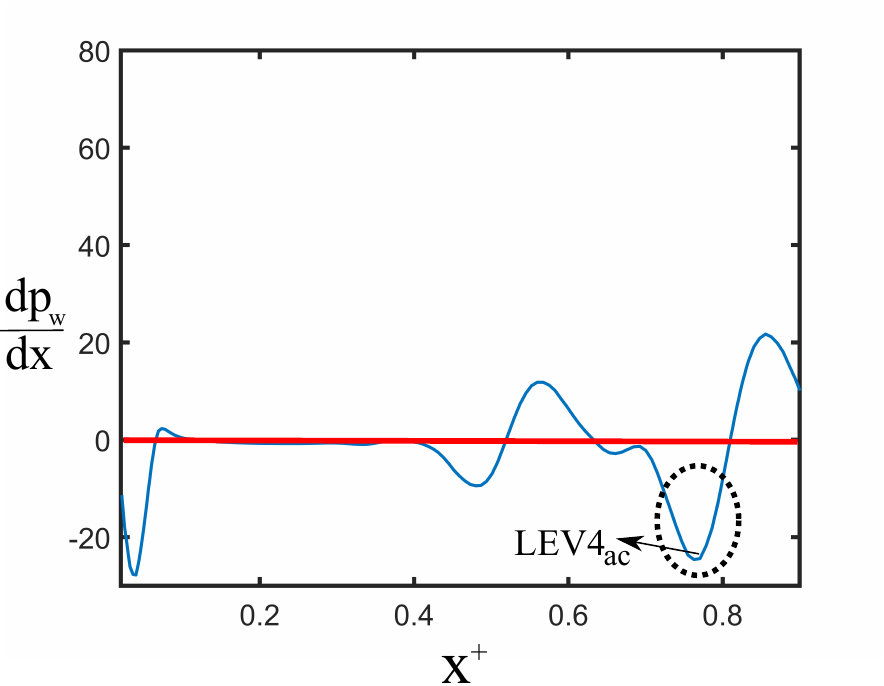}%
			}
		\end{minipage}\\
		\caption{Span-averaged $dp_{w}/dx$ over the foil surface in streamwise direction at first three quarter phases of an oscillation cycle. The cases correspond to $\phi=$ 90$^\circ$ (a,b,c), $\phi=$ 180$^\circ$ (d,e,f), $\phi=$ 225$^\circ$ (g,h,i) and $\phi=$ 270$^\circ$ (j,k,l).}
		\label{fig:Pressure_f04}
	\end{figure}

	Figure \ref{fig:Pressure_f04} depicts span-averaged profiles of pressure gradient on foil in the chordwise direction ($dp_{w}/dx$), for different $\phi$. Three time-instants corresponding to the first three quarter phases of an oscillation cycle, i.e. $t^+=$ 0 (leftmost column), 0.25 (middle column) and 0.5 (rightmost column), are examined here to evaluate the streamwise compression of vortical flow \cite{56}, and its potential contribution to the growth of the secondary $LEV$. A subsequent connection to the streamwise hairpin-like evolution (through mechanism $``1"$) will also be established. The distribution of $dp_{w}/dx$ for $\phi=$ 90$^\circ$ at $t^+=$ 0 (see Figure \ref{fig:Pressure_f04}(a)), indicates that the local minima coincides with the advecting $LEV1_{ac}$ (Figure \ref{fig:Lambda2_Present_Phi}(a)). A rapid increase in $dp_{w}/dx$ towards the upstream direction features a streamwise flow compression, very similar to the observations of Obabco and Casell \cite{56}. This yields a favorable pressure gradient (indicated by an overshoot towards a positive value) for the onset of $LEV1_c^{s}$. This onset is marked as a sharp drop at $X^+\approx0.16$. As $LEV1_{ac}$ and $LEV1_{c}^s$ advect downstream, Figure \ref{fig:Pressure_f04}(b) shows a local minima of relatively low magnitude, which corresponds to $LEV1_{c}^s$. This minima eventually approached 0 at $t^+=$ 0.5 (see Figure \ref{fig:Pressure_f04}(c)). Drop in the magnitude of pressure gradient minima in the region featuring secondary $LEV$ coincides with the eventual streamwise hairpin-like evolution (through mechanism $``1"$). 
	
	The profiles of $dp_{w}/dx$ at $\phi=$ 180$^\circ$ (see Figure \ref{fig:Pressure_f04}(d-f)) provide another crucial quantitative evidence for the lack of secondary $LEV$ transformation to dominant secondary hairpin-like structures via mechanism $``1"$. This is consistent with the circulation strength assessment evaluated recently by Verma $\&$ Hemmati \cite{57}. A local minima in $dp_{w}/dx$ is observed in Figure \ref{fig:Pressure_f04}(d), which corresponds to $LEV2_{ac}$. This is relatively smaller in magnitude compared to the local minima at $\phi=$ 90$^\circ$. A sharp rise towards a positive $dp_{w}/dx$ in the upstream chord direction is further evident although a sharp (peak) drop is not noticeable. A plateaued $dp_{w}/dx$ region eventually becomes evident at $t^+=$ 0.25, which coincides with the location of diffused $LEV2_{c}^s$ in Figure \ref{fig:Lambda2_Present_Phi}(f). The smoother drop in $dp_{w}/dx$ indicates that the streamwise flow compression is weaker, which hence leads to a weak secondary $LEV$ growth. This reflects an important association between kinematics and the evolution of secondary structures, where the reduced capacity of mechanism $``1"$ becomes clear with an increase in $\phi$ to 180$^\circ$. On a similar note, a transition from mechanism $``1"$ to $``2"$ coincides with the discussed changes in streamwise distribution of $dp_{w}/dx$. With transition of kinematics towards an onset of pitch domination at $\phi=$ 225$^\circ$ and 270$^\circ$, a similar trend is persistent with respect to the growth of secondary $LEV$. No substantial drop is seen upstream of the primary $LEVs$ ($LEV3_{ac}$ and $LEV4_{ac}$) in Figures \ref{fig:Pressure_f04}(g), \ref{fig:Pressure_f04}(h), \ref{fig:Pressure_f04}(j) and \ref{fig:Pressure_f04}(k), respectively. Coincidentally, absence of the secondary hairpin-like arrangement through mechanism $``1"$ is evident in these kinematic regime (see Figure \ref{fig:Lambda2_Present_Phi}(i,h)). It is important to note that there are several other features that can be extracted and discussed from the results in Figure \ref{fig:Pressure_f04}. Such discussions, however, fall outside the scope of the current study.}

Based on the observations at $St_c=$ 0.32, a transition in mechanisms that govern the growth of secondary structures is clear with respect to the change in heave-dominated kinematics, towards an onset of pitch domination.  %However, it is still unclear if the mechanisms follow similar transition with respect to kinematics, at increasing $St_c$. This is now investigated below. 
{This explanation provides us with another important aspect of vortex dynamics around oscillating foils (representing flapping wings and tail-fins), which relates to the potentially similar role of these mechanisms for increasing $St_c$. It holds importance, because a plausible answer to this question expands the applicability of these mechanisms to a broader set of kinematics.}

\subsection{Influence of increasing $St_c$}

\subsubsection{$St_c =$ 0.40}

Flow observations at $St_c=$ 0.4 and $\phi=$ 90$^\circ$ demonstrates mechanism $``1"$ of the secondary structure formation, similar to $St_c =$ 0.32. The wake topology is not shown or discussed here for brevity {(see supplementary videos)}. However, in order to qualitatively reveal a progression of secondary structure growth towards  kinematics dominated by decreasing heave domination, wake topologies corresponding to $\phi=$ 180$^\circ$ and 225$^\circ$ are shown in Figure \ref{fig:Lambda2_f05}(a-d) and \ref{fig:Lambda2_f05}(e-h), respectively.  The wake at $\phi=$ 180$^\circ$ reflects growth of the primary $LEV5_{ac}$, in a paired configuration with a weaker secondary vortex $LEV5_{c}^s$ (see Figures \ref{fig:Lambda2_f05}(a) and \ref{fig:Lambda2_f05}(b)). At the same instant, we also observe formation of a strong secondary hairpin-like vortex ($HP3^{2'}$) downstream of the trailing edge, that evolved in the previous oscillation cycle via mechanism $``2"$  (i.e. deformation and dislocation growth on $TEV$). At $t^+ =$ 0.5 (see Figure \ref{fig:Lambda2_f05}(c)) thin strands of coherent streamwise structures evolve via core vorticity outflux from $LEV5_{c}^s$, which then form weak hairpin-like structures. These are marked as $HP3^1$ at $t^+ =$ 0.75 (see Figures \ref{fig:Lambda2_f05}(d)). The size of $HP3^1$ appears much smaller (visually) when compared with $HP3^{2\prime}$. This is also consistent with flow observations at $\phi=$ 180$^\circ$ at $St_c=$ 0.32. 

Wake topology at $\phi=$ 225$^\circ$, however, demonstrates some contrasting features with reference to the growth of secondary structures at $St_c=$ 0.32. This is evident in Figures \ref{fig:Lambda2_f05}(e-h). Formation of thin secondary hairpin-like filaments are now evident at $\phi =$ 225$^\circ$. These streamwise pairs are marked as $HP4^{2\prime}$ in Figure \ref{fig:Lambda2_f05}(e), which were formed in the previous shedding cycle. Despite the advection of paired primary and secondary rollers ($LEV6_{ac}$ and $LEV6_{c}^s$), an outflux of streamwise vorticity does not appear to be imminent at later stages of the oscillation cycle. However, at $t^+ =$ 0.75 (see Figure \ref{fig:Lambda2_f05}(h)), a paired configuration of $TEV6_c$ and $LEV6_{ac}$ is noticeable, which leads to an elliptic instability between the counter-rotating rollers. This consequently promotes the formation of secondary hairpin-like structures marked as $HP4^{2^\prime}$ (Figure \ref{fig:Lambda2_f05}(e)) at $t^+ =$ 0. 

\begin{figure}
	\centering
	\begin{minipage}{0.5\textwidth}
		\centering
		\hspace{-0.15in}\includegraphics[width=5.0cm,height=1.0cm]{Omega_Legend-eps-converted-to.pdf}%
	\end{minipage}\\
	\vspace{0.1in}
	\begin{minipage}{0.25\textwidth}
		\centering
		\subcaptionbox{\hspace*{-2.75em}}{%
			\hspace{-0.05in}\includegraphics[width=3.5cm,height=3.4cm]{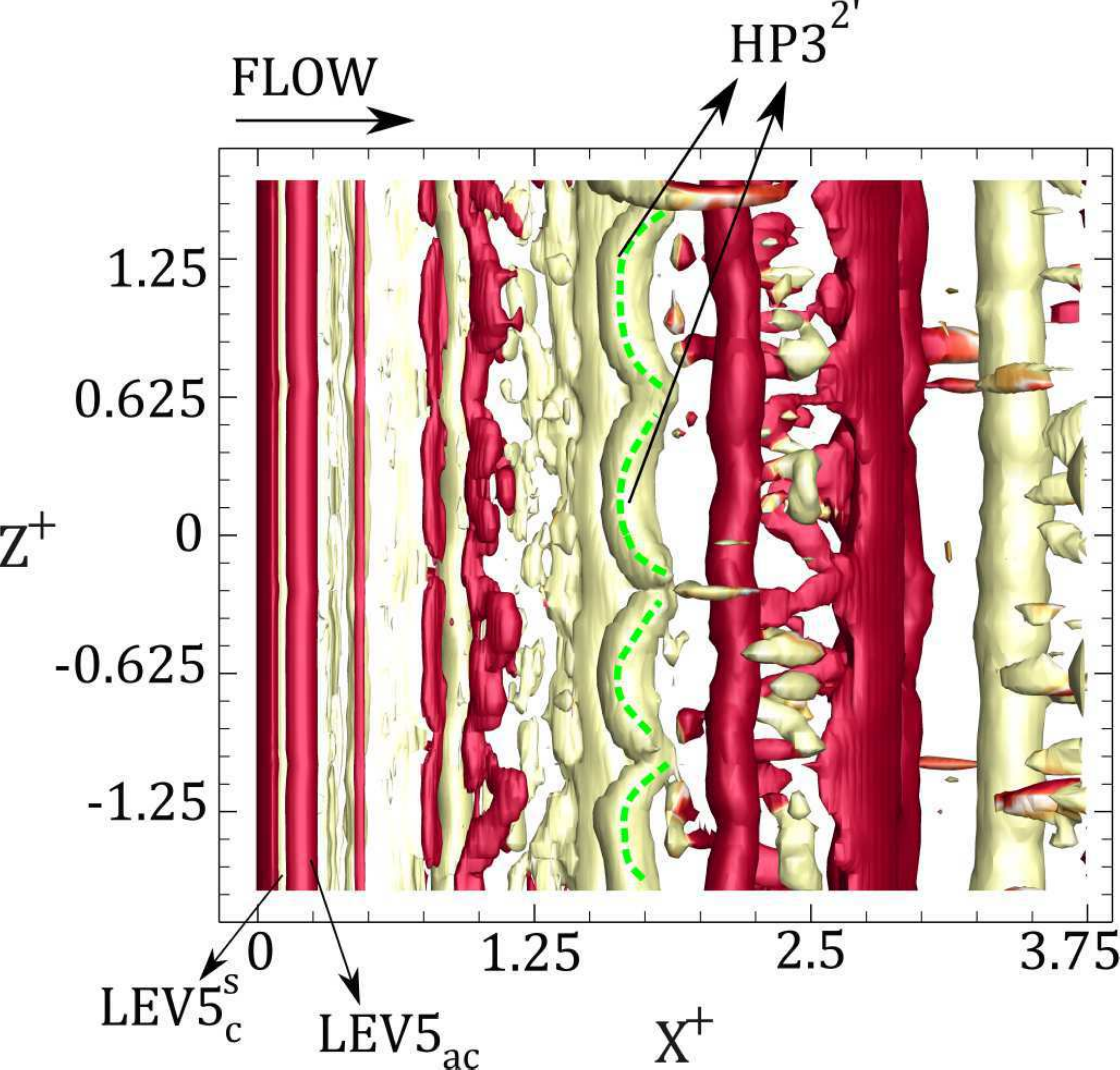}%
		}
	\end{minipage}\hfill
	\begin{minipage}{0.25\textwidth}
		\centering
		\subcaptionbox{\hspace*{-2.75em}}{%
			\hspace{-0.05in}\includegraphics[width=3.2cm,height=3.4cm]{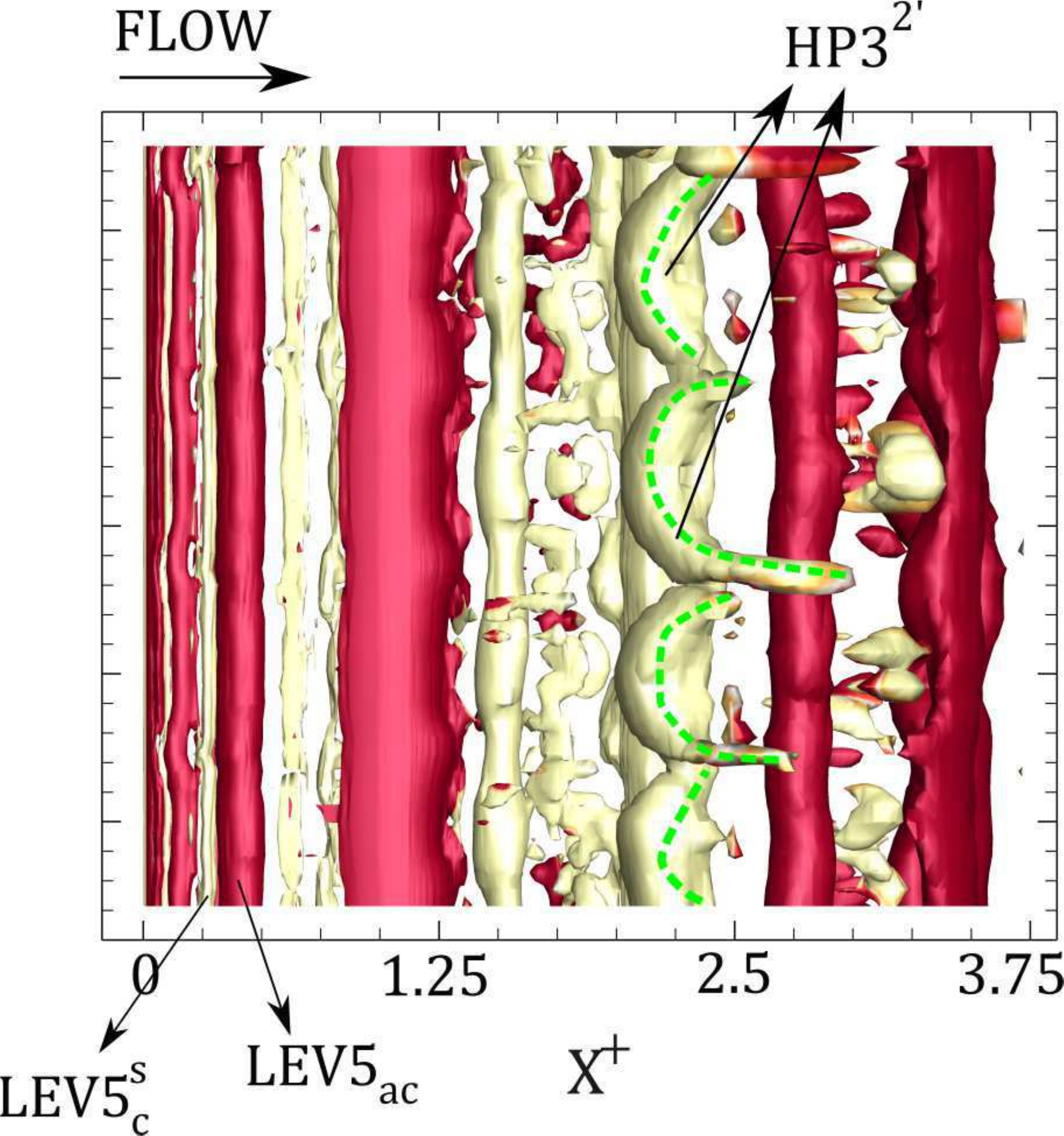}%
		}
	\end{minipage}\hfill
	\begin{minipage}{0.25\textwidth}
		\centering
		\subcaptionbox{\hspace*{-2.75em}}{%
			\hspace{-0.05in}\includegraphics[width=3.2cm,height=3.4cm]{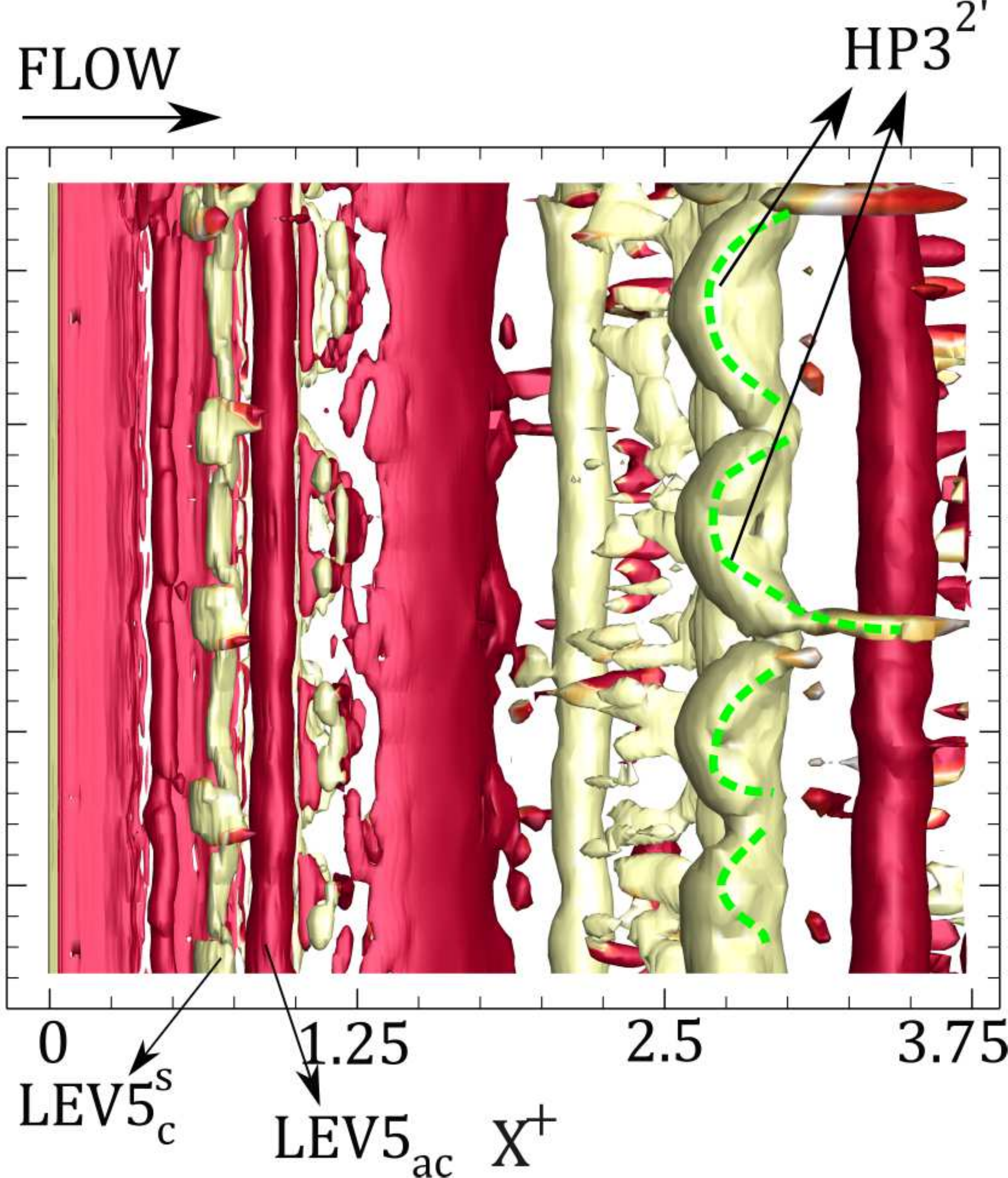}%
		}
	\end{minipage}\hfill
	\begin{minipage}{0.25\textwidth}
		\centering
		\subcaptionbox{\hspace*{-2.75em}}{%
			\hspace{-0.05in}\includegraphics[width=3.2cm,height=3.4cm]{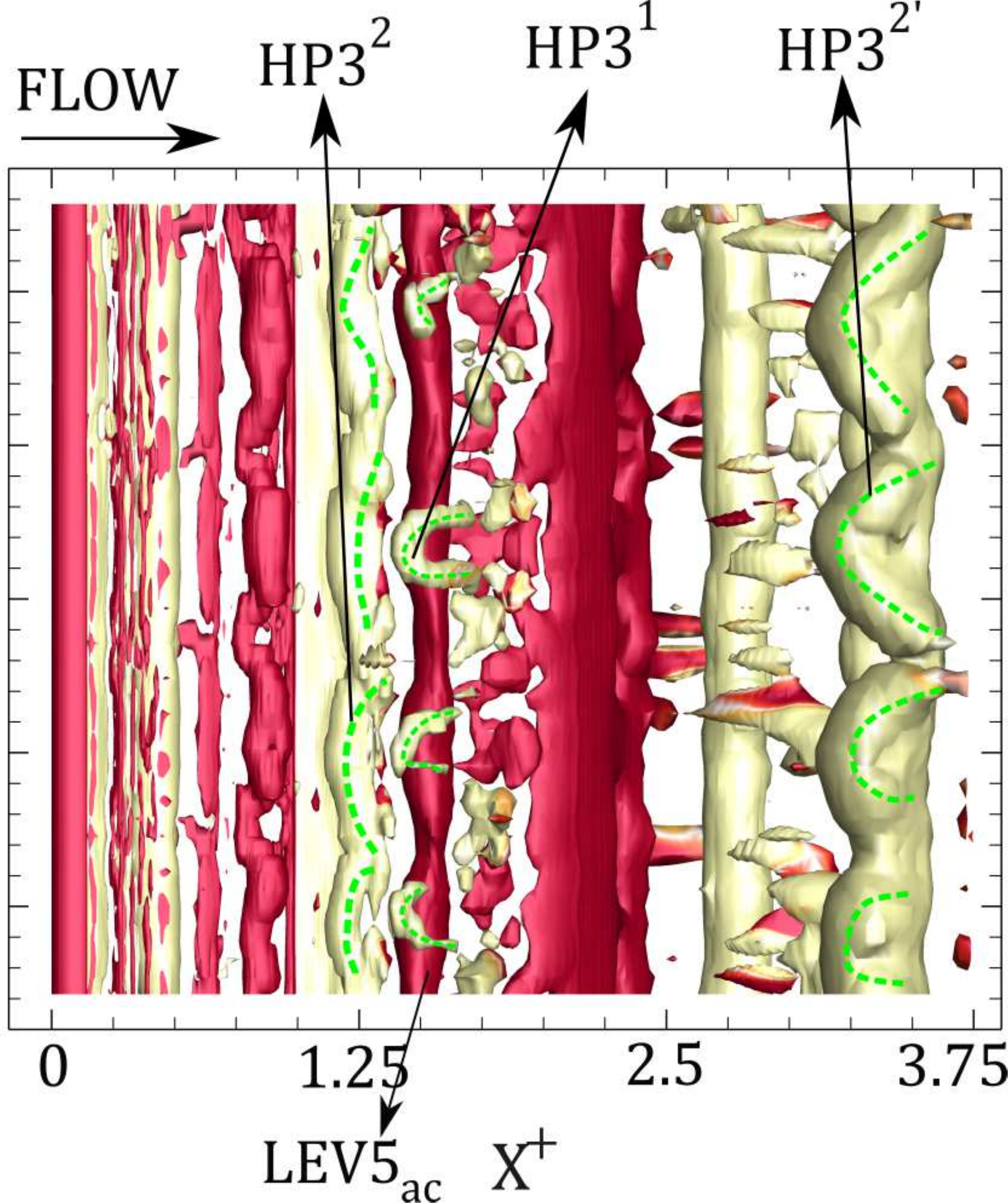}%
		}
	\end{minipage}\\
	\begin{minipage}{0.25\textwidth}
		\centering
		\subcaptionbox{\hspace*{-2.75em}}{%
			\hspace{-0.05in}\includegraphics[width=3.6cm,height=3.4cm]{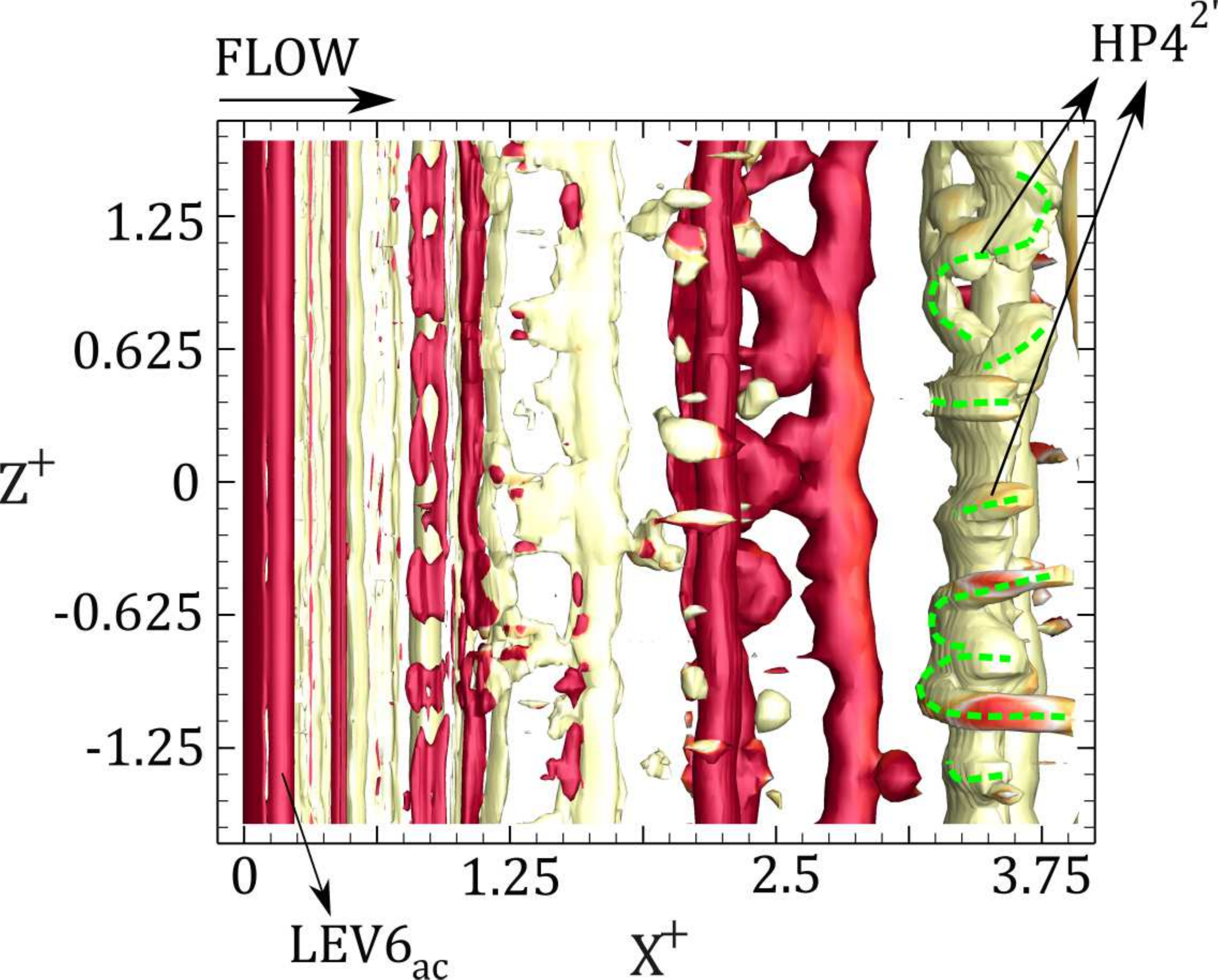}%
		}
	\end{minipage}\hfill
	\begin{minipage}{0.25\textwidth}
		\centering
		\subcaptionbox{\hspace*{-2.75em}}{%
			\hspace{-0.00in}\includegraphics[width=3.2cm,height=3.4cm]{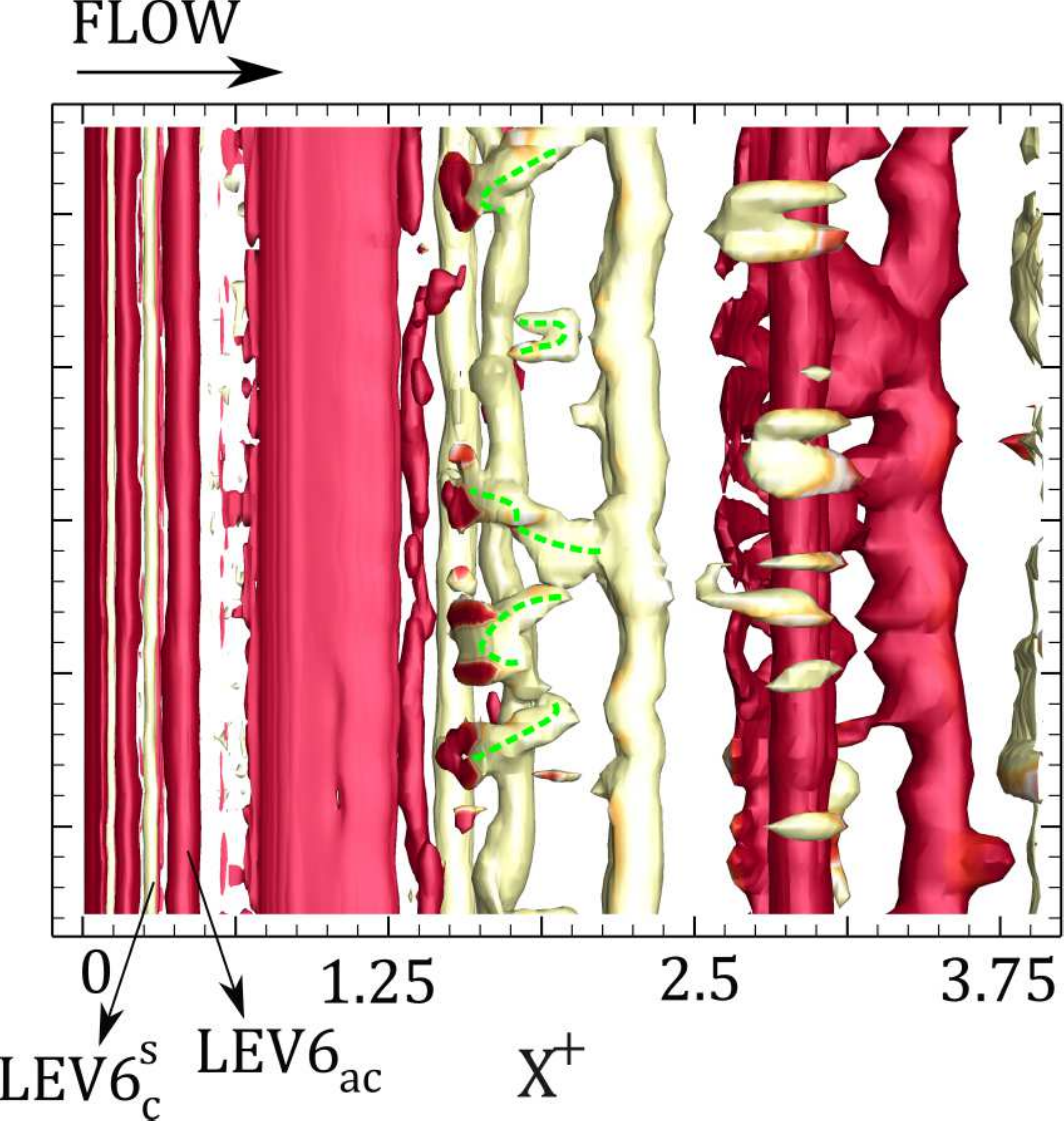}%
		}
	\end{minipage}\hfill
	\begin{minipage}{0.25\textwidth}
		\centering
		\subcaptionbox{\hspace*{-2.75em}}{%
			\hspace{-0.05in}\includegraphics[width=3.2cm,height=3.4cm]{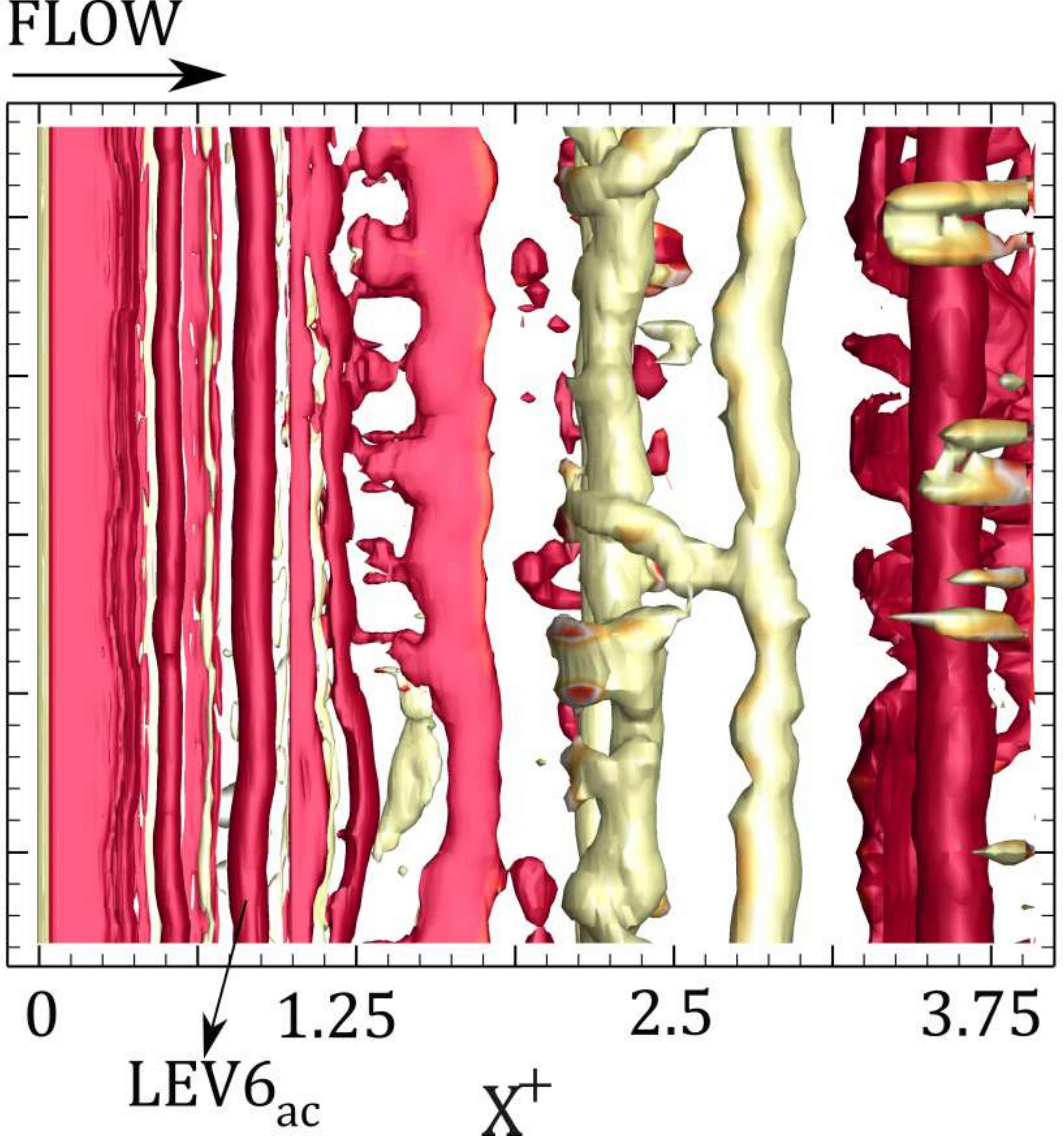}%
		}
	\end{minipage}\hfill
	\begin{minipage}{0.25\textwidth}
		\centering
		\subcaptionbox{\hspace*{-2.75em}}{%
			\hspace{-0.05in}\includegraphics[width=3.2cm,height=3.4cm]{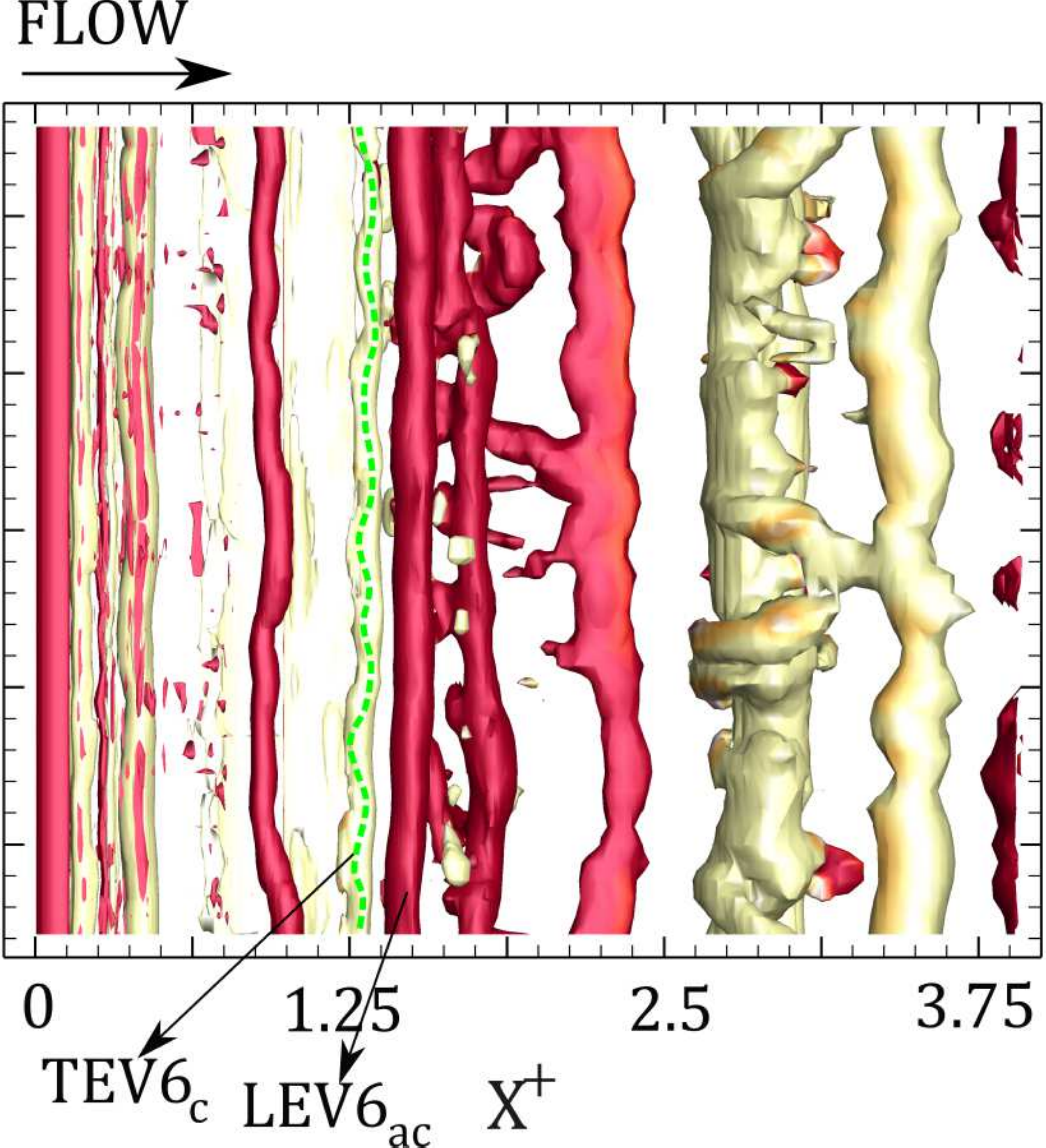}%
		}
	\end{minipage}\\
	\caption{Wakes corresponding to (a-d) $\phi=$ 180$^\circ$ and (e-h) $\phi=$ 225$^\circ$, at $St_{c}=$ 0.40. The oscillation time varies from $t^+=$ (a,b) 0 to (d,h) 0.75. Each stage is represented using iso-surfaces of $\lambda_2(= -0.05)$, which are colored based on $|\omega_{z}^+|=$ 5.}
	\label{fig:Lambda2_f05}
\end{figure} 

Based on the observations discussed at $St_c=$ 0.40, it is evident that the wake topology at a decreased heave domination (i.e. at $\phi=$ 225$^\circ$) is now characterized by growth of secondary hairpin-like structures through mechanism $``2"$. The wake at $\phi=$ 270$^\circ$, however, remains devoid of any secondary hairpin-like formation (wake evolution not shown here for brevity). {On comparing with the results discussed at $St_c=$ 0.32, we also observe a consistent transition of growth mechanisms which concern evolution of secondary hairpin-like structures. The transition follows a route characterized by mechanism $``1"$ at the beginning of heave domination (i.e. at $\phi=$ 90$^\circ$). This is followed by mechanism $``2"$ during peak heave domination at $\phi=$ 180$^\circ$. As the heave domination begins to decrease ($\approx$ 225$^\circ$), mechanism $``2"$ is observed to be persistent which is then succeeded by an absence of secondary structure at the onset of pitch domination (at $\phi=$ 270$^\circ$). We will next evaluate the applicability of this characteristic transition route at higher $St_c$.}

\subsubsection{$St_c=$ 0.48}

\begin{figure}
	\centering
	\begin{minipage}{0.5\textwidth}
		\centering
		\hspace{-0.15in}\includegraphics[width=5.0cm,height=1.0cm]{Omega_Legend-eps-converted-to.pdf}%
	\end{minipage}\\
	\vspace{0.1in}
	\begin{minipage}{0.5\textwidth}
		\centering
		\subcaptionbox{\hspace*{-2.75em}}{%
			\hspace{-0.05in}\includegraphics[width=4.8cm,height=4.4cm]{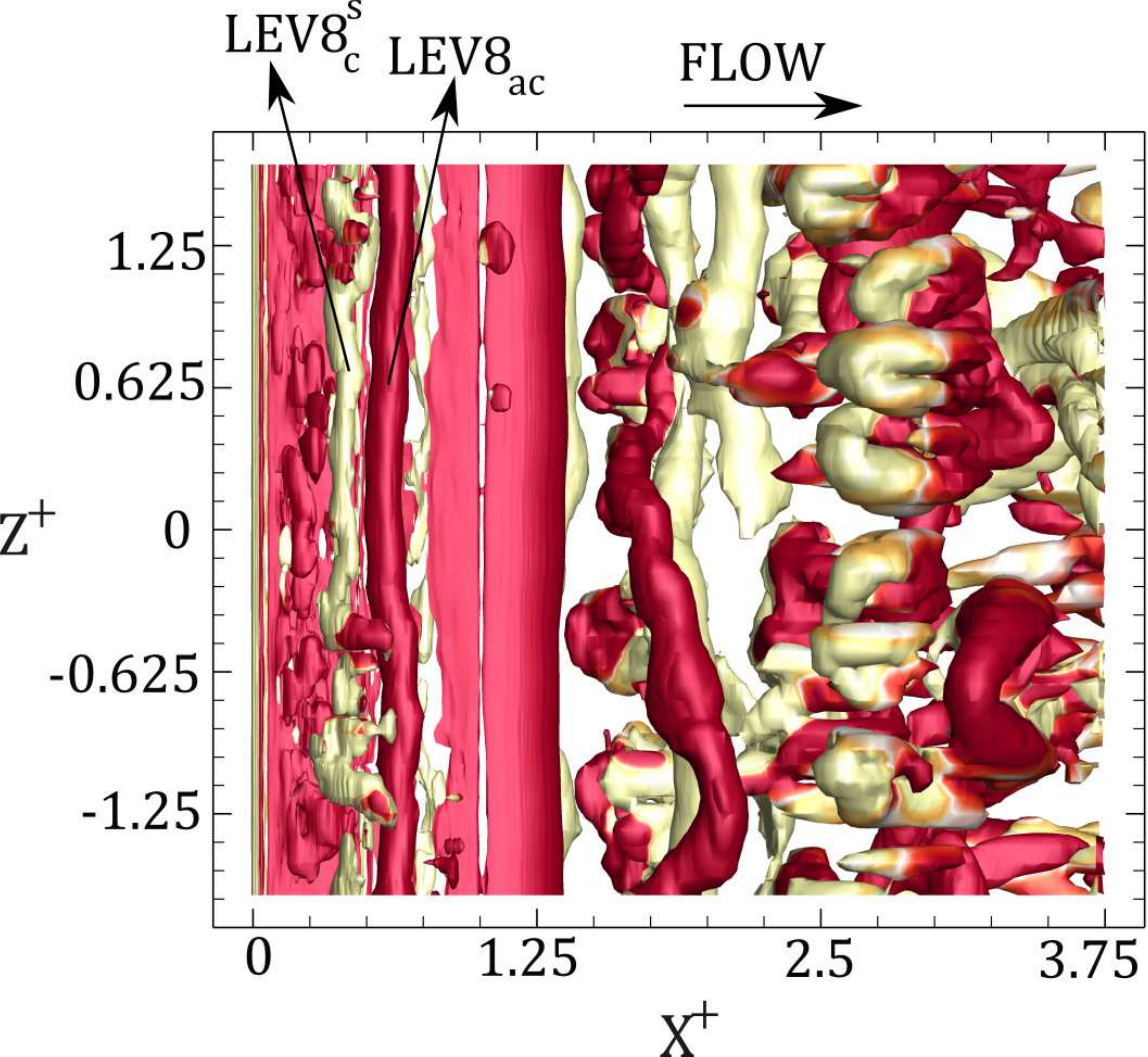}%
		}
	\end{minipage}\hfill
	\begin{minipage}{0.5\textwidth}
		\centering
		\subcaptionbox{\hspace*{-2.75em}}{%
			\hspace{-0.05in}\includegraphics[width=4.8cm,height=4.4cm]{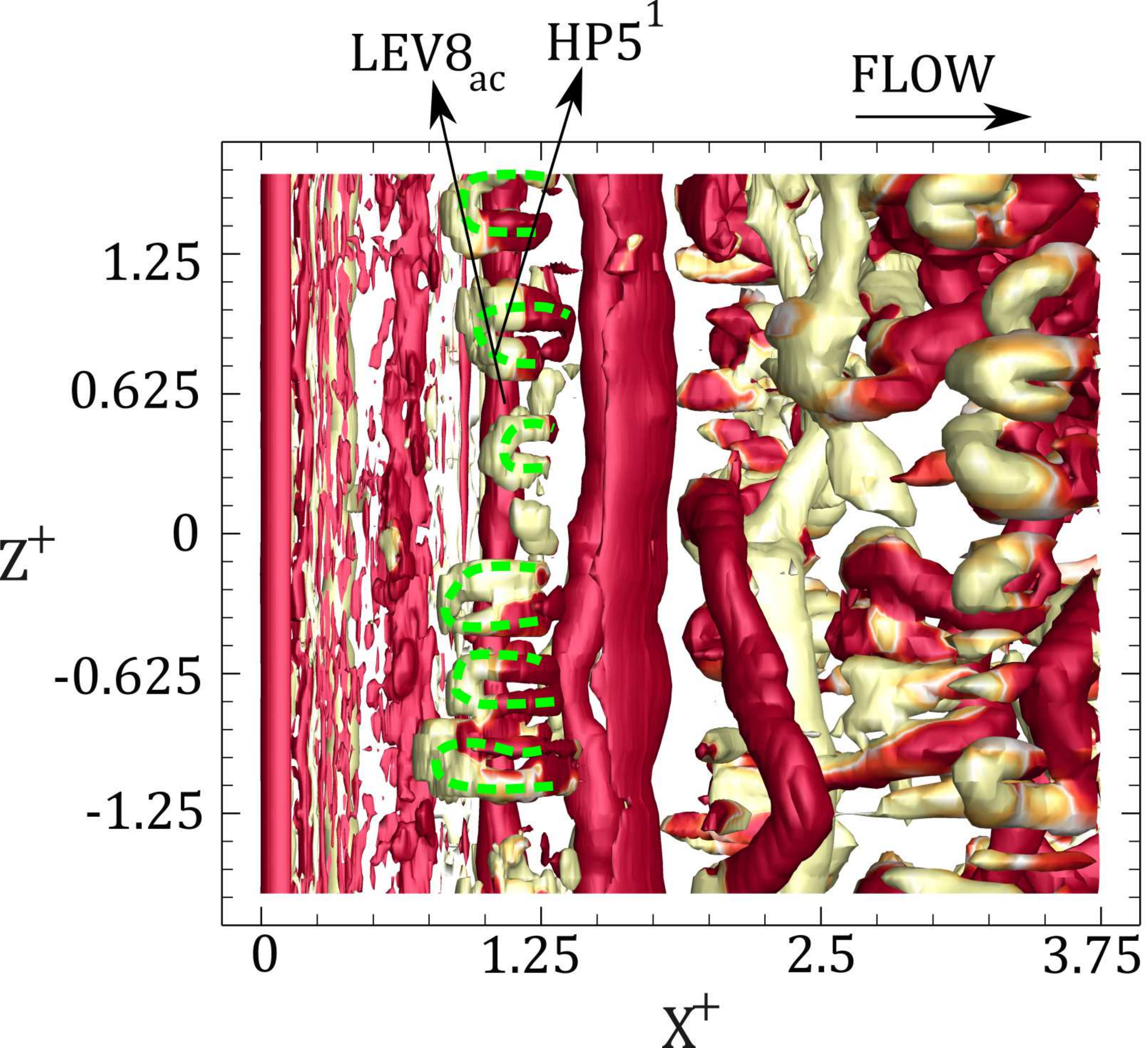}%
		}
	\end{minipage}\\
	\begin{minipage}{0.5\textwidth}
		\centering
		\subcaptionbox{\hspace*{-2.75em}}{%
			\hspace{-0.05in}\includegraphics[width=4.8cm,height=4.4cm]{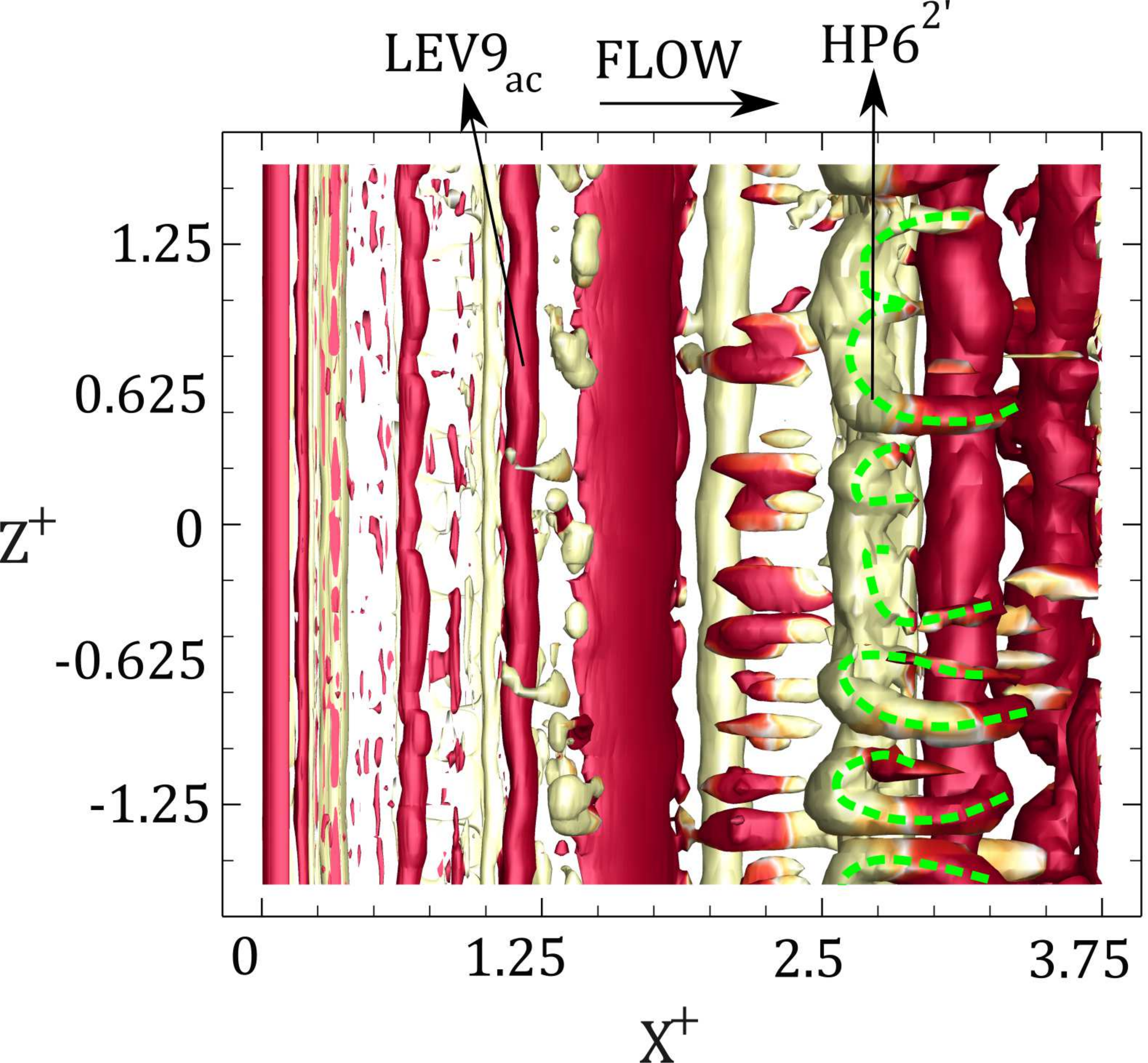}%
		}
	\end{minipage}\hfill
	\begin{minipage}{0.5\textwidth}
		\centering
		\subcaptionbox{\hspace*{-2.75em}}{%
			\hspace{-0.05in}\includegraphics[width=4.8cm,height=4.4cm]{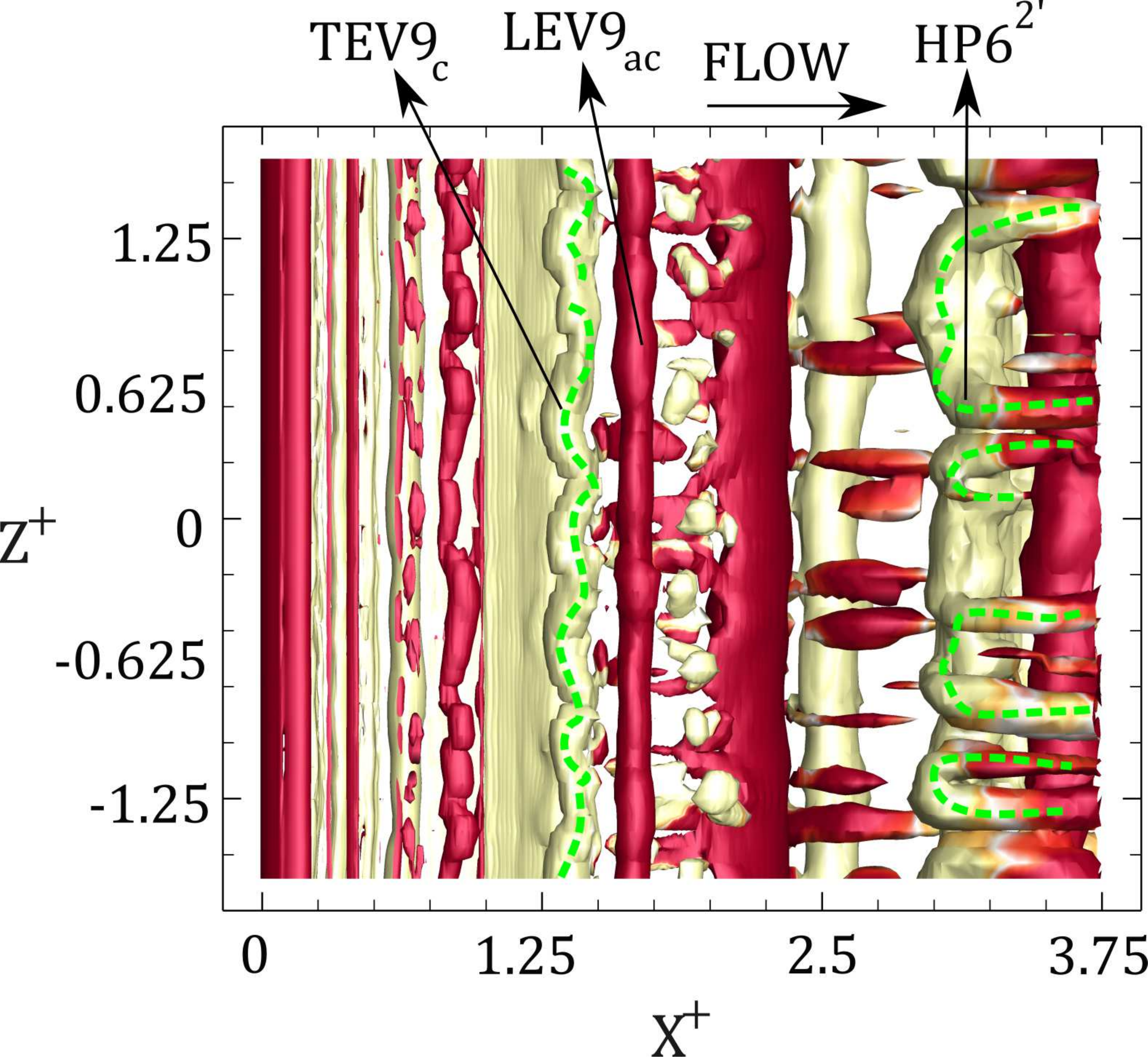}%
		}
	\end{minipage}\\
	\begin{minipage}{0.5\textwidth}
		\centering
		\subcaptionbox{\hspace*{-2.75em}}{%
			\hspace{-0.05in}\includegraphics[width=4.8cm,height=4.4cm]{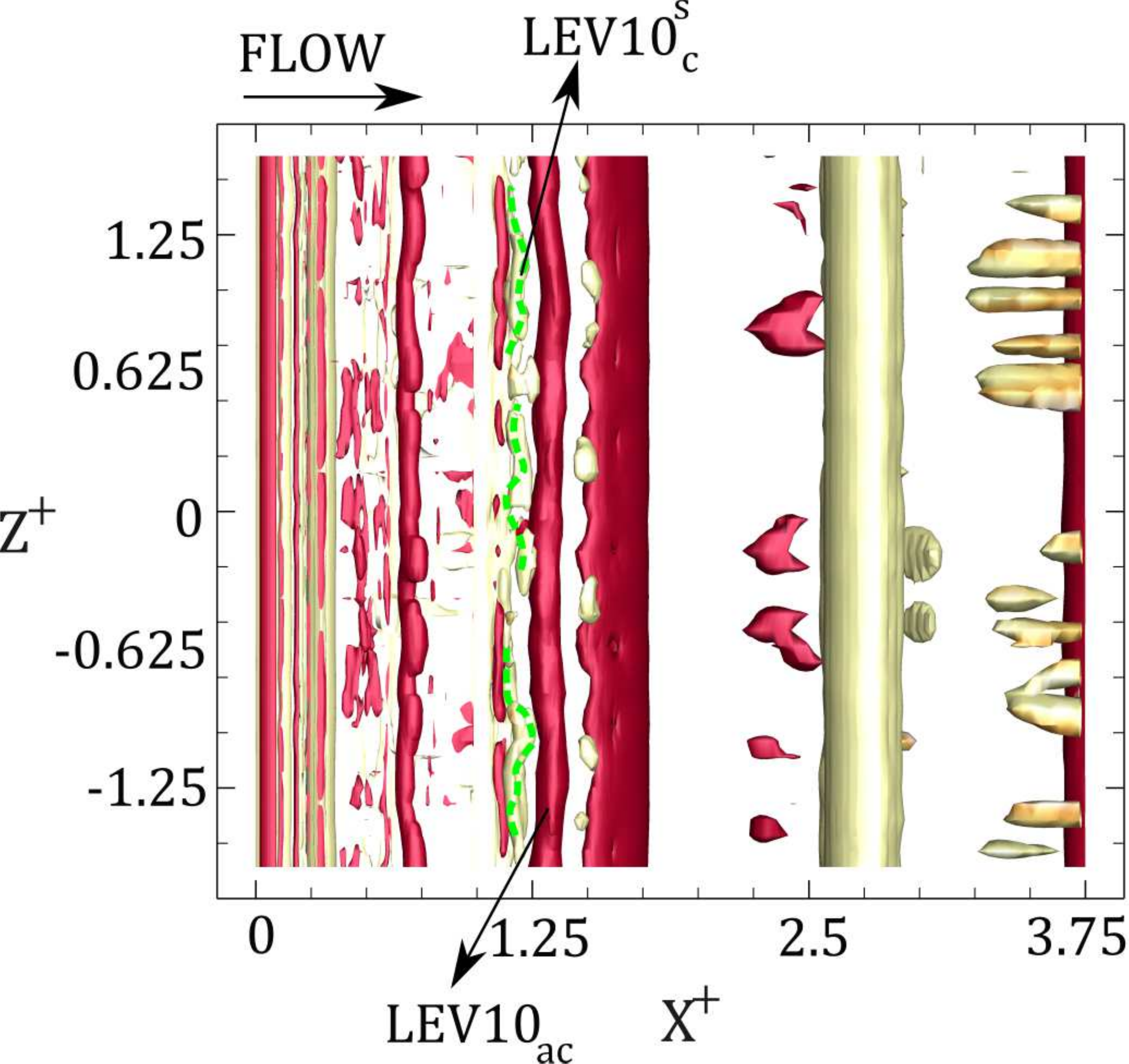}%
		}
	\end{minipage}\hfill
	\begin{minipage}{0.5\textwidth}
		\centering
		\subcaptionbox{\hspace*{-2.75em}}{%
			\hspace{-0.05in}\includegraphics[width=4.8cm,height=4.4cm]{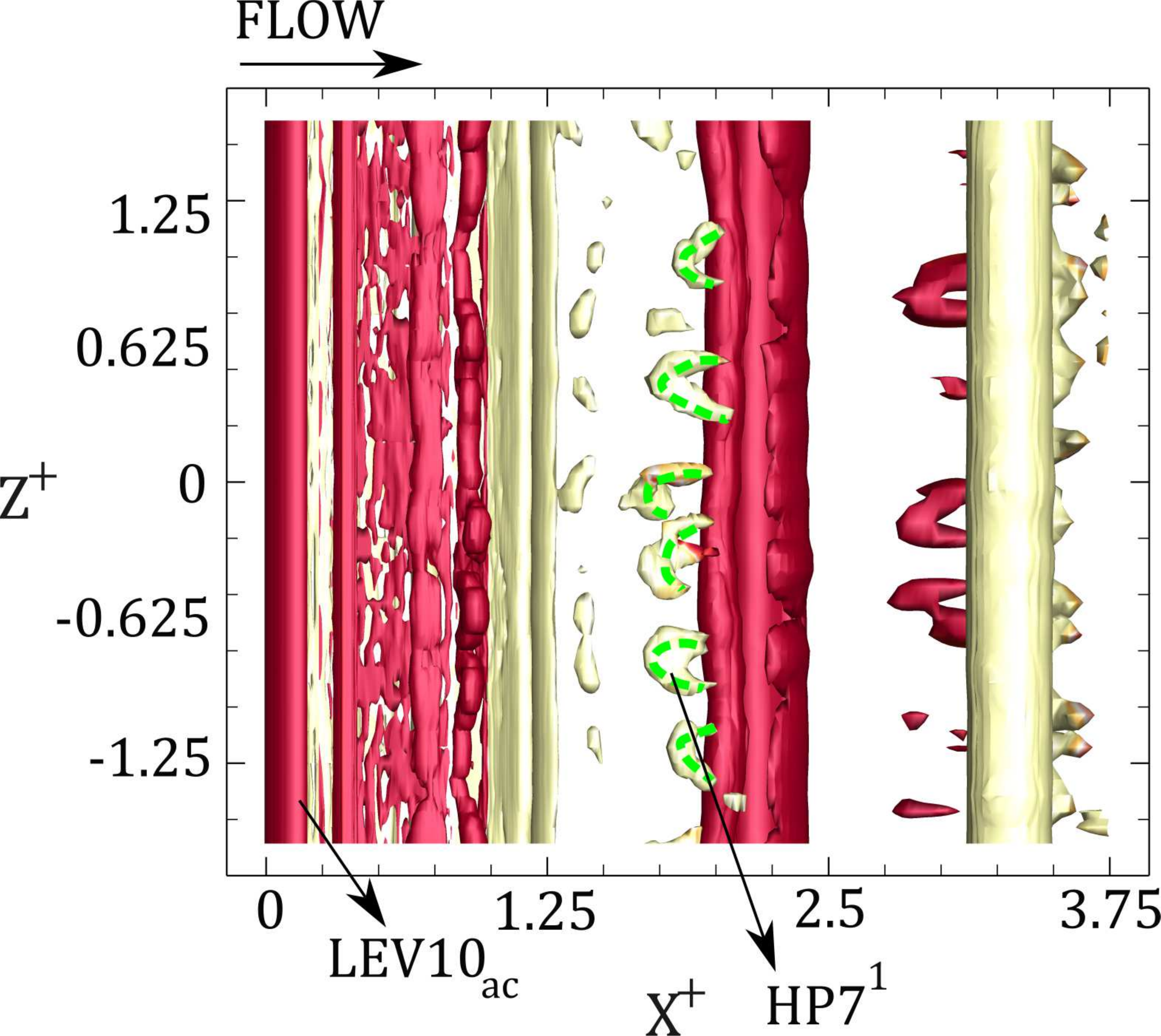}%
		}
	\end{minipage}\\
	\caption{Wakes corresponding to (a,b) $\phi=$ 180$^\circ$, (c,d) $\phi=$ 180$^\circ$, and (e,f) $\phi=$ 270$^\circ$, at $St_c =$ 0.48. The oscillation time corresponds to $t^+=$ 0.5 (a,c,e) and 0.75 (b,d,f), respectively. Each stage is represented using iso-surfaces of $\lambda_2(= -0.05)$, which are colored based on $|\omega_{z}^+|=$ 5.}
	\label{fig:Lambda2_f06}
\end{figure} 

An increase in $St_c$ to 0.48 coincides with the secondary hairpin-like formation within the entire $\phi$ range (90$^\circ$ to 270$^\circ$), characterized by a transition from heave domination to an onset of pitch-dominated kinematics. At $\phi=$ 90$^\circ$, mechanism $``1"$ contributes towards the secondary hairpin-like formation (see supplementary videos), similar to the observations at a lower $St_c$. Figure \ref{fig:Lambda2_f06} provides instantaneous three-dimensional wake topologies at an increasing $\phi$ from 180$^\circ$ to 270$^\circ$. The time instants were chosen at $t^+ =$ 0.5 and 0.75, respectively. At $\phi=$ 180$^\circ$ (see Figures \ref{fig:Lambda2_f06}(a) and \ref{fig:Lambda2_f06}(b)), a paired secondary ($LEV8_{c}^s$) and primary ($LEV8_{ac}$) roller configuration is evident near the trailing edge of the foil (at $t^+=$ 0.5 in Figure \ref{fig:Lambda2_f06}(a)). A subsequent evolution of hairpin arrangement ($HP5^1$) is observable at $t^+ =$ 0.75 in Figure \ref{fig:Lambda2_f06}(b), which is now governed by mechanism $``1"$. Therefore, mechanism $``2"$, which governed the growth of secondary hairpin-like structures at lower $St_c(=$ 0.32 and 0.40), has now transitioned to mechanism $``1"$, characterized by an elliptic instability and subsequent outflux of streamwise vorticity through a secondary $LEV$. This also reflects that mechanism $``1"$ remains sustained at peak heave domination (i.e. at $\phi=$ 90$^\circ$ and 180$^\circ$), as $St_c$ increases to 0.48. 

{In order to identify if mechanism $``1"$ transitions to $``2"$, with decreasing heave domination ($\phi >$ 180$^\circ$), we now investigate the wake at $\phi=$ 225$^\circ$. This will also help establish if the consistent route of transition concerning the growth mechanisms of secondary structures, remains evident with increasing $St_c$, while kinematics change from heave to domination.} Figures \ref{fig:Lambda2_f06}(c) and \ref{fig:Lambda2_f06}(d) provides another indication of qualitatively stronger secondary hairpin-like formation at $\phi=$ 225$^\circ$, which were either absent (at $St_c =$ 0.32), or remained sufficiently small in the wake (at $St_c=$ 0.40). The presence and growth of hairpin arrangement, marked by $HP6^{2\prime}$ in Figures \ref{fig:Lambda2_f06}(c) and \ref{fig:Lambda2_f06}(d) is attributed to mechanism $``2"$. $LEV-TEV$ pair is formed by structures $TEV9_c$ and $LEV9_{ac}$, where the elliptic instability features are well developed on the former roller. These features or dislocations outgrow to form secondary structures, i.e. $HP6^{2\prime}$. {Indeed, the growth mechanism transitions to $``2"$ as heave domination decreases at $\phi=$ 225$^\circ$, which was suggested earlier at $St_c <$ 0.48.}
%The presence of elliptic instability on $TEV9_{c}$ is clearly evident at $t^+=$ 0.75, which then leads to development of $HP_{5}^2$ structures downstream in the wake. 

Although there exists a consistent transition mechanism from $``1"$ to $``2"$, we notice that the wake at $\phi=$ 270$^\circ$ (Figure \ref{fig:Lambda2_f06}(e) and \ref{fig:Lambda2_f06}(f)) 
%still lacks dominant secondary structure growth. However, 
now depicts spanwise hairpin arrangement of substantially smaller spatial scales, while dominant secondary structure growth is still absent. In order to track the formation mechanism of these hairpin-like structures, Figure \ref{fig:Lambda2_f06}(e) shows the presence of primary ($LEV10_{ac}$) and secondary ($LEV10_c^{s}$) rollers in a paired configuration near the trailing edge. The coinciding elliptic instability induce the core vorticity outflux, which contributes to the process of mechanism $``1"$ in onset of secondary hairpins (marked as $HP7^1$) evident in Figure \ref{fig:Lambda2_f06}(f). 

Hence, it is interesting to note that with the onset of pitch-dominated kinematics at $St_c=$ 0.48, mechanism $``1"$ starts to dominate the wake evolution rather than $``2"$. A possible reasoning is increasing strength and timing for development of $TEV$ structures at $\phi=$ 270$^\circ$, since the kinematics coincide with a relative increase in trailing edge amplitude compared to the leading edge \cite{33}. During this prolonged time for  development of $TEVs$, the primary $LEV$ and small scale hairpin-like structures (through mechanism $``1"$) advect past the trailing edge, thereby failing to experience any significant interaction with $TEVs$. 

\subsubsection{$St_c=$ 0.56}

\begin{figure}
	\centering
	\begin{minipage}{0.5\textwidth}
		\centering
		\hspace{-0.15in}\includegraphics[width=5.0cm,height=1.0cm]{Omega_Legend-eps-converted-to.pdf}%
	\end{minipage}\\
	\vspace{0.1in}
	\begin{minipage}{0.3\textwidth}
		\centering
		\subcaptionbox{\hspace*{-2.75em}}{%
			\hspace{-0.00in}\includegraphics[width=4cm,height=3.4cm]{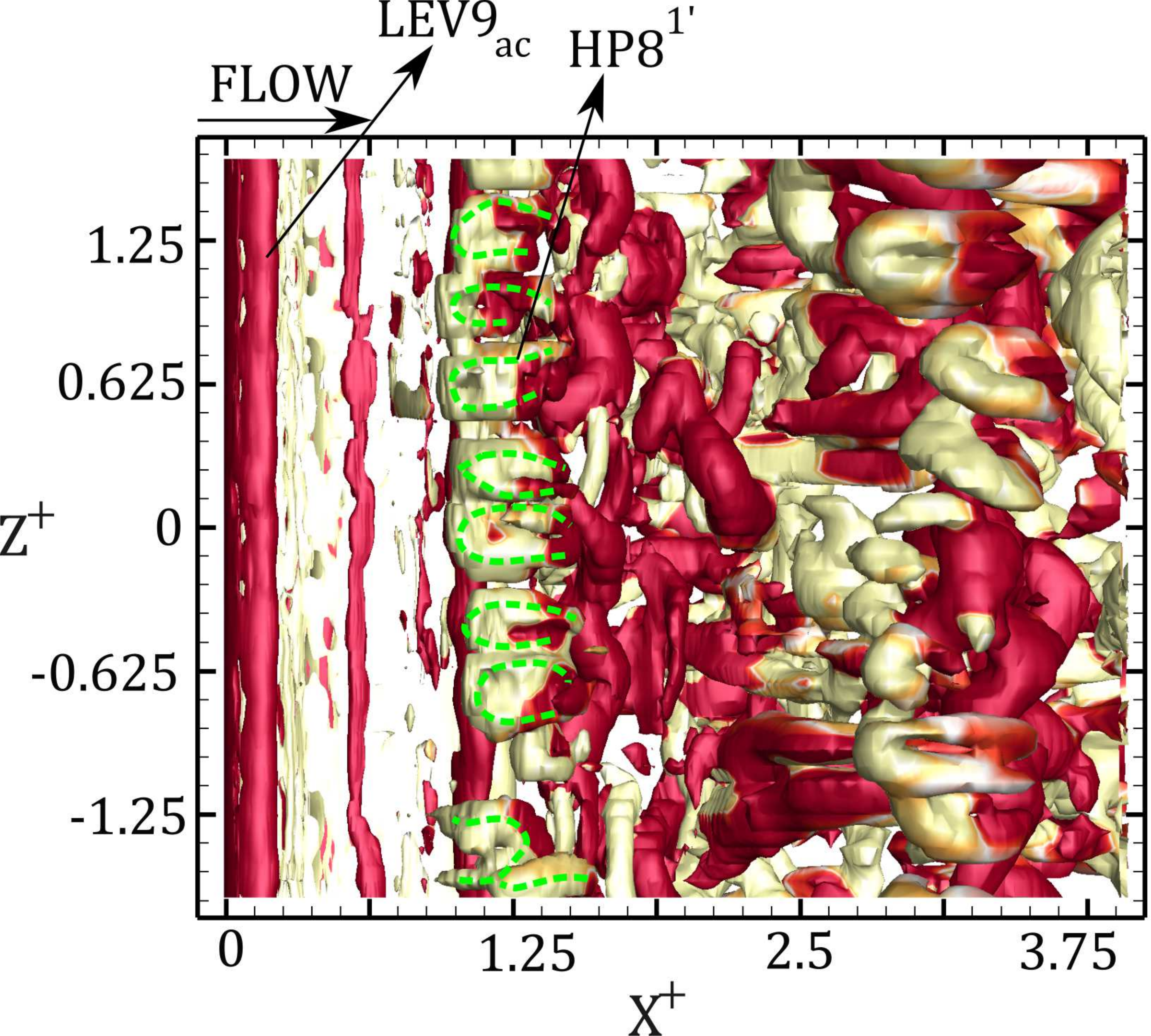}%
		}
	\end{minipage}\hfill
	\begin{minipage}{0.3\textwidth}
		\centering
		\subcaptionbox{\hspace*{-2.75em}}{%
			\hspace{-0.00in}\includegraphics[width=4cm,height=3.4cm]{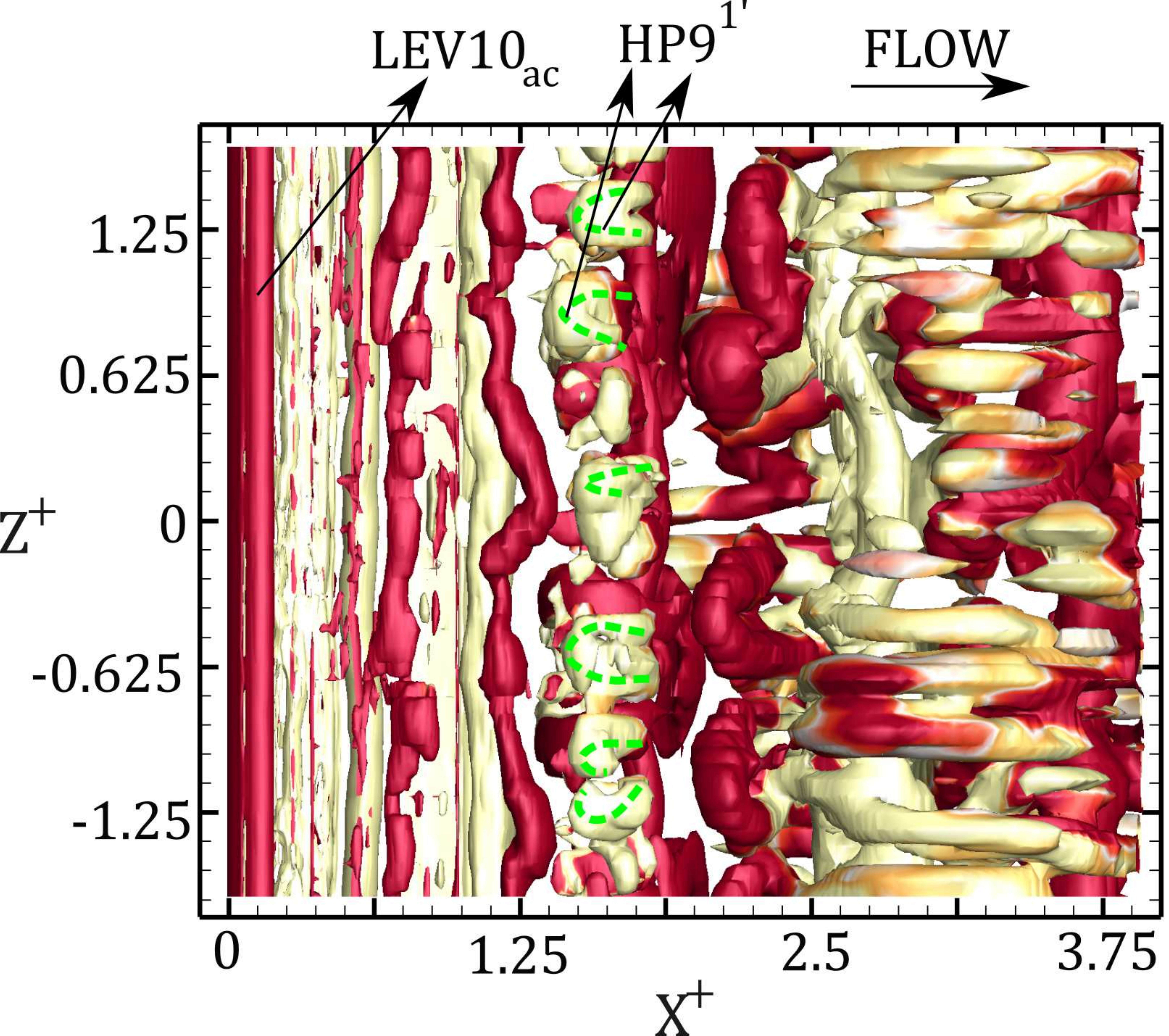}%
		}
	\end{minipage}\hfill
	\begin{minipage}{0.3\textwidth}
		\centering
		\subcaptionbox{\hspace*{-2.75em}}{%
			\hspace{-0.00in}\includegraphics[width=4cm,height=3.4cm]{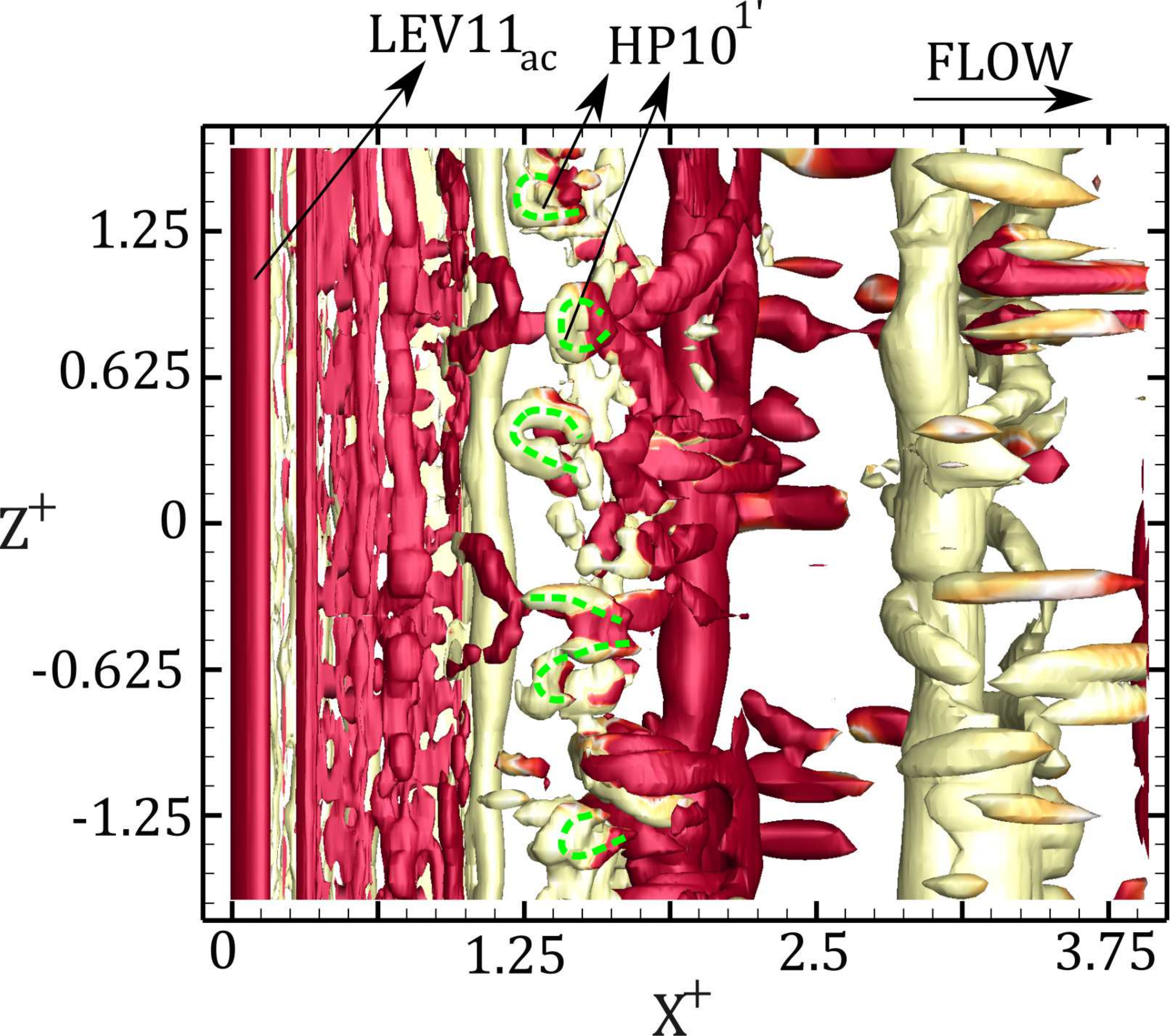}%
		}
	\end{minipage}\\
	\caption{Wakes corresponding to (a) $\phi=$ 90$^\circ$, (b) $\phi=$ 180$^\circ$ and (c) $\phi=$ 270$^\circ$ at $St_{c}=$ 0.56. The oscillation time corresponds to $t^+=$ 0.75. Each stage is represented using iso-surfaces of $\lambda_2(= -0.05)$, which are colored based on $|\omega_{z}^+|=$ 5.}
	\label{fig:Lambda2_f07}
\end{figure} 

As $St_c$ further increases to 0.56, wake topologies corresponding to $\phi=$ 180$^\circ$-270$^\circ$ now demonstrate dominant hairpin-like growth (see Figure \ref{fig:Lambda2_f07}). Besides $\phi=$ 90$^\circ$, the transition from heave- to the onset of pitch-dominated kinematics coincides with the presence of secondary hairpin-like structures along the spanwise direction (marked as $HP8^{1^\prime}$, $HP9^{1^\prime}$ and $HP10^{1^\prime}$ for $\phi=$ 180$^\circ$, 225$^\circ$, and 270$^\circ$, respectively). The origin of these secondary hairpin-like structures is also entirely attributed to mechanism $``1"$. 

As we compare the findings at higher $St_c$ (i.e. at 0.48 and 0.56), it is clear that  while the route to transition from mechanism $``1"$ to $``2"$ remains persistent, a simultaneous progression of the secondary structure growth also occurs for $\phi$. This coincide with a weaker heave domination. This characteristic was not observable at $St_c <$ 0.48. Also, mechanism $``1"$ completely dominates the range of $\phi$ at $St_c >$  0.48. This suggests that the dependence of mechanisms $``1"$ and $``2"$ on transitioning nature of kinematics is lost at $St_c =$ 0.56.

\subsection{Representative transition models on $\phi-St_c$ phase-space}

The increase in $\phi$ and $St_c$ depicted transition of mechanisms that contributed to the growth of secondary hairpin-like structures. It is useful to map these growth mechanisms on a $\phi-St_c$ phase-space diagram such that a consistent pattern in their apparent transition can be established. Figure \ref{fig:Phase_Space} shows a schematic that lays out the mechanisms observed at varying $\phi$ and $St_c$ in the previous section. Two patterns are evident. The first demonstrates that the transition from mechanism $``1"$ to an absence of the secondary wake structures occurs via an intermediate stage that is governed by mechanism $``2"$. We term this transition model as Series $A$, which either coincides with a change in kinematics from heave domination (at $\phi=$ 90$^\circ$) to the onset of pitch dominated motion (at $\phi =$ 270$^\circ$), or a decrease in $St_c$. Note that the absence of secondary wake structures for Series A model, corresponding to $\phi=$ 180$^\circ$, is not captured within the range of  decreasing $St_c$ considered here, and may be expected at $St_c <$ 0.32.} 

Regarding the second pattern, we see that at the onset of pitch domination (i.e. $\phi=$ 270$^\circ$), there is a vivid transition from mechanism $``1"$ to an absence of secondary wake structures. However, in contrast to Series A model, this characterizes a lack of intermediate stage or mechanism $``2"$ within the decreasing range of $St_c$. Hence, we define this transition model as Series $B$. 
%Both models are well highlighted in Figure \ref{fig:Phase_Space}. %At $St_c$ lower than 0.48, an absence of secondary hairpin formation is also demonstrated in Figure \ref{fig:Phase_Space}, which specifically coincides with onset of pitch domination. At $St_c \ge$ 0.48, it is also clear that the secondary hairpin formation is prominent within the entire $\phi$ range, although, pairing mechanism $`1'$ dominates the formation.

\begin{figure}
\centering

%\subcaptionbox{\hspace*{-2.75em}}{%
\hspace{-0.20in}\includegraphics[width=12cm,height=7.24cm]{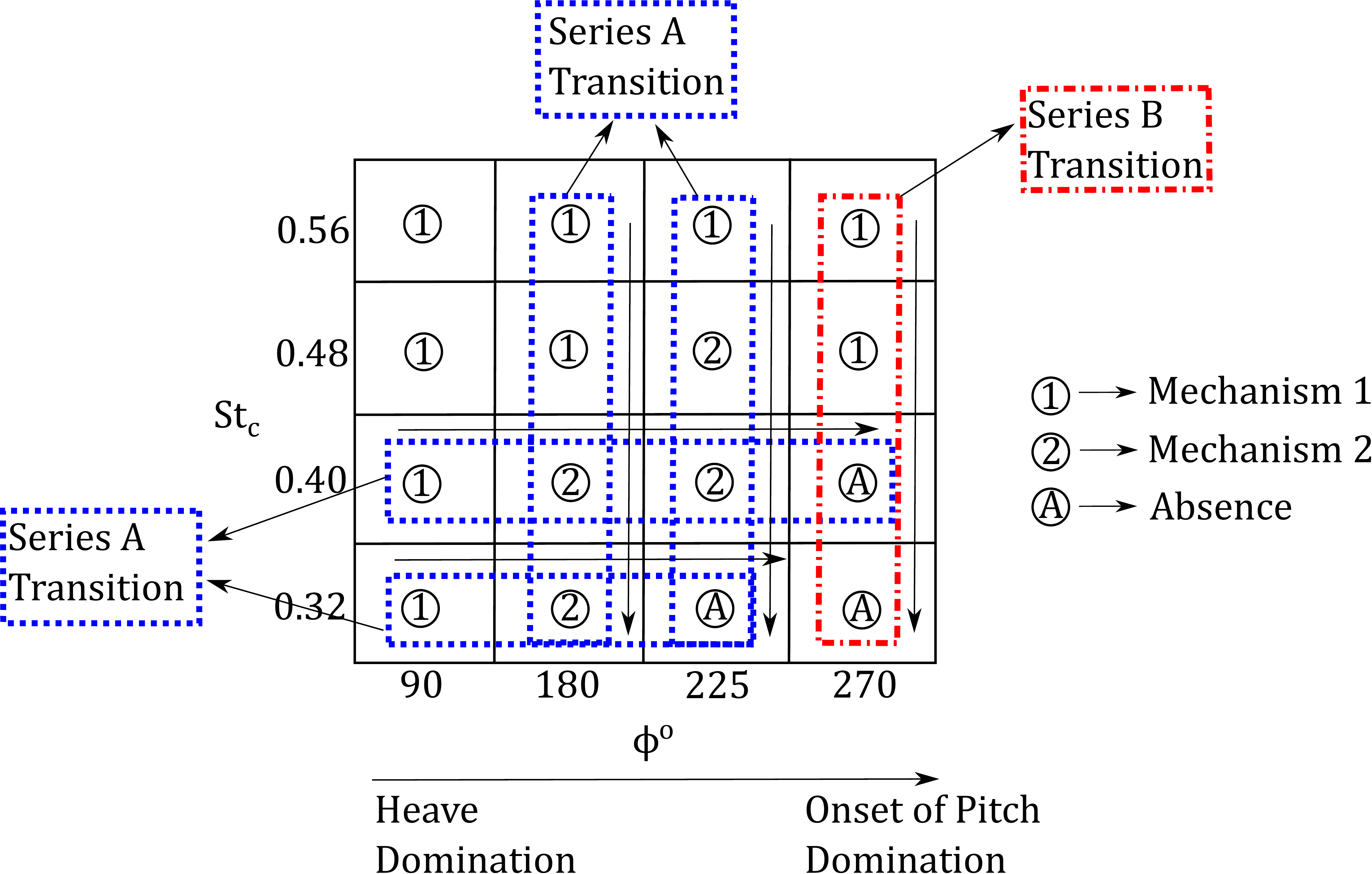}%
\caption{Phase-map representing dominance of growth mechanisms for secondary hairpin-like structures at increasing $\phi-St_c$.}
\label{fig:Phase_Space}
\end{figure} 

\subsubsection{Insights from Circulation}

The $\phi-St_c$ map in Figure \ref{fig:Phase_Space} reveals two major transition models that characterizes the change in growth of secondary wake structures from mechanism $``1"$ to either $``2"$, or directly to their absence. These patterns demand quantitative justification based on changes in the physical features of vortical structures that primarily govern the onset of secondary hairpins in the wake. The vortical structure that is pertinent to both mechanisms $``1"$ and $``2"$ is the primary $LEV$. We therefore discuss the quantitative circulation ($\Gamma$) variation of the primary $LEV$ within one oscillation cycle. The recent study by Verma $\&$ Hemmati \cite{57} confirmed that as kinematics transition from heave domination to an onset of pitch dominated regime (i.e. from $\phi=$ 90$^\circ$ to 270$^\circ$), a  decrease in $\Gamma^+$ for the primary $LEV$ was evident. This substantially reduces the intensity of core vorticity outflux through the interaction of foil boundary and primary $LEV$, thus leading to an absence of secondary wake structures. This justification supports the Series $A$ model represented at $St_c =$ 0.32 and 0.40, respectively. Here, we present supplementary $\Gamma^+$ assessments in order to provide physical reasoning behind Series $A$ and $B$ transition models observed with respect to a decreasing $St_c$ from 0.56 to 0.32 (see Figure \ref{fig:Phase_Space}), and for transitioning kinematics from heave to pitch domination.
%The dominance of secondary hairpin formation became high as $St_c$ increased above 0.32 for entire range of $\phi$ considered in this study. %This demands a closer looks at the quantitative circulation ($\Gamma$) variation of the primary $LEV$, which can provide an apparent reasoning behind consistent secondary hairpin formation at $St_c \ge$ 0.48. 

\begin{figure}
\centering
\begin{minipage}{0.5\textwidth}
	\centering
	\subcaptionbox{\hspace*{-2.75em}}{%
		\hspace{-0.00in}\includegraphics[width=6cm,height=4.54cm]{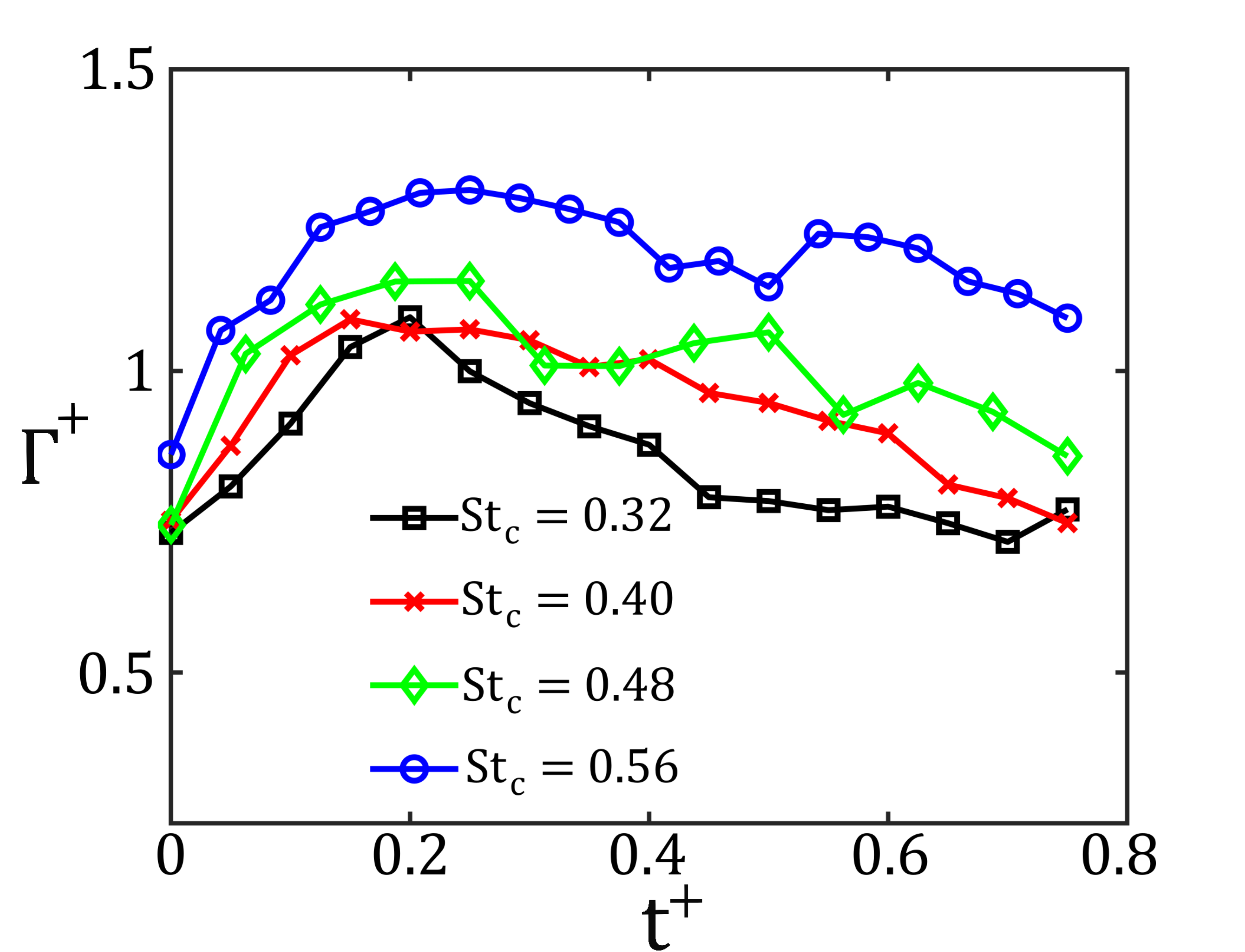}%
	}
\end{minipage}\hfill
\begin{minipage}{0.5\textwidth}
	\centering
	\subcaptionbox{\hspace*{-2.75em}}{%
		\hspace{-0.00in}\includegraphics[width=6cm,height=4.54cm]{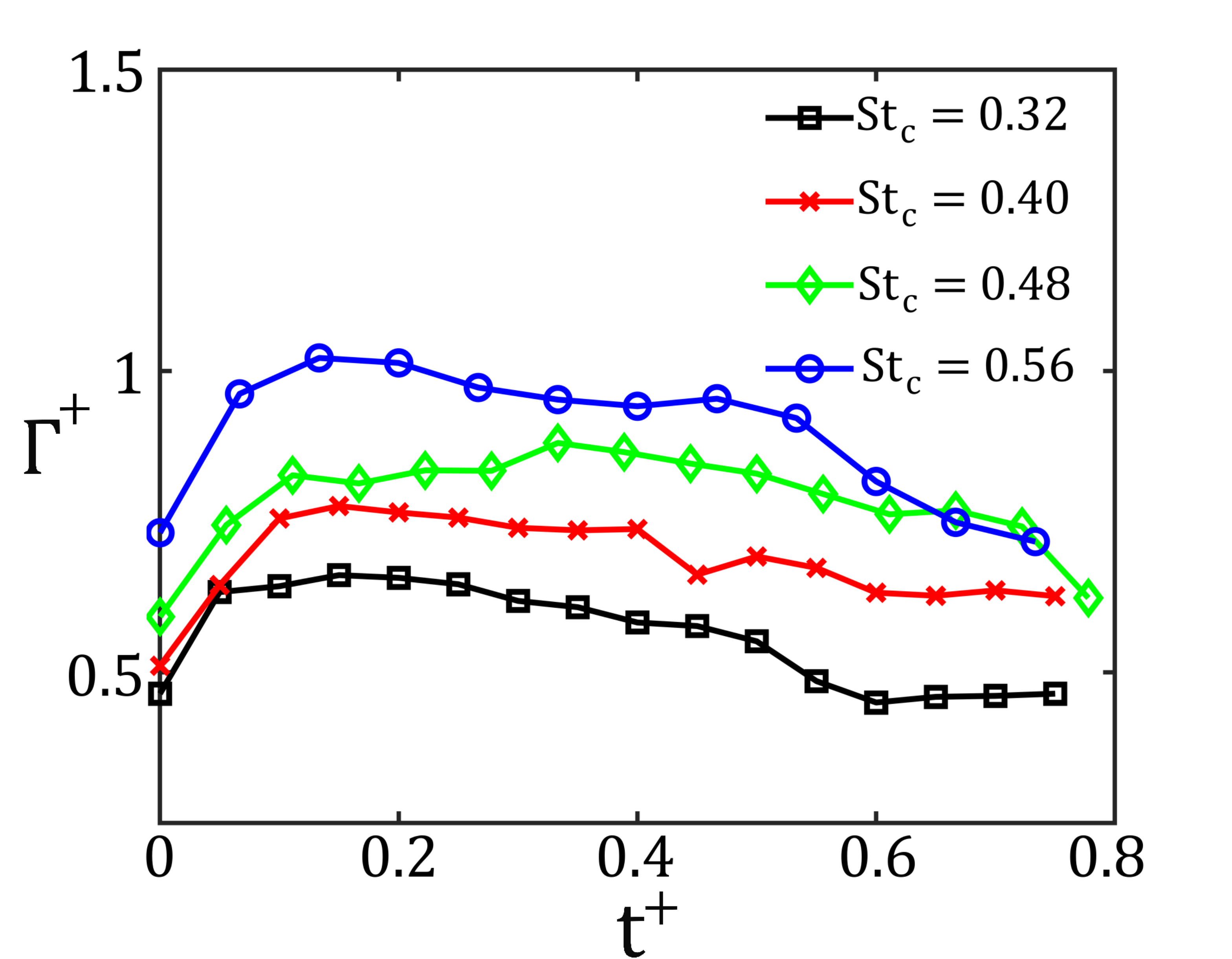}%
	}
\end{minipage}\\
\begin{minipage}{\textwidth}
	\centering
	\subcaptionbox{\hspace*{-2.75em}}{%
		\hspace{-0.00in}\includegraphics[width=6cm,height=4.54cm]{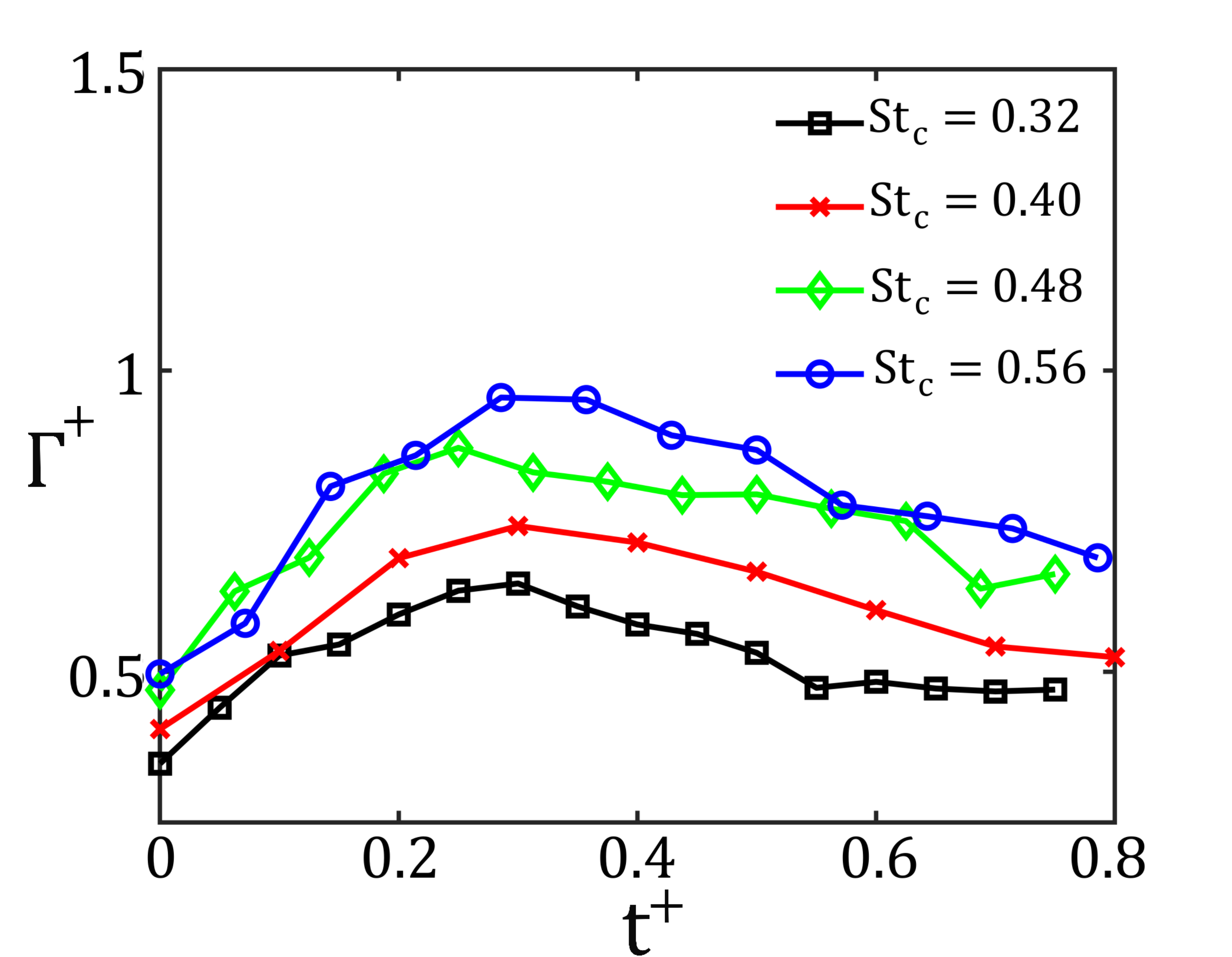}%
	}
\end{minipage}\\
\caption{Variation of $\Gamma^+$ at increasing $St_c$ within 1 oscillation cycle for $\phi$ corresponding to (a) 180$^\circ$, (b) 225$^\circ$ and (c) 270$^\circ$.}
\label{fig:Circ_All}
\end{figure}

Figure \ref{fig:Circ_All}(a-c) presents the instantaneous $\Gamma^+$ variation within one oscillation cycle corresponding to $\phi=$ 180 (Figure \ref{fig:Circ_All}(a)), 225$^\circ$(Figure \ref{fig:Circ_All}(b)) and 270$^\circ$(Figure \ref{fig:Circ_All}(c)). The variation of $\Gamma^+$ for $\phi=$ 90$^\circ$ is not shown since it depicts formation of secondary hairpin-like structures via mechanism $``1"$, across the entire $St_c$ range (see Figure \ref{fig:Phase_Space}). A consistent increase in peak $\Gamma^+$ magnitude of primary $LEV$ with increasing $St_c$ is noticeable despite the transition of kinematics from heave to pitch domination. %This nature is observed to be consistent across the $\phi$ range from 180$^\circ$ to 270$^\circ$. 
A stronger primary $LEV$ can therefore lead to a higher outflux of streamwise vorticity in two ways. The first way involves coherent streamwise vorticity outflux from secondary $LEV$ (i.e. mechanism $``1"$), while the second way present a strong enough interaction with the developing $TEV$ near the trailing edge. The latter involves secondary hairpin's evolution via mechanism $``2"$. This observation also remains consistent with recent literature \cite{32,51} that suggested a greater $LEV$ deformation for a higher $St_c$.

\section{CONCLUSIONS}
\label{Section:4}

The association between coupled kinematics and the wake three-dimensionality behind an oscillating foil is numerically investigated in this study. The foil kinematics are modified based on changing phase offset ($\phi$) between coupled heave and pitch motion, and an increasing reduced frequency ($St_c$) at $Re =$ 8000. The considered range of $\phi$ is largely representative of a heave-dominated kinematics, which also enables a dominant leading edge vortex formation that undergoes deformation on account of the spanwise instability. It concurrently plays a role in the evolution of secondary wake structures. 

At lower $St_c$, spanwise undulations of the primary roller remain evident due to the developed instability. 
%But, the quantitative estimation of its wavelength ($\lambda_{z}$) reflects substantial changes with increasing $\phi$ from 90$^\circ$ to 270$^\circ$. 
The wake corresponding to $\phi=$ 90$^\circ$ and 180$^\circ$, where the relative leading to trailing edge amplitude is greatest within the heave-dominated kinematics considered here\cite{33}, depicts 
%a shorter instability wavelength of the characteristic undulation ($\lambda_{z}<1$), along with 
a dominant growth of secondary hairpin-like structures. These are essentially an outcome of core vorticity outflux from either the formation of a secondary $LEV$, that neighbors a primary $LEV$ ($\phi=$ 90$^\circ$), or a $TEV$ interacting with a primary $LEV$ shed behind the foil ($\phi=$ 180$^\circ$).  Contrary to these cases, it is observed that at $\phi=$ 225$^\circ$ and 270$^\circ$, 
%the primary roller is characterized by a longer instability wavelength ($\lambda_{z}>1$), while 
the wake coincides with the absence of hairpin-like structures. This range of $\phi$ also coincides with a decreasing leading edge amplitude relative to the trailing edge, thus presenting the onset of pitch domination at $\phi=$ 270$^\circ$. This transition in mechanisms that govern the growth of secondary hairpin-like structures is consistent with increasing $St_c$. 

The transition in growth mechanisms is characterized by two major routes (Series $A$ and $B$) that reflect either a change in kinematics from heave to pitch domination, or a decrease in $St_c$. These routes confirm that a transition from mechanism $``1"$ to an absence of secondary hairpin-like structure either occurs directly or via an intermediate stage (i.e. mechanism $``2"$). Quantitative assessment on circulation of primary $LEVs$ indicates that the reduction in their strength coincides with a lower possibility of core vorticity outflux, as $\phi$ increases from 90$^\circ$ to 270$^\circ$. This explains the disappearance of secondary wake structures for $\phi=$ 225$^\circ$ and 270$^\circ$ at lower $St_c$. However, as $St_c$ increases, the higher $\Gamma^+$ associated with primary $LEV$ promotes the formation of secondary hairpin-like structures within the entire range of $\phi$. Overall, a close association between the kinematics, evolving vortex instability, and growth of secondary wake structures is observed for the first time in literature for oscillating foils in couple heaving and pitching kinematics. 

\section{ACKNOWLEDGMENTS}
This study has received support from the Future Energy Systems, and Compute Canada clusters were implemented for the required computational analysis.

\section{DATA AVAILABILITY}

The data that supports the findings of this research are available within the article.

% If in two-column mode, this environment will change to single-column format so that long equations can be displayed. 
%\begin{widetext}
%$$\mbox{put long equation here}$$
%\end{widetext}

% Figures should be put into the text as floats. 
% Use the graphics or graphicx packages (distributed with LaTeX2e).
% Put the label that you will use with \ref{} command in the braces of the \label{} command.
%
% \begin{figure}
% \end{figure}

% \begin{table}
% \end{tabular}
% \end{table}

%\begin{acknowledgments}
% Put your acknowledgments here.
%\end{acknowledgments}


\end{document}